\pdfoutput=1 

\documentclass[a4paper, 11pt, final]{article}

\usepackage{amssymb} % provides various useful mathematical symbols
\usepackage{amsthm} % provides extended theorem environments
\usepackage{amsmath,empheq}
\usepackage{bbm}

\DeclareMathAlphabet{\mathcal}{OMS}{cmsy}{m}{n}
\DeclareMathAlphabet\mathbfcal{OMS}{cmsy}{b}{n}

\DeclareFontFamily{U}{dutchcal}{\skewchar\font=45 }
\DeclareFontShape{U}{dutchcal}{m}{n}{<-> s*[1.0] dutchcal-r}{}
\DeclareFontShape{U}{dutchcal}{b}{n}{<-> s*[1.0] dutchcal-b}{}
\DeclareMathAlphabet{\mathcald}{U}{dutchcal}{m}{n}
\SetMathAlphabet{\mathcald}{bold}{U}{dutchcal}{b}{n}
\DeclareMathAlphabet\mathcalz{T1}{pzc}{mb}{it}

\usepackage[nice]{nicefrac} % nicer-looking fractions
\usepackage{siunitx} % standard units, e.g., scientific notation

\usepackage{newtxtext, newtxmath} % change font to Adobe Times New Roman

\usepackage{anyfontsize}
\usepackage[utf8]{inputenc}
\usepackage[T1]{fontenc}
\usepackage{microtype} % Slightly tweak font spacing for aesthetics

\providecommand{\JEL}[1]{\textit{\textbf{JEL: }} #1}
\providecommand{\keywords}[1]{\textbf{\textit{Keywords--- }} #1}

\usepackage{setspace} 

\usepackage{parskip} % each new line automatically spaces previously paragraph correctly
\usepackage{etoolbox}
\usepackage{titlesec}
\titleformat{\section}{\normalfont\Large\bfseries}{\thesection}{1em}{}
\titleformat{\subsection}{\normalfont\large\bfseries}{\thesubsection.}{1em}{}
\titleformat{\subsubsection}{\normalfont\normalsize\itshape}{\thesubsubsection.}{1em}{}

\usepackage{abstract}
\renewenvironment{abstract}
 {\normalfont
  \begin{center}
  \bfseries \abstractname\vspace{-.5em}\vspace{0pt}
  \end{center}
  \list{}{
    \setlength{\leftmargin}{0cm}%
    \setlength{\rightmargin}{\leftmargin}%
  }%
  \item\relax}
 {\endlist}

\usepackage{authblk} % package for author affiliations
\usepackage[bottom]{footmisc} % Makes footnotes stick to bottom of the page

\usepackage{multicol} % for multi-tab enumerated lists

\usepackage{enumitem} % for more flexibility in lists; see https://mirror.ufs.ac.za/ctan/macros/latex/contrib/enumitem/enumitem.pdf

\usepackage{graphicx} % More advanced figure inclusion
\usepackage{float} % For specifying table/figure locations, i.e. [ht!]
\usepackage{subcaption}
\usepackage{afterpage} % For encapsulating a floating figure on a single page   

\usepackage{printlen}

\usepackage[labelfont=bf]{caption}
\usepackage{color, colortbl}
\definecolor{LightGray}{rgb}{0.93,0.914,0.914}    
\usepackage{soul} % for highlighting stuff
\usepackage{longtable,rotating} % For long tables that span multiple pages
\usepackage{booktabs} % used for making more professional-looking tables
\usepackage{multirow} % used for cell-merging in complex tables
\usepackage{arydshln} % for drawing other line types in tables
\usepackage{bigdelim} % for curly braces; see https://tex.stackexchange.com/questions/164546/rotate-text-next-to-right-brace-in-multirow-environment

\usepackage[most]{tcolorbox}            % for boxed environments in colour

\PassOptionsToPackage{hyphens}{url} % for hyphenating URLs

\usepackage{fancyhdr}
\fancyheadoffset{0cm}
\makeatletter
\newcommand{\quickwordcount}[1]{
  \immediate\write18{texcount -quiet -incbib -sub=none -utf8 -1 -sum -merge -encoding=utf8 #1.tex > #1-words}%
  \immediate\openin\somefile=#1-words
  \read\somefile to \@@localdummy
  \immediate\closein\somefile
  \setcounter{wordcounter}{\@@localdummy}
  \@@localdummy
}
\makeatother

\usepackage{silence}
\usepackage[style=apa,backend=biber,natbib,hyperref]{biblatex}
\DeclareLanguageMapping{british}{british-apa}
\defbibenvironment{bibliography}{\enumerate}{\endenumerate}{\item}

\usepackage[colorlinks=true,allcolors=gray]{hyperref} 
\let\orgautoref\autoref

\renewcommand{\autoref}[1]
{%
\def\equationautorefname{Eq.}%
\def\sectionautorefname{Sec.}%
\def\subsectionautorefname{Subsec.}%
\def\figureautorefname{Fig.}%
\def\subfigureautorefname{Fig.}%
\orgautoref{#1}%
}

\usepackage[nameinlink,capitalise]{cleveref}
\usepackage{algorithm,algpseudocode}

\makeatletter
\newlength{\trianglerightwidth}
\algnewcommand{\LineCommentCont}[1]{\Statex \hskip\ALG@thistlm%
  \parbox[t]{\dimexpr\linewidth-\ALG@thistlm}
{\leftskip=\algorithmicindent
  \hangindent=\algorithmicindent 
  \hangafter=1%
  \strut\makebox[\algorithmicindent][c]{$\triangleright$}#1\strut}
  } % \trianglerightwidth
\makeatother

\usepackage[left=1.8cm, right=1.8cm, bottom=2.5cm, top=2.5cm]{geometry}

\usepackage{listings}
\usepackage{xcolor}
\begin{document}

%TC:ignore

% more Hyperref options (must be after \begin{document}
\renewcommand{\figureautorefname}{Fig.}
\onehalfspacing

%--------------------------------------------------------%
%	TITLE PAGE
%--------------------------------------------------------%

%--------------------------------------------------------%
%	TITLE
%--------------------------------------------------------%

% Article title
\newcommand{\MainTitleText}{Deriving the term-structure of loan write-off risk under IFRS 9 by using survival analysis: A benchmark study}

\title{\fontsize{20pt}{0pt}\selectfont\textbf{\MainTitleText
}}

%--------------------------------------------------------%
%	AUTHORS
%--------------------------------------------------------%   

\author[,a,b]{\large Arno Botha \thanks{ ORC iD: 0000-0002-1708-0153; Corresponding author: \url{arno.spasie.botha@gmail.com}}}
\author[,a]{\large Mohammed Gabru \thanks{ ORC iD: ; email: \url{mogabru@gmail.com}}}
\author[,a]{\large Marcel Muller \thanks{ ORC iD: ; email: \url{marcelcelliers@gmail.com}}}
\author[,a,b]{\large Janette Larney \thanks{ ORC iD: 0000-0003-0091-9917; email: \url{janette.larney@nwu.ac.za}}}
\affil[a]{\footnotesize \textit{Centre for Business Mathematics and Informatics \& Unit for Data Science and Computing, North-West University, Potchefstroom, South Africa}}
\affil[b]{\footnotesize \textit{National Institute for Theoretical and Computational Sciences (NITheCS), Potchefstroom, South Africa}}
\renewcommand\Authands{, and }

% Today's date
 	%\date{Submitted: \usvardate\today}
    
%by specifying the below, we essentially "rewrite" the command \maketitle, which is normally called in main.tex.
%this is done primarily to abuse the \date command above

\makeatletter
\renewcommand{\@maketitle}{
    \newpage
     \null
     \vskip 1em%
     \begin{center}%
      {\LARGE \@title \par
      	\@author \par}
     \end{center}%
     \par
 } 
 \makeatother
 
 \maketitle

%--------------------------------------------------------%
%	ABSTRACT
%--------------------------------------------------------%    
{
    \setlength{\parindent}{0cm}
    \rule{1\columnwidth}{0.4pt}
    \begin{abstract}
        The estimation of marginal loan write-off probabilities is a non-trivial task when modelling the loss given default (LGD) risk parameter in credit risk. We explore two types of survival models in estimating the overall write-off probability over default spell time, where these probabilities form the term-structure of write-off risk in aggregate. These survival models include a discrete-time hazard (DtH) model and a conditional inference survival tree. Both models are compared to a cross-sectional logistic regression model for write-off risk. All of these (first-stage) models are then ensconced in a broader two-stage LGD-modelling approach, wherein a loss severity model is estimated in the second stage. In expanding the model suite, a novel dichotomisation step is introduced for collapsing the write-off probability into a 0/1-value, prior to LGD-calculation. A benchmark study is subsequently conducted amongst the resulting LGD-models. We find that the DtH-model outperforms other two-stage LGD-models admirably across most diagnostics. However, a single-stage LGD-model still had the best results, likely due to the peculiar `L-shaped' LGD-distribution in our data. Ultimately, we believe that our tutorial-style work can enhance LGD-modelling practices when estimating the expected credit loss under IFRS 9.
    \end{abstract}
     
     % Insert keywords here
     
     % Insert JEL codes here
     \JEL{C44, C63, G21.}
    
    \rule{1\columnwidth}{0.4pt}
}

\noindent Word count (excluding front matter and appendices):  16871 %\quickwordcount{ms} 

\subsection*{Disclosure of interest and declaration of funding}
\noindent This work is not financially supported by an institution or study grant, and has no conflicts of interest that may have influenced the outcome of this work.

%TC:endignore

%--------------------------------------------------------%
%	CONTENT
%--------------------------------------------------------%

\newpage

\section{Introduction}
\label{sec:intro}

% Introduce credit risk and ECL under IFRS 9
In banking, a fundamental task is the estimation of the loss associated with a borrower who may default, i.e., credit risk. The level (and evolution) of credit risk dictates the amount by which a financial asset ought to be adjusted regularly, as governed by the International Financial Accounting Standard (IFRS) 9 from the \citet{ifrs9_2014}. These regular adjustments are based on a statistical model of the asset's \textit{expected credit loss} (ECL). In turn, the ECL represents the probability-weighted sum of cash shortfalls that are expected to be forfeited over a specific time horizon; see \citet[\S 5.5.17--18, \S B5.5.28--31, \S B5.5.41--44]{ifrs9_2014}. This ECL-amount should be unbiased and be determined across a range of possible outcomes that can influence the asset's value over time. Furthermore, estimating the ECL should consider the time value of money, past events, current conditions, and forward-looking information (e.g., macroeconomic forecasts). Changes in the ECL-amount are then reserved within a central loss provision, which ultimately absorbs future write-off amounts.

% LGD & workout process
One of the fundamental risk parameters within an ECL-model is that of the \textit{loss given default} (LGD), which is the fraction of the outstanding balance at the default point that is expected to be lost. There are four common methods for estimating this LGD-quantity, as described by
\citet{schuermann2004we}, \citet[pp.~217-222]{VanGestel2009book}, and \citet[\S 10]{baesens2016credit}; though we shall focus on the \textit{workout} method that uses a bank's own internal data. This method aims to mimic the real-world resolution process through which defaulted loans typically progress, with the goal of nursing the strained relationship between bank and borrower back to health. 
\citet[pp.~11-13]{finlay2010book} and \citet[\S2.2]{botha2021Proc} describe a few remedial actions that, if successful, can induce a full recovery in a defaulted loan. In this case, the loan is said to have `cured' from default and a zero loss is typically assigned. However, these remedial actions might fail and the bank might be left with little choice but to initiate legal proceedings and/or foreclose on any available collateral towards recovering as much of the defaulted debt as soon as possible. The resulting receipts (or recoveries) are then offset against the last-known balance of the loan, whereafter any non-zero remainder is written off as a credit loss. As described by \citet{larney2025LGD_discount}, the workout process therefore culminates in either a cured or written-off outcome, with the remaining cases considered as unresolved (or right-censored). Considering write-offs, the workout method takes as input the resulting series of collected receipts, and calculates the discounted sum thereof over the workout period back to the default point. The percentage change between this discounted sum and the default balance is then defined as the empirical loss rate, or the actual LGD.

% Bimodality & skew; Two-stage LGD-modelling approach
As discussed and illustrated by \citet{schuermann2004we}, \citet{calabrese2010bank}, \citet{loterman2012benchmarking}, \citet[\S 10]{baesens2016credit}, and \citet{larney2025LGD_discount}, the distribution of the resulting loss rates typically have two defining characteristics. These include a heavy right-skewed tail and bimodality in the distribution, with one of the modes always centred at 0 due to zero-loss cures. The extent of the skew would directly depend on the prevalence of zero-loss cured outcomes, which is anecdotally about 70-80\% of resolved defaults for residential mortgages, as an example.
% Two-stage approach
These characteristics have inspired a particular modelling strategy called \textit{two-stage} LGD-modelling that has become quite popular; see \citet{leow2012LGD} and \citet{gurtler2013LGD} for an overview thereof. In this strategy, the LGD of a defaulted loan $i$ is decomposed into a write-off component $w_i$ and a loss severity component $l_i$, each forming a separate `stage'. The write-off component $w_i$ represents the probability of writing/charging off this $i$ given a few input variables $\boldsymbol{x}_i$. The loss severity component $l_i$ signifies the realised loss rate of $i$ in the event of write-off, i.e., the loss given write-off; itself also modelled as a function of input variables. Each component is then estimated separately using a particular modelling technique with input variables, e.g., logistic regression for $w_i$ and linear regression for $l_i$. Thereafter, the estimates of $w_i$ and $l_i$ are combined towards producing a final LGD-estimate, reconstituted as $\mathrm{LGD}=w_i\cdot l_i$.

% Write-off term structure
In this paper, we shall define the write-off component $w_i$ as a function of time in default, at the very least. For a defaulted loan $i$, let $t=\tau_d, \dots, \tau_r$ index the time in default from the default point $\tau_d$ up to the resolution time $\tau_r$. The collection of write-off probabilities for $i$ is then referred to as the \textit{term-structure} of write-off risk given inputs $\boldsymbol{x}_i$, denoted as $\boldsymbol{w}(\boldsymbol{x}_i)=\left\{ w\left(t,\boldsymbol{x}_i\right) \right\}_{t=\tau_d}^{\tau_r}$. Modelling this term-structure across many defaulted loans is the primary focus of this paper.
% Premise and problem formulation
The premise hereof is that write-off risk changes with $t$ in a non-linear fashion, as we shall demonstrate later. If this relationship is ignored, then the estimation of LGD may very well become inaccurate and biased. In turn, this inaccuracy can negatively affect the ECL-amount under IFRS 9, and attenuate a bank's loss provision. Given such an inaccurate ECL-amount, a bank may unnecessarily hold excess provisions (too high an ECL-amount), which poses an opportunity cost. On the other hand, a bank may also hold too little provisions (too low an ECL-amount), thereby risking insolvency. The aim of producing unbiased ECL-estimates under IFRS 9 therefore becomes a matter of prediction accuracy, which is a non-trivial task in LGD-modelling.

% Contribution
In this paper, our objective and main contribution is to explore and model this write-off probability using a few techniques towards building bespoke LGD-models. In particular, we shall investigate a particular class of modelling techniques called \textit{survival analysis}, which includes discrete-time hazard (DtH) models and conditional inference survival tree (ST) models. We compare both of these more dynamic survival techniques to a classical cross-sectional model for the write-off probability using logistic regression.
As far as we know, this particular analysis and the use of DtH and ST models for write-off risk have not yet been explored in literature, hence our contribution. For each model, we also introduce a novel dichotomisation step wherein the derived write-off probability is first collapsed into a 1 or a 0 value before calculating the LGD. Finally, we embed these models into a wider two-stage LGD-modelling approach, whereafter a loss severity model (given write-off) is built in the second stage. This loss severity model is itself estimated using a Tweedie compound Poisson \textit{generalised linear model} (GLM), following some experimentation, as will be defended later. The resulting two-stage LGD-models are then compared to a few simple single-stage GLM-based LGD-models in the interest of completeness.

% Diagnostics
Our model comparison uses a variety of diagnostics, which we believe further differentiates our benchmark study from others. These (reusable) diagnostics include time-dependent measures of discriminatory power, prediction accuracy, and the extent to which predictions agree with observations when aggregated in various ways. One of these aggregation ways include by default spell time $t$, thereby forming the overall term-structure of write-off risk; itself a rather unique view inspired by survival analysis. Lastly, we perform a distributional analysis between the empirical and expected distributions of the LGD-predictions emanating from each model.
% Data
In conducting our comparative study, we shall use a rich dataset of residential mortgage loans from a large South African bank. This dataset spans January 2007 up to December 2022, during which time mortgages were continuously originated, ultimately containing 653,317 loan accounts.
% Implication of work
We believe that our work can guide practitioners in modelling the aggregated term-structure of write-off risk over time, thereby enabling greater accuracy in LGD-modelling. In so doing, the central aim of IFRS 9 can be better realised in producing timeous and accurate ECL-estimates towards sizing a bank's loss provision.

% Layout paragraph
This paper is structured as follows. In \autoref{sec:background}, we summarise the literature on two-stage LGD-modelling, with a particular focus on using survival analysis as modelling technique. Basic survival modelling concepts are discussed and illustrated in \autoref{sec:method} towards formulating our setup mathematically, structuring our data, estimating the empirical write-off term-structure that our models shall try to recover, and specifying our models.
We then evaluate the various write-off risk models in \autoref{sec:results} across five main diagnostics. These results are complemented in \autoref{sec:results_LGD} by a distributional comparison of the downstream LGD-models, whereafter we conclude the study in \autoref{sec:conclusion}. 
% Appendices
Given its rather exotic nature, the basics of ST-models are discussed in the appendix. Other ancillary material in the appendix include the correct specification of survival data; an illustration of a sampling representativeness measure (i.e., the \textit{resolution rate}); a description of the selected input variables within our various models; and a short (but novel) optimisation procedure for dichotomising models that output probabilities. Finally, this work is accompanied by an open-source R-codebase, as maintained by \citet{gabru2026WriteOffSurvSourcecode}.

\section{Reviewing two-stage LGD-modelling and the use of survival analysis}
\label{sec:background}

% Two-stage approach
The two-stage approach to LGD-modelling has its genesis in the work of \citet{leow2012LGD}, who demonstrated the approach using residential mortgage data from the UK. In particular, the authors decomposed the LGD into the product of a \textit{repossession} model and a \textit{haircut} model. 
The repossession probability model estimates the likelihood of repossessing the property as a function of a few input variables, e.g., loan-to-value (LTV), time on book, type of security, and a previous default indicator. The haircut model estimates the ratio between the sale price and the market value of the repossessed asset at default, i.e., the loss severity given repossession. They compared this two-stage model against a single-stage LGD-model based on ordinary least squares (OLS). The authors found that the two-stage model can faithfully reproduce the empirical LGD-distribution with its peak at 0, whereas the single-stage model struggles to do so.
% Loterman2012
This finding was corroborated by the work of \citet{loterman2012benchmarking}, who compared 24 different techniques (including single-stage and two-stage approaches) for estimating the LGD. Most curiously, they found that sophisticated single-stage LGD-models (e.g., support vector machines and artificial neural networks) perform similarly to a two-stage LGD-model with simpler linear component models (e.g., logistic and linear regression).
Furthermore, their LGD-decomposition within the two-stage approach foregoes the idea of repossession and rather focusses on whether the LGD was either equal to or greater than 0, i.e., a \textit{write-off} risk component. This convention is a useful generalisation across product type, particularly since they have used six different datasets. For this reason, we too shall adopt a write-off risk component rather than repossession, given its generalisability and the fact that `loss' is typically associated with a write-off event in credit risk modelling.

% Survival analysis: PD-modelling
Originating from the biostatistical literature, \textit{survival analysis} is a rather powerful class of techniques; as discussed by \citet{singer1993time}, \citet{kleinbaum2012survival}, \citet{kartsonaki2016survival}, and \citet{schober2018survival}. These techniques generally analyse the length of time until reaching some well-defined endpoint, should the event occur. Survival analysis therefore do not only predict the probability of an event occurring, but also its timing. Furthermore, survival analysis is able to use all available data, including those unresolved/right-censored cases. So far, its use in the literature of credit risk modelling has largely been restricted to modelling another risk parameter instead of the LGD, i.e., the \textit{probability of default} (PD). In this regard, see \citet{banasik1999not}, \citet{stepanova2002survival}, \citet{bellotti2009macro}, \citet{bellotti2013}, \citet{bellotti2014stresstesting}, \citet{dirick2017time}, \citet{djeundje2019dynamic}, \citet{breeden2022multihorizon}, \citet{botha2025recurrentEvents}, and \citet{botha2025discTimeSurvTutorial}. Nonetheless, and as we shall demonstrate, the use of survival analysis in LGD-modelling is still relatively scant.

\subsection{Using survival analysis in LGD-modelling: Cox proportional hazards (CPH) models}
\label{sec:background_survAnalysis}

% Survival analysis: LGD-modelling (Zhang2012, Witzany2012)
One of the earliest works that explored the use of survival analysis in LGD-modelling is that of \citet{witzany2012survival}. In particular, they fit a \textit{Cox proportional hazards} (CPH) model, which can leverage unresolved/right-censored defaults. Two other models were developed within a broader comparative study: 1) a single-stage ordinary least squares (OLS) model; and 2) a two-stage model consisting of a logistic regression model for estimating write-off probabilities, coupled with an average loss severity given write-off. Using a few custom diagnostics (e.g., a modified coefficient of determination), they found that the single-stage OLS-model is again outperformed by the two-stage model. Yet both of these models are significantly outperformed by the single-stage CPH-model that directly estimates the LGD as the surviving proportion of the default balance. %However, the element of time, i.e., the random variable $T$, is treated rather strangely in that it represents not the latent lifetimes of default spells, but instead denotes the percentage recovered over the lifetime of a default spell, i.e., the recovery rate. We assume that doing so is necessary when building a single-stage LGD-model. 
\citet{zhang2012comparisons} also compared a few survival models to an OLS-model within both a single-stage and two-stage setup. Their survival models included a CPH-model and an \textit{accelerated failure time} (AFT) model in estimating the recovery rate. Since the latter AFT-model cannot handle zero-values in the outcome variable, the authors essentially adopted a two-stage approach towards estimating both survival models. They first had to classify complete write-offs (i.e., a recovery rate of zero) using a separate logistic regression model, whereafter both survival models were built on the non-zero recovery rates. In contrast to \citet{witzany2012survival}, the authors ultimately achieved mixed results in that the OLS-model performed the best in certain metrics, whilst the reverse is true for survival models in other metrics. Nonetheless, these results bode well for using survival analysis in general, and serve as a premise for our work.

% Fenech2016
The work of \citet{fenech2016modelling} examined the determinants of loan recovery outcomes using survival analysis on defaulted American commercial loans. The authors compared nonparametric (Kaplan–Meier), semi-parametric (CPH-models), and several parametric Cox-models. They found that the dynamics of debt recovery were best captured when assuming that the baseline hazard function follows a log–logistic distribution. In particular, their results revealed a "hump-shaped" hazard function, which peaked at 23 months post-default, after which the recovery likelihood declined again. Our study will partly corroborate this result in that the write-off probabilities over time have a similar right-skewed shape.

% Wood2017
\citet{wood2017addressing} extended the application of survival analysis in two-stage LGD-modelling, though they swapped write-off risk for repossession risk in the LGD decomposition.
Within the repossession component, they developed a series of CPH-models, each following the Fine-Gray fashion in contending with the competing nature of repossession vs cure events. Each CPH-model embeds a different probation period $k$ in defining (re)default over loan life. In particular, a defaulted loan that resumes payment is said to cure only after $k$ periods have lapsed, during which time default criteria must not apply. The premise of imposing such a probation period within the data is to subdue `noise' in the repayment histories of those high-risk loans, which can otherwise exit and re-enter the default state multiple times over loan life. Put differently, $k$ should be sufficiently large so that a new default spell is a truly new and independent spell, instead of being an extension of the previous spell. 
The authors then performed a simulation study and showed that the cumulative incidence function (of the repossession event) differs substantially over time and across different $k$-values, where $k\in\{1,3,6,9,12 \}$ months. 
We similarly contend with a probation period in our data and set $k=6$ months, though future work can certainly experiment in this regard towards loss-optimising this $k$-value.

% Joubert2018
In \citet{joubert2018making}, the authors used CPH-models in modelling the write-off risk component, as motivated by the inherent ability of CPH-models to use right-censored information. They built two separate CPH-models for modelling the survival probability of either the write-off or cure event over time. Both survival curves were combined using the cumulative incidence function within a competing risks setup. These CPH-models were then compared to a logistic regression (LR) model in estimating write-off risk, whereafter all model outputs were multiplied with a simple loss severity model towards obtaining overall LGD-estimates. The authors found that the \textit{mean squared error} (MSE) of the LGD-values produced by the CPH-model was substantially lower than that of the LR-model, when compared to the empirical LGD-values. The authors unfortunately did not provide other diagnostics of the various models, except for the MSE and decile-graphs between the empirical and expected LGD-values. Nonetheless, the results of \citet{joubert2018making} will inform the design of our own comparative study of LGD survival models, which will include an expanded set of model diagnostics.

% Joubert2018 and Joubert2021
\citet{joubert2018directly} and \citet{joubert2021adapting} introduced a new approach to using survival analysis in single-stage LGD-modelling. Their approach refines the exposure-weighted approach from \citet{witzany2012survival} to one that is weighted by the default balance using frequency weights. Furthermore, their approach can cater for negative cash flows, and it can embed the dynamics of recovering more than the default balance.
% Limitation
However, and similar to \citet{witzany2012survival} and \citet{zhang2012comparisons}, all of these authors formulate the survival probability $S(t)$ as the surviving proportion of the default balance up to a given time $t$. 
In so doing, one implicitly assumes that each unit of currency is an independent `life' that has survived the write-off event. But consider that a single recovery (or receipt) $R_t$ at time $t$ has, in fact, many (inter-dependent) units of currency in its make-up; particularly since they all derive from the same obligor. This variance-related aspect is yet unstudied to the best of our knowledge.
%Furthermore, this approach assumes that all units will eventually be lost, given infinite time; which is a central assumption to any survival analysis.
Our work differs conceptually in that we explicitly model the time-dependent write-off probability separately from the loss severity, in following a two-stage approach to LGD-modelling.

%A key limitation of this formulation lies in its treatment of each unit of currency as if it were an independent “life” in the survival process. While such independence is a standard assumption in survival analysis applied to individual lives, it becomes inappropriate in the LGD context: the different currency units within a repayment are in fact fully dependent, as they all derive from the same obligor and the same loan contract. To illustrate, the first rand and the hundredth rand of a R100 repayment will always be realised together, and the R100 received this month from a borrower is fully dependent on the, say, R100 received from that same borrower last month. By ignoring this complete intra-loan dependence, the currency-unit approach exaggerates the number of independent observations and understates the true uncertainty around recoveries

% Li2023
Using unsecured consumer loans, \citet{li2023predicting} improved LGD prediction by leveraging time-varying scores as input variables that emanate from separate survival models. Their premise is that no consensus exists on the best LGD-modelling technique, and that one should rather spend effort on crafting inputs (such as these scores) towards obtaining better results. As such, they used a CPH-model in developing application scores, which reflect the PD at the point of loan application. Then, they built a multiplicative hazard (MH) model, which is based on a counting process formulation and generalises the CPH-model towards dealing with recurrent default events. This MH-model produces behavioural scores, which signify the PD at each time point during loan life up to the default point. Both score types are then used as input variables, together with others such as macroeconomic variables, which are consumed within four kinds of LGD-models: 1) Tobit regression; 2) regression trees; 3) logit-transformed regression; and 4) beta regression. While the authors found that the inclusion of these scores resulted in better LGD-models, their performance metrics (e.g., the coefficient of determination) still indicated a general trend of poor performance, with Tobit regression scoring the best results. Based on the known correlation between the PD and the LGD, they ultimately showed that the LGD prediction accuracy benefits from using time-dependent inputs, such as these survival scores.

\subsection{Extensions to survival analysis in credit risk modelling}
\label{sec:background_extensions}

% Larney2023
As for extensions, \citet{larney2023modelling} developed a \textit{promotion time cure} (PTC) model in estimating the time to write-off amongst competing risks. Emerging from cancer studies, this type of survival model recognises that a certain proportion of defaulted loans will never experience write-off; i.e., they are immune.
In particular, the defaulted account is said to have $N$ number of unobservable and competing causes (or "carcinogenic cells") for the main write-off event such that the activation of any of these causes can trigger write-off. 
Examples of such latent causes may include the distressed borrower's discretionary expenditure, or undisclosed debts.
For $N\geq1$ causes, the failure time is then defined as $T=\min{\left(Z_1,\dots,Z_N\right)}$, where $Z_1,\dots,Z_N$ represent the time required for the $j^\mathrm{th}$ cause to be realised. A non-susceptible or cured account is said to have $N=0$ causes, where $N$ is commonly assumed to have a Poisson distribution.
% Oliveira2014
%Though the context differs slightly, \citet{oliveira2014recovery} made exactly this assumption for the $N$ causes of cure when they explored a PTC-model. A Weibull distribution was fitted to the time-to-recovery variable.
% Latent factor
\citet{larney2023modelling} also contended that the time to write-off within LGD-modelling is influenced by latent factors, e.g., unobservable elements of the economy. Accordingly, they combined their gamma PTC-model with a frailty component to account for such factors. Doing so can model the hazard function more flexibly, which may lead to improved prediction accuracy. In fact, they demonstrated the superiority of such a PTC-model with gamma frailty using US corporate loans, having compared it against other parametric frailty types, including a PTC-model with no frailty. These frailty PTC-models can ultimately help characterise the unexplained heterogeneity in write-off times. %However, the practical use of such models may be hampered somewhat due to their inherent complexity. 
However, assuming that certain accounts have write-off immunity may itself become a assumption subject to stress during economic crises. This assumption may very well invite criticism from regulators and auditors alike when using PTC-models practically. This waxing and waning of write-off immunity is yet unstudied to the best of our knowledge, and we shall therefore leave PTC-models outside of our study scope for the time being.

% Botha2025-Tutorial
In \citet{botha2025discTimeSurvTutorial}, the authors explored the use of \textit{discrete-time hazard} (DtH) models for predicting the loan-level PD, which culminated in a data-driven tutorial. Their work included a rigorous survey of survival modelling in credit risk, a rich description of the necessary data structure, DtH-models themselves, and a demonstration thereof using residential mortgages. They used the familiar \textit{generalised linear models} (GLM) framework with a logit link function for estimating these DtH-models, which bodes well for the practical adoption of such models.
These DtH-models were then assessed using a range of applicable diagnostics such as time-dependent varieties of both ROC-analyses and Brier scores. Both of these diagnostics were succinctly reformulated for the credit risk modelling context, and implemented in the R-programming language. The authors also compared the empirical vs expected term-structure of default risk, or the collection of average default probabilities over spell time.
Overall, they found that the predictions of DtH-models agree quite closely with reality, depending on the quality of the input variables, which was itself varied. These results augur well for the use of DtH-models in estimating write-off probabilities, as we shall explore in this paper.

%Frydman2022
As another survival modelling technique, consider a \textit{binary survival tree}, which extends recursive partitioning methods to right-censored data; i.e., marrying survival analysis with decision trees. Such a tree is grown by recursively splitting the data into two "daughter nodes" for a particular input variable. During this process, a splitting rule is used that maximises the difference in the survival probabilities between these two nodes, as remarked by \citet{frydman2022random}. In so doing, dissimilar cases are pushed apart, and each node eventually contains homogenous cases with similar survival probabilities over time. This process continues until reaching a saturation point whereat each node contains at least $n_0>0$ cases, and the most extreme nodes are called terminal nodes (or leaves).
In fact, the authors explored an ensemble-based extension of this technique, called a \textit{random survival forest} (RSF), which was applied within the context of PD-modelling with competing risks, i.e., default vs early settlement. 
Using Polish data on car leases, the authors compared this RSF-model to a classical competing risks CPH-model, itself built using the Fine-Gray method. They found that the former model had greater prediction power than the latter classical model, at least based on the time-dependent Brier scores of these models.

% Blumenstock2022
Another work that leverages survival trees is that of \citet{blumenstock2022deep}, who developed four survival models amongst two competing risks (default and early settlement) within the context of PD-modelling. These models included a classical CPH-model with/without using the Fine-Gray method, as well as an RSF-model, and a deep learning-based model called "DeepHit". All four models seek to estimate the cumulative incidence function for a particular competing risk. The authors used the time-dependent concordance statistic (Harrell's $c$) as their metric of choice in comparing the prediction performance across the four models. The DeepHit-model outperformed all others, unsurprisingly; however, the RSF-model was outperformed by only a relatively small margin in most of the experiments. This result attests of the robustness of a tree-based method such as RSF in holding its own against the more complex DeepHit-model.
The authors also address the typical concern with machine learning models (i.e., their opaqueness and lack of explainability) by using "permutation importance" towards establishing the extent to which model predictions are influenced by individual input variables. These permutation importance scores evaluate the decrease in model performance after `corrupting' a particular input by adding random noise.
Overall, the works of \citet{frydman2022random} and \citet{blumenstock2022deep} demonstrated that advanced tree-based methods (e.g., RSF) can perform admirably in PD-modelling. This success may very well translate to LGD-modelling, or at least elements thereof, though there is yet little research on this aspect. In this regard, \citet{ptak2024random} are probably the first who developed an RSF-based LGD-model, and compared it to a CPH-model in predicting the survival probability. Their diagnostics differ from ours in that they used the classical concordance $c$-statistic and the coefficient of determination ($R^2$), itself calculated using a bootstrap analysis. The authors found that the RSF-model produced a higher $c$-statistic, which again bodes well for tree-based methods in LGD-modelling.
Therefore, our study too shall include such a tree-based method, and we review the fundamentals of a particular type of survival tree in \autoref{app:appendix} called a \textit{conditional inference} survival tree.

% Jacobs2024 & Betz2021
The use of survival analysis has also surfaced in other aspects of LGD-modelling. In particular, \citet{jacobs2024modeling} proposed a modelling framework wherein the ultimate LGD is linked with a time-to-resolution concept using a joint fractional logit-survival model. Their framework produces both unconditional LGD forecasts for the purposes of pricing and capital modelling, and it produces conditional forecasts for loans currently in default.
Using European small business loans, \citet{betz2021time} developed a Bayesian hierarchical modelling framework wherein the default resolution time (DRT) is intrinsically linked with the eventual loss severity. They used a Weibull AFT-model for the DRT and a finite mixture model for the loss severity, thereby allowing for correlated shocks between resolution times and losses. Their results show that longer DRTs are associated with higher loss rates, and that economic crises exacerbate both, which can lead to severe underestimation in standard models (with a bias up to 20\%-points).
Ultimately, both of these studies showcase the importance of explicitly modelling the link between time and loss severity. Our work is cognisant hereof and therefore both components within our two-stage LGD-modelling approach will include a time factor.

% Summary
As shown, the use of survival analysis in LGD-modelling is still relatively new within literature, at least when compared to its maturity in PD-modelling. While the two-stage approach is followed in some cases, its definition/decomposition often involves a repossession component instead of an explicit write-off component; which we believe to be more fundamental to ECL-estimation under IFRS 9. Our comparative study generally follows the design of \citet{joubert2018making} and \citet{loterman2012benchmarking}, though we amend both the list of diagnostics and selection of techniques. In particular, our diagnostics assesses various aspects of the models, including their discriminatory power, prediction accuracy, and the extent to which predictions agree with reality when aggregated in various ways. Our study therefore comprises a greater variety of diagnostics, which renders the benchmark study as more valuable to practitioner and regulator alike.
Regarding techniques, we include a few single-stage LGD-models as a baseline against which a variety of two-stage LGD-models are compared. In this regard, we believe our study to be the first to include DtH-models in this comparison. While RSF-models were positively compared in previous works, little attention has been paid to one of its classical siblings -- the humble and more explainable (conditional inference) survival tree, which we include in the comparison. Lastly, and whereas some studies have focused on the competing risks nature of LGD-modelling, we shall restrict our attention to building cause-specific (write-off) survival models in the interest of simplicity.

%Across credit asset classes, recent research has highlighted the importance of explicitly modelling the link between time and loss severity. \citet{jacobs2024modeling} addresses corporate debt, proposing a joint fractional logit–survival model that links ultimate LGD with time-to-resolution (TTR). The framework produces both unconditional LGD forecasts for pricing and capital models, and conditional forecasts for loans currently in default. A key contribution is the regulatory and practitioner orientation, showing how LGD can be estimated consistently within Basel/IFRS 9 requirements while accounting for capital structure and censoring effects. \textcolor{red}{AB: Need to critique and link back to our article towards answering the "so what?" question.}

%\citet{betz2021time} develop a Bayesian hierarchical joint model that links default resolution time (DRT) and LGD for European SME loans. Using a Weibull survival model for DRT and a finite mixture model for LGD, the approach accounts for censoring and allows for correlated shocks between resolution times and losses. Results show longer DRTs are associated with higher LGDs, and that crises exacerbate both, leading to severe underestimation in standard models (bias up to 20 percentage points). The model provides unbiased LGD estimates for both unresolved and resolved defaults, aligning with forward-looking IFRS 9 requirements. \textcolor{red}{AB: Need to critique and link back to our article towards answering the "so what?" question.}

\section{The estimation of lifetime write-off risk using survival analysis in discrete time}
\label{sec:method}

We present some basic concepts and notation in \autoref{sec:survival_concepts} and discuss the structuring of credit data into a format that is conducive to survival analysis for LGD-modelling. We briefly formulate discrete-time survival analysis in \autoref{sec:survFundamentals}, and present the empirical term-structure of write-off risk, which our eventual modelling will strive to recover. Finally, the techniques are presented in \autoref{sec:survModels} towards modelling the write-off risk over an account's lifetime. Many of the aforementioned subsections (and appendix material) may be viewed as particular steps within a broader data-driven tutorial to modelling the write-off risk component in LGD-modelling.

\subsection{Basic concepts \& notation towards structuring credit data for LGD survival modelling}
\label{sec:survival_concepts}

\begin{figure}[ht!]
    \centering
    \includegraphics[width=1\linewidth, height=0.36\textheight]{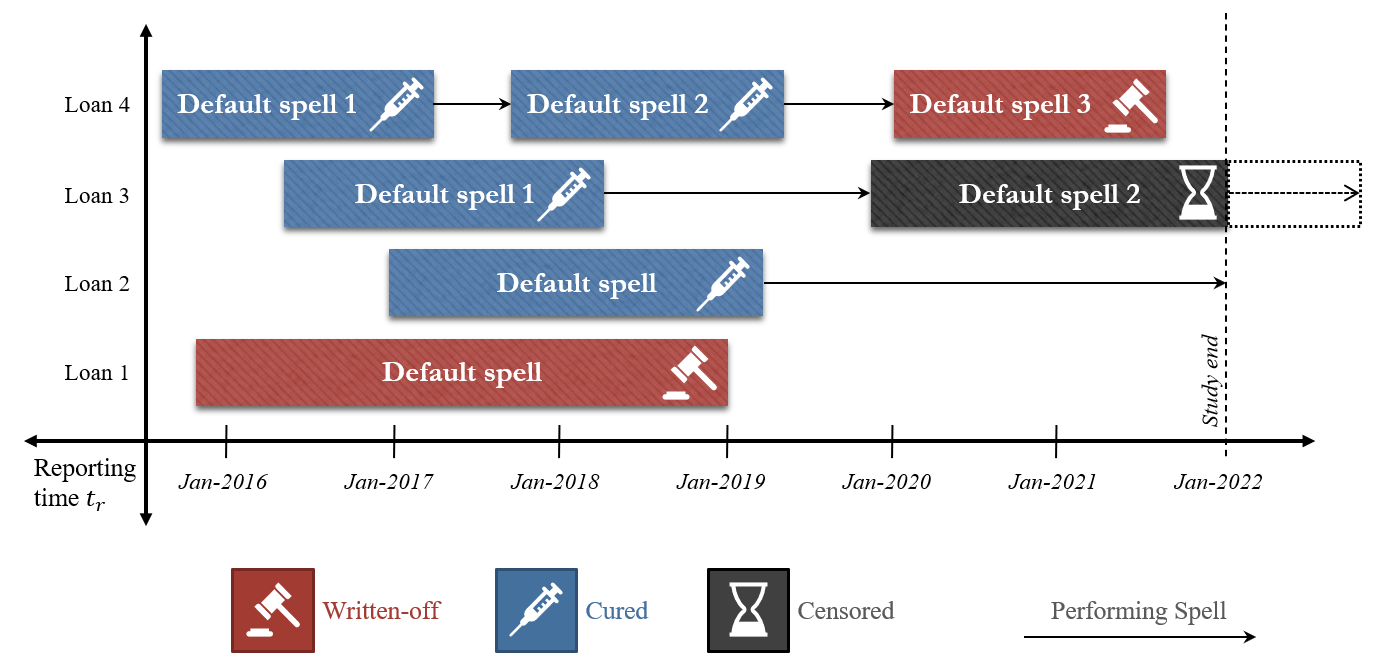}
    \caption{Demonstrating the resolution types of default spells over time for a few hypothetical loans.}
    \label{fig:DefSpells}
\end{figure}

% Basic concepts of spells
We shall start with some notation towards formalising the use of survival analysis in estimating write-off risk. As discussed by \citet{botha2025discTimeSurvTutorial}, the lifetime of a loan can be bifurcated into either performance or default spells, as demonstrated in \autoref{fig:DefSpells} for a few hypothetical loans. A \textit{spell} denotes a multi-period time span during which the repayment performance of a loan is monitored up to the time of the spell's resolution. A default spell starts at the default time $\tau_d$ and ends at a resolution time $\tau_r>\tau_d$, indexed by $t=\tau_d,\dots,\tau_r$. Such a default spell may cure and the loan may re-default later during its life, which implies a `multi-spell' setup in tracking the loan over its lifetime. This multi-spell setup is also known as recurrent survival analysis, as discussed by \citet{willett1995}, \citet[\S1.1]{jenkins2005survival}, and explored by \citet{botha2025recurrentEvents}. Moreover, the cure-outcome competes with the write-off outcome at every time point during a spell's lifetime, which has bearing on the overall modelling of write-off risk.

% Basic notation and resolution outcomes
Consider a portfolio of $N_d$ defaulted loans, wherein any loan $i=1,\dots,N_d$ may have $j=1,\dots,n_i$ number of default spells, where $n_i$ denotes the maximum number of spells for loan $i$. Let $(i,j)$ refer to a specific subject-spell that represents the history of a single default spell $j$ of $i$, as accompanied by a spell resolution outcome. Some spells may lack such an outcome, in which case they are said to be \textit{right-censored}. We denote these right-censored spells with $c_{ij}\in\{0,1\}$ in that $c_{ij}=1$ for a right-censored spell, and $c_{ij}=0$ otherwise. The few outcomes into which $(i,j)$ may resolve are coalesced into a single nominal variable $\mathcal{R}_{ij}^\mathrm{D}$, defined as 
\begin{align} \label{eq:spellResolution_Types}
    \mathcal{R}_{ij}^\mathrm{D} = 
    \begin{cases}
        1: \text{Written-off} \quad & \text{if} \ c_{ij}=0 \ \text{and write-off criteria applies} \\
        2: \text{Cured} \quad & \text{if} \ c_{ij}=0 \ \text{and curing criteria applies} \\
        3: \text{Censored} \quad & \text{if} \ c_{ij} = 1
    \end{cases} \, .
\end{align}

% Timing
Similar to \citet{botha2025discTimeSurvTutorial} regarding performing spells, we observe a default spell $(i,j)$ from its entry time $\tau_d(i,j)\geq1$ up to one of two points. These endpoints include either the resolution time $\tau_r(i,j)$ for $c_{ij}=0$, or the censoring time $C_{ij}<\tau_r(i,j)$ for $c_{ij}=1$. We shall contend with the overall spell stop time $\tau_s(i,j)$, which is simply the minimum between $\tau_r(i,j)$ and $C_{ij}$. During an on-going spell, we measure time discretely using an integer-valued counter variable, called the \textit{spell period} and is denoted by $t_{ij}=\tau_d(i,j),\dots,t_{ijk},\dots,\tau_s(i,j)$ for spell $j$ of defaulted loan $i$. This notation implies that our data follows the \textit{counting process} style in using the lexicon from survival analysis, as discussed by \citet[pp.~20-23]{kleinbaum2012survival}. We shall denote the overall spell age of $(i,j)$ as $T_{ij}$, which represents the observable follow-up time, defined as $T_{ij}=\tau_s(i,j)-\tau_d(i.j)$. The $(i,j)$-part will now be dropped from the notation of certain quantities $\left\{\tau_d, \tau_s, \tau_r \right\}$ in the interest of simplicity, though its connection to a particular spell remains implied. 
% Loan period
Furthermore, the loan period $t_i$ tracks the overall history (or age) of $i$, e.g., the variable "time on book". This variable may differ from the spell period $t_{ij}$, which itself measures the time spent in the spell at each point of its duration.
% Event history indicator
Lastly, let $e_{ijt}$ be the event history indicator that flags whether the main (write-off) event transpired at a specific point $t_{ij}$, where zero-values indicate either competing risks or right-censoring. In the words of \citet{singer1993time}, this $e_{ijt}$ may be described as a "chronology of event indicators" since its collection over time represents one of two vectors, which differ only in the last element: either $(0,0,\dots,1)$ for a written-off subject-spell, or $(0,0,\dots,0)$ for a censored subject-spell. Whilst certainly not ideal, our setup encodes competing risks as right-censored cases, which is known as the \textit{latent risks} approach to handling them; see \citet{putter2007tutorial}.

% Thrust of survival analysis
From \citet[\S 1.3]{jenkins2005survival}, survival analysis examines a random non-negative variable $T\geq0$ that denotes the latent lifetimes of default spells. However, and given the observed lifetimes $T_{ij},i=1,\dots,N_d,j=1,\dots,n_i$, we cannot truly examine the aforementioned latent lifetimes because of the confounding possibilities of \textit{censoring} and \textit{truncation}. Put simply, a completed spell $(i,j)$ ending in write-off would suggest $T=T_{ij}$, whereas a right-censored spell suggests $T>T_{ij}$, which hampers the estimation of $T$. \citet{witzany2012survival} imagined this problem using a right-angled "triangle of data", wherein increasingly recent cohorts of loans become progressively more right-censored over time. Naturally, our mortgage data is duly affected by right-censoring.
A left-truncated spell further complicates the interpretation of $T_{ij}$, since the starting point of $(i,j)$ predates that of the overall sampling window; see \citet[\S 1.2.1]{jenkins2005survival} and \citet[pp.~132-134]{kleinbaum2012survival}. Practically, our mortgage data exhibits extensive left-truncation and we encode its presence by adjusting the starting time of the first affected spell. Specifically, $\tau_d$ is set to the starting loan age; itself calculated as the difference between the current date and the origination date.
These ideas ultimately culminate in the longitudinal dataset credit dataset $\mathcal{D} = \left\{i, t_i, j, t_{ij}, \tau_d, \tau_s, \mathcal{R}_{ij}^\mathrm{D}, T_{ij}, e_{ijt} \right\}$, which is illustrated in the appendix for a few hypothetical loans.

\begin{figure}[ht!]
    \centering
    \includegraphics[width=0.9\linewidth, height=0.53\textheight]{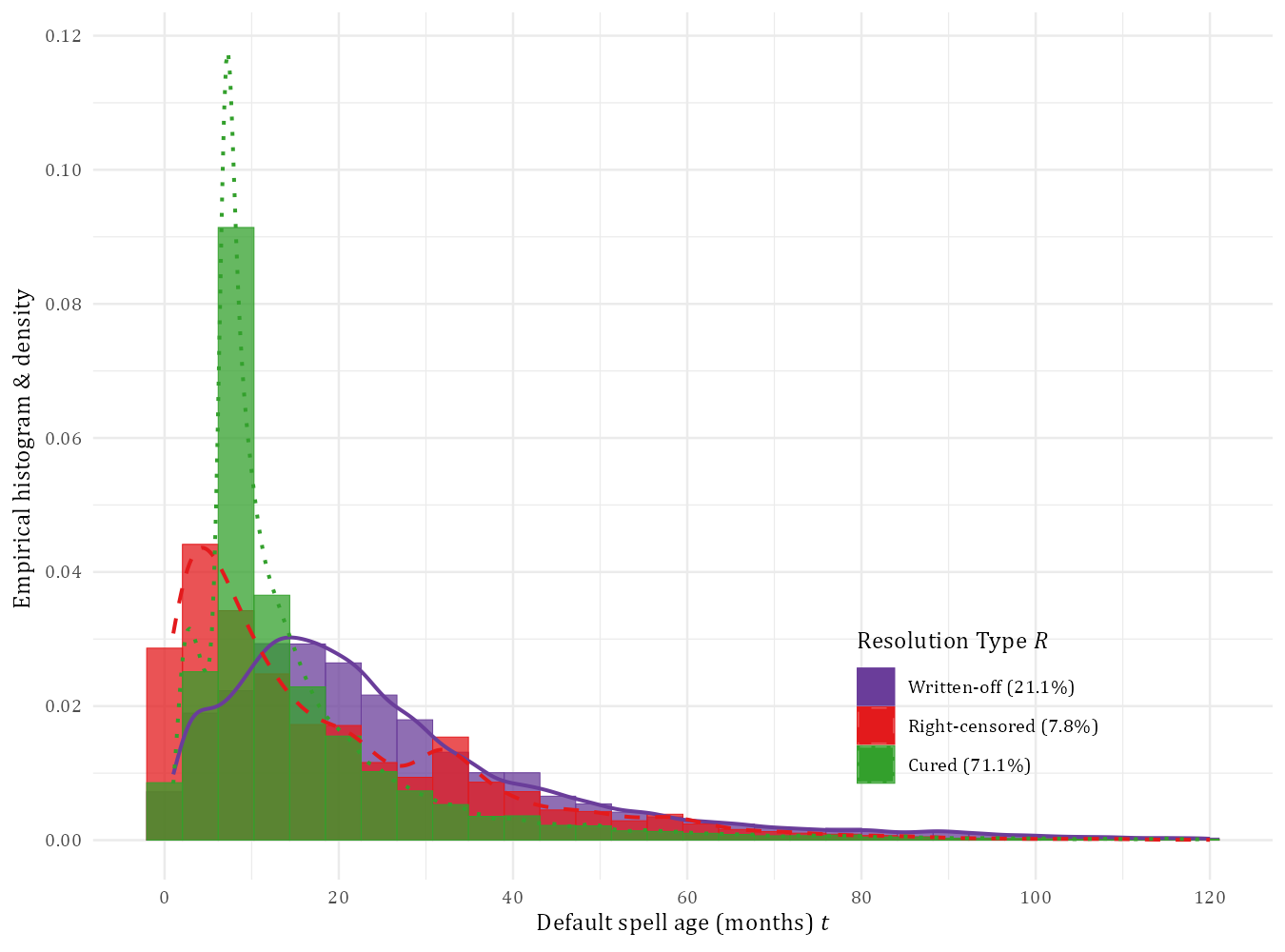}
    \caption{Histograms of failure times (or spell ages) for $T_{ij}\leq 300$ by resolution type, having used residential mortgage data. Empirical density estimates are overlaid.}
    \label{fig:WOffFailureTimes_Dist}
\end{figure}

% Failure time histogram
Using the mortgage data, the distributions of observed lifetimes are graphed in \autoref{fig:WOffFailureTimes_Dist} per resolution type $\mathcal{R}^\mathrm{D}$, which reveal quite a few insights. Firstly, it is evident that the distributions have different shapes, even though all of them are right-skewed. Consider the distribution of the right-censored outcome $\mathcal{R}^\mathrm{D}=3$, which is markedly different from that of the main write-off outcome $\mathcal{R}^\mathrm{D}=1$. This difference suggests that the censoring times $C_{ij},i=1,\dots,N_d,j=1,\dots,n_i$ are independent from the write-off resolution times $\tau_r(i,j)$. In turn, it suggests that censoring is \textit{non-informative} in that it does not affect the occurrence or timing of the main write-off event. Note that non-informative censoring is a necessary assumption in survival analysis, as discussed by \citet[p.~ 42]{kleinbaum2012survival} and \citet{schober2018survival}.
Secondly, \autoref{fig:WOffFailureTimes_Dist} can help us understand the effect and prevalence of competing risk events, whose occurrence precludes the main event from taking place. E.g., write-off occurred only in about 21\% of spells, whilst 71\% thereof ended in the competing cure event.
Thirdly, the right-skewed nature of the main write-off event affirms the intuition that most default spells are rather short-lived, with only a few surviving beyond 60 months (itself an arbitrarily chosen point for illustration purposes). All of these results render the histogram of failure times into a rather useful (and reusable) diagnostic in practice, and we advocate its use.

% Sampling
In sampling data from the credit dataset $\mathcal{D}$ for survival analysis, we retain the entire spell history over all of its periods $t_{ij}=\tau_d,\dots,\tau_s$, lest the resulting survival estimates become compromised due to missing spell periods. The row-observations in $\mathcal{D}$ are therefore \textit{clustered} around a common characteristic -- the loan ID -- before sampling randomly amongst the resulting $N_d$ clusters. We resample all of the default spells into the training set $\mathcal{D}_T$ across the entire lifetimes of randomly selected loans, which constitutes 70\% of $N_d$. This sampling fraction is set using convention, though can certainly be investigated in future work. The remaining spell histories are relegated into the non-overlapping validation set $\mathcal{D}_V$. This clustered random sampling scheme is also briefly discussed and illustrated by \citet[\S 6]{baesens2016credit} within the PD-modelling context. 
We affirm that the process by which the sets $\mathcal{D}_T$ and $\mathcal{D}_V$ are created is representative of the trends in the main dataset $\mathcal{D}$. This process uses the \textit{resolution rate} that is calculated for each dataset, as described by \citet{botha2025discTimeSurvTutorial} and detailed in the appendix (\autoref{app:resolutionRate}).
\subsection{Estimating the empirical term-structure of write-off risk using discrete-time survival analysis}
\label{sec:survFundamentals}

As discussed by \citet{botha2025discTimeSurvTutorial}, we discretise continuous time into a sequence of distinct contiguous intervals, given the typical structure of credit data, and we do so towards formulating discrete-time survival analysis. Let this sequence of \textit{unique ordered failure times} (excluding censored cases) be defined as $\left(0,t_{(1)}\right], \left(t_{(1)},t_{(2)}\right], \dots, \left(t_{(k-1)},t_{(k)}\right], \dots \left(t_{(m-1)},t_{(m)}\right]$, where $t_{(1)} < t_{(2)},< \dots < t_{(k)} < \dots < t_{(m)}$ up to some maximum time point $m$. Within each of these intervals $\left({t_{(k-1)}}, {t_{(k)}}\right]$, one may tally the number of certain events at month-end. In particular, let $f_k$ represent the number of spells that have failed (or been written-off) at $t_{(k)}$, and let $c_k$ denote the number of spells that became right-censored during said interval. The risk set at $t_{(k)}$ is said to contain $n_k$ number of spells that are \textit{at risk} of ending immediately prior to $t_{(k)}$, where such spells have a spell age of at least $T_{ij}>t_{(k)}$. This $n_k$ may be defined as the summation of all failure times $f_q$ and censoring times $c_q$, beyond those remaining times from and beyond $t_{(k)}$, as indexed by $q\geq k$; i.e.,
\begin{equation}\label{eq:risk_set}
    n_k = \left( f_q + c_q \right) + \left( f_{q+1} + c_{q+1} \right) + \dots + \left( f_m + c_m \right) =     
    \sum_{q=k }^{m} (f_q + c_q) \, , 
\end{equation}
assuming that $f_0=0$ and that $n_0$ is the initial population count. The aforementioned event history indicator $e_{ijt}$ from \autoref{sec:survival_concepts} can now be more formally defined as $e_{ijk}=\mathbb{I}\left(t_{(k-1)} <T_{ij} \leq t_{(k)}\right)$, which equals 1 if subject-spell $(i,j)$ was written-off during $\left(t_{(k-1)},t_{(k)} \right]$, and 0 otherwise.
For more detail on these quantities in discrete-time survival analysis, see \citet{singer1993time}, \citet[pp.~15-17,~\S 4.1]{jenkins2005survival}, \citet[\S7]{allison2010survival}, \citet[pp.~15-16,~57-58,~81-82]{crowder2012credit}, \citet{kartsonaki2016survival}, and \citet{suresh2022survival}.

% Random variable and Kaplan-Meier
We now consider the lifetime $T_{ij}$ of each spell $(i,j)$ to be a realisation from a non-negative random variable $T$ that represents the latent lifetimes of default spells. Up to some period $t_{(k)}$, let $F\left( t_{(k)} \right)=\mathbb{P}(T\leq t_{(k)})$ denote the cumulative lifetime distribution, i.e., the probability of experiencing write-off during the long time frame $\left( t_{(0)}, t_{(k)} \right]$. The complement thereof $S\left( t_{(k)} \right) = 1-F\left( t_{(k)} \right)=\mathbb{P}\left( T>t_{(k)}\right)$ represents the classical survivor function. An associated probability mass function exists $f\left( t_{(k)} \right)=\mathbb{P}\left(T=t_{(k)}\right)$ that represents the \textit{marginal write-off probability}, i.e., the probability of $T$ assuming a specific event time. In discrete-time, and drawing inspiration from \citet[pp.~17-20]{jenkins2005survival}, \citet[pp.~15-16]{crowder2012credit}, and \citet{suresh2022survival}, we relate $f\left( t_{(k)} \right)$ to $S\left( t_{(k)} \right)$ as 
\begin{equation}
    f\left(t_{(k)} \right) = \mathbb{P}\left( T=t_{(k)} \right) = \mathbb{P}\left( t_{(k-1)} < T \leq t_{(k)} \right) = S\left(t_{(k-1)} \right) - S\left(t_{(k)} \right) \, . \label{eq:PDF_discTime} 
\end{equation}
By convention, zero-length lifetimes are not possible such that $S\left( t_{(0)} \right)=1$, while $f(t)=0$ whenever $t$ does not equal any of the ordered failure time $t_{(k)}$.

% Hazard Function
The survivor function $S\left( t_{(k)} \right)$ also features in another useful quantity, i.e., the \textit{discrete hazard} $h\left( t_{(k)} \right)$. This quantity may be interpreted as the proportion of the risk set just prior to $t_{(k)}$ that was written-off during the contiguous interval $\left( t_{(k-1)}, t_{(k)} \right]$, i.e., exiting the spell during the $k\mathrm{th}$ interval. In following \citet[pp.~17-20]{jenkins2005survival}, \citet[pp.~15-16]{crowder2012credit}, and \citet{botha2025discTimeSurvTutorial}, the hazard function is the \textit{conditional write-off probability}, having survived hitherto, and is expressed in discrete time as 
\begin{equation} \label{eq:hazardfunction_discTime}
    h\left(t_{(k)}\right) = \mathbb{P}\left(  t_{(k-1)} < T \leq t_{(k)} \, | \, T > t_{(k-1)} \right) = 1 - \frac{S\left(t_{(k)} \right)}{S\left(t_{(k-1)} \right)} \, , \quad \text{with} \ 0 \leq h\left( t_{(k)} \right) \leq 1.
\end{equation}
Naturally, this conditional write-off probability $h\left(t_{(k)}\right)$ is related to the marginal variant thereof $f\left(t_{(k)}\right)$ as 
\begin{equation} \label{eq:discHaz_eventProb}
    f\left( t_{(k)} \right) = S\left( t_{(k-1)} \right)\cdot h\left( t_{(k)} \right) \quad \implies \quad h\left( t_{(k)} \right)=\frac{f\left( t_{(k)} \right)}{S\left( t_{(k-1)} \right)}\,.
\end{equation}
Most importantly, the collection $\left\{f\left(t_{(0)}\right),\dots, f\left(t_{(m)}\right) \right\}$ constitutes the empirical \textit{term structure} of write-off risk over spell time, which is the focus of this paper.

\begin{figure}[!ht]
    \centering
    \includegraphics[width=0.7\linewidth,height=0.43\textheight]{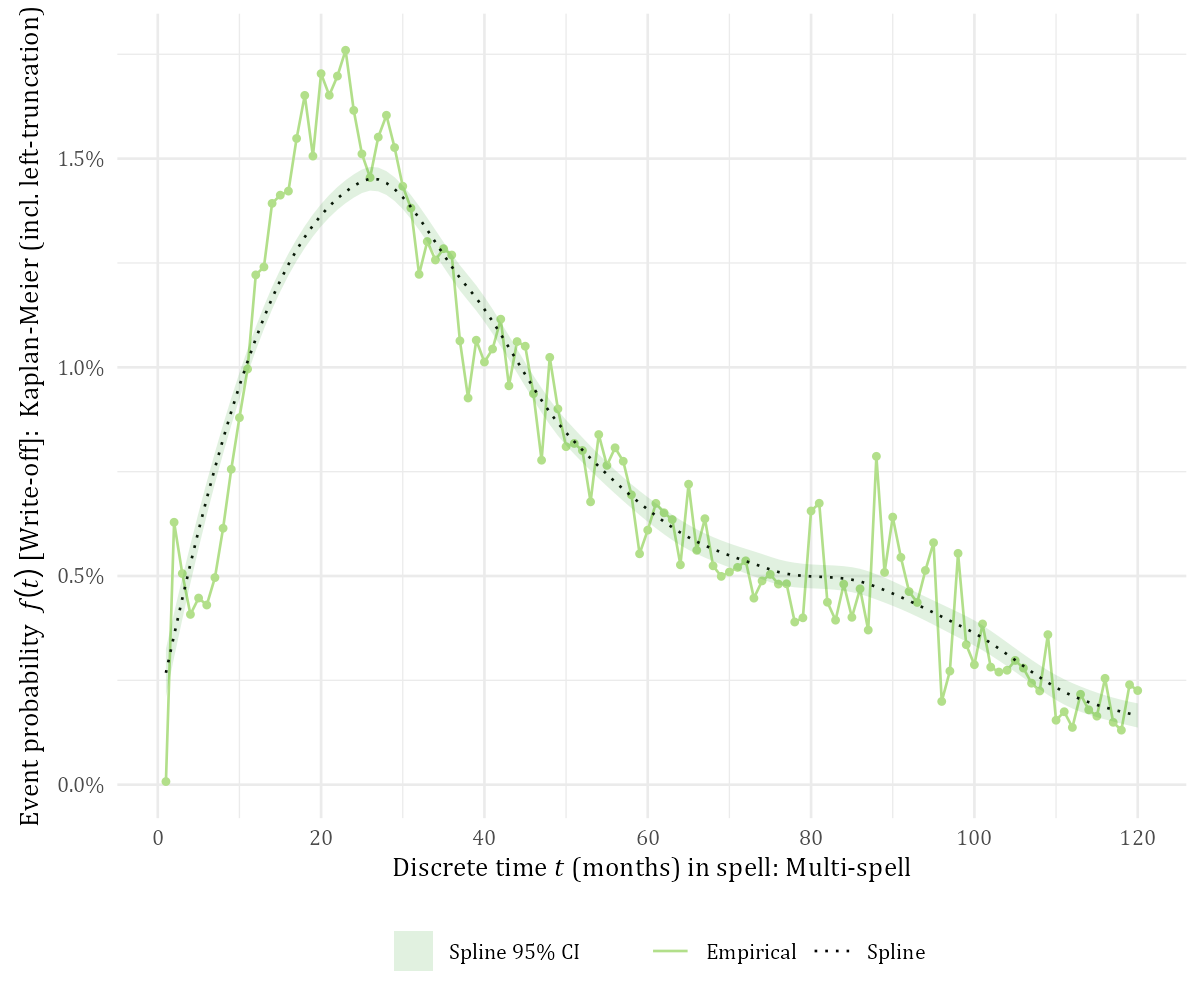}
    \caption{The empirical term-structure of write-off risk, as constituted by the discrete event probabilities $f(t),t=t_{(1)},\dots,t_{(m)}$ over spell time. Its estimation relies upon the KM-estimator from \autoref{eq:KaplanMeier} using residential mortgage data. A LOESS-smoother with a 95\% confidence interval is overlaid merely to summarise the visual trend.}
    \label{fig:KaplanMeier_EventProb}
\end{figure}

% Estimating S: Kaplan-Meier
In estimating these survival quantities, consider the well-known \textit{Kaplan-Meier} (KM) estimator from \citet{kaplan1958credit} in estimating $S\left( t_{(k)} \right)$. This KM-estimator is defined as 
\begin{equation} \label{eq:KaplanMeier}
   \hat{S}\left( t_{(k)} \right)=\prod_{s \, : \, t_{(s)} \, \leq \, t_{(k)}}{\left( 1-\frac{f_s}{n_s} \right)} = \prod_{s \, : \, t_{(s)} \, \leq \, t_{(k)}}{\left( 1-h_s \right)} \, ,
\end{equation}
where $h_k=f_k / n_k$ is the estimated discrete hazard during the $k^\mathrm{th}$ time interval, as discussed by \citet[pp.~15,~55,~77,~81]{crowder2012credit} and \citet{kartsonaki2016survival}.
Given estimates of $S(t), t=t_{(0)},\dots,t_{(m)}$, one may use \autoref{eq:PDF_discTime} to derive the empirical term-structure of write-off risk, i.e., the collection $\left\{ f\left(t_{(k)} \right) \right\}_{k=1}^{m}$. It is this collection that can form the empirical (or `actual') baseline term-structure against which expected varieties thereof, as produced by competing models, can eventually be compared. In fact, we illustrate this empirical term-structure in \autoref{fig:KaplanMeier_EventProb} over spell time using the same residential mortgage data. It is evident that the probabilities become increasingly unstable at greater spell ages (i.e., towards the right-hand side of the graph), which is attributed to the increasing sparsity of data during those periods. 
The event probability is also lower during extremely early ages, which we believe attests of internal write-off policies that have not yet taken effect due to on-going collection efforts.
Furthermore, the right-skewed nature of this curve suggests that earlier failure (write-off) is generally more probable than later failure. The credit system is therefore prone to "wear in" such that the hazard rate decreases over time, which is similar to infant mortality studies, as described by \citet[pp.~14]{crowder2012credit}. The "hump-shaped" hazard function that underpins \autoref{fig:KaplanMeier_EventProb} also corroborates the work from \citet{fenech2016modelling}, i.e., a right-skewed log-logistic distribution had the best fit.

\subsection{Three competing models for estimating write-off risk}
\label{sec:survModels}

We shall now formulate our various models in \crefrange{sec:survModels_LR}{sec:survModels_ST} for estimating the write-off probability as a function of a set of input variables $\boldsymbol{x}_{ij}$ specific to each spell $(i,j)$. These competing models include a classical (cross-sectional) logistic regression (LR) model, a more dynamic discrete-time hazard (DtH) model, and a conditional inference survival tree (ST).
There are also two versions of the LR and DtH models: a Type A and a Type B, as explained in \autoref{sec:survModels_Dichotomising}.

\subsubsection{A logistic regression (LR) model}
\label{sec:survModels_LR}

A classical cross-sectional dataset is used for this modelling technique, whereby each row in the modelling dataset represents a single subject-spell $(i,j)$. By implication, the data is structured as $\left\{i,j,y_{ij}, \boldsymbol{x}_{ij} \right\}$, where $y_{ij}\in\{0,1\}$ denotes the written-off outcome of each $(i,j)$, i.e., those cases for which $\mathcal{R}^\mathrm{D}_{ij}=1$. Each $y_{ij}$-value is a realisation from an underlying Bernoulli random variable $Y_{ij}$ with its conditional mean denoted as $\mu_{ij}=\mathbb{E}\left(Y_i |\, \boldsymbol{x}_{ij} \right)$.
Furthermore, the vector $\boldsymbol{x}_{ij}=\left\{x_{ij1},\dots,x_{ijp} \right\}$ represents the $p$ input variables of each $(i,j)$. These inputs may include both account-level and portfolio-level information, as well as macroeconomic covariates; though all inputs are observed only once per spell at its start. In formalising our LR-model, we employ the well-known \textit{generalised linear models} (GLM) framework, and define its linear predictor as $\eta_{ij} = \alpha + \boldsymbol{\beta}^\mathrm{T}\boldsymbol{x}_{ij}$, where $\boldsymbol{\beta}=\{\beta_1, \dots, \beta_p\}$ is a vector of estimable coefficients. This linear predictor is then modelled using
\begin{equation} \label{eq:LogitLink}
    g(\mu_{ij})=\log{\left( \frac{\mu_{ij}}{1-\mu_{ij}} \right)} =\eta_{ij} \implies \log{\left( \frac{w(\boldsymbol{x}_{ij})}{1-w(\boldsymbol{x}_{ij})} \right)} = \eta_{ij} 
\end{equation}
as the logit link function, where $w(\boldsymbol{x}_{ij})$ denotes the conditional write-off probability $\mathbb{P}\left( Y_{ij} = 1 | \, \boldsymbol{x}_{ij} \right)$ given the inputs $\boldsymbol{x}_{ij}$ of each $(i,j)$. The values within $\boldsymbol{\beta}$, together with that of the intercept $\alpha$, are found by maximising the log-likelihood function, as implemented within the \texttt{glm()} function in the R-programming language; see script 4c in the accompanying codebase.

\subsubsection{A discrete-time hazard (DtH) model}
\label{sec:survModels_DtH}

% Model specification
As described in \autoref{app:dataStructure}, the modelling dataset for this technique is in the counting process style, whereby each record represents a point in the history of the subject-spell $(i,j)$. 
In estimating our DtH-model, we reuse the previous GLM-framework with a logit link function, which follows the work of \citet{singer1993time} and \citet{botha2025discTimeSurvTutorial}. Accordingly, the discrete-time hazard probability from \autoref{eq:hazardfunction_discTime} is specified as
\begin{equation} \label{eq:discTimeHazard_logit}
    h\left(t| \, \boldsymbol{E}_{ij}, \boldsymbol{x}_{ij}, \boldsymbol{x}_{ij}(t), \boldsymbol{x}(t) \right) = \frac{1}{ 1 + \exp{\left( -\left[ \boldsymbol{\alpha}^\mathrm{T} \boldsymbol{E}_{ij} +  \boldsymbol{\beta}^\mathrm{T}\boldsymbol{x}_{ij} + \boldsymbol{\gamma}^\mathrm{T} \boldsymbol{x}_{ij}(t) + \boldsymbol{\delta}^\mathrm{T} \boldsymbol{x}(t) \right] \right) } } \, .
\end{equation}
In \autoref{eq:discTimeHazard_logit}, the vector $\boldsymbol{E}_{ij}=\left\{E_{ij1},\dots,E_{ijm} \right\}$ contains indicator variables, which flag a particular period $t\in\left[ 1, \dots, m\right]$ during the discretely-valued history of $(i,j)$, up to the observed maximum $m$. These period indicators are accompanied by  the estimable coefficients $\boldsymbol{\alpha}=\left\{\alpha_1,\dots,\alpha_m\right\}$; which together with the period indicators, represent the baseline hazard.
Moreover, the vector $\boldsymbol{\beta}=\left\{\beta_1,\dots,\beta_p \right\}$ contains estimable coefficients for $p$ spell-level time-fixed input variables, as denoted by  $\boldsymbol{x}_{ij}=\left\{ x_{ij1},\dots,x_{ijp}\right\}$. All of these particular inputs are measured at the start of each $(i,j)$. Similarly, the coefficients $\boldsymbol{\gamma}=\left\{ \gamma_1, \dots, \gamma_{p'} \right\}$ accompany the $p'$ time-dependent variables $\boldsymbol{x}_{ij}(t)=\left\{x_{ij1}(t),\dots,x_{ijp'}(t) \right\}$ of each $(i,j)$ over time, whereas the coefficients $\boldsymbol{\delta}=\left\{ \delta_1, \dots, \delta_{p^*} \right\}$ are associated with the $p^*$ portfolio-level time-dependent variables $\boldsymbol{x}(t)=\left\{ x_1(t), \dots, x_{p^*}(t) \right\}$, e.g., macroeconomic variables.
These various coefficient vectors $\left\{ \boldsymbol{\alpha}, \boldsymbol{\beta}, \boldsymbol{\gamma}, \boldsymbol{\delta} \right\}$ are found using the same \texttt{glm()} function in R by maximising the log-likelihood function, though given a different data structure.

% Model space
Using \autoref{eq:discTimeHazard_logit}, we formulate two DtH-models that differ only in the breadth of their input spaces (or set of input variables): a basic and an advanced DtH-model, labelled respectively as DtH-Basic and DtH-Advanced. As will be explained later, the full variable selection process is followed for the DtH-Advanced model. A smaller subset of this input space is chosen for the DtH-Basic model, based on expert judgement. See the 4b script series in the accompanying codebase for more details.

\subsubsection{A conditional inference survival tree (ST) model}
\label{sec:survModels_ST}

A conditional inference survival tree (ST) is fit using data that is structured very similarly to that of the LR-model. In particular, each row again represents a single spell $(i,j)$, though we also observe the spell age $T_{ij}$, which implies the dataset $\left\{i,j,y_{ij},T_{ij}, \boldsymbol{x}_{ij} \right\}$.
We use the \texttt{ctree()}-function from the \texttt{partykit} R-package from \citet{hothorn2015partykit} in fitting an ST-model. In creating two daughter nodes, an ST-model uses a splitting criterion that is based on the log-rank score statistic, which quantifies the association between the survival outcome $(T_{ij},y_{ij})$ and each candidate input variable. This statistic arises from a permutation-based test of the null hypothesis, which itself asserts independence between the survival response and the input variable. In turn, the variable selection procedure becomes unbiased and is not driven by the scale or number of possible cut points of an input variable. Only the variable $x_{ij}$ with the strongest association with the survival response is selected for further splitting. 
Assuming that such a variable is numeric, the cut point $c$ is selected that maximises the corresponding log-rank statistic, thereby partitioning the data into two daughter nodes, i.e., $x_{ij} < c$ and $x_{ij}\geq c$. A Kaplan-Meier curve is then respectively fit from the observations within each terminal node. From this survivor function, the corresponding discrete hazard and marginal write-off probabilities are derived analogously to \autoref{sec:survFundamentals}.
For more details on conditional inference ST-models, see the appendix (\autoref{app:SurvivalTrees}).

% Tuning process
Regarding the hyperparameters of our ST-model, we followed a manual tuning process using the validation set $\mathcal{D}_V$. This process is based on selecting the values of the hyperparameters that optimised various diagnostics, as will be discussed in \autoref{sec:results}. The final hyperparameters include the following: a maximum tree depth of four; a minimum number of 1000 observations needed to attempt a split; and a minimum number of 50 observations within a terminal node. We also impose a minimum of 99\% ($1-p$-value) when maximising the splitting criterion; i.e., a split is only attempted when the $p$-value of the overarching permutation test is smaller than 1\%. For more details on the fitting process, see script 4f in the accompanying codebase. Furthermore, we provide the ST-model with the same input space of the DtH-Advanced model. However, the resulting tree contains only a few of these input variables, chiefly due to the specified tree depth. See \autoref{app:InputSpace} for details on the final input space.

%\textcolor{red}{For a subject-spell assigned to a particular terminal node, the  Kaplan--Meier estimator within that node yields an estimated survivor function over spell time. From this survivor function, the corresponding discrete hazard and marginal write-off probabilities are derived analogously to \autoref{sec:survFundamentals}. The resulting node-level probabilities therefore constitute a spell-specific term-structure of write-off risk, which can be aggregated across accounts to produce the model-implied term-structure used in the comparative diagnostics.}

\subsubsection{Dichotomising the probabilistic models into discrete classifiers: Type A and B}
\label{sec:survModels_Dichotomising}

Each write-off risk model produces a probability score as output, which is denoted by $w(t,\boldsymbol{x}_i)\in[0,1]$ for defaulted loan $i$ with inputs $\boldsymbol{x}_i$ at spell time $t$. These probability scores are typically multiplied with an estimate of the loss severity towards calculating the LGD. The model that outputs such `raw' probability scores shall be called a Type A model. One can however dichotomise these scores into 0/1-decisions using a cut-off value $c$, such that the indicator function $\mathbb{I}\left(w(t,\boldsymbol{x}_i) > c\right)$ outputs 1 if $w(t,\boldsymbol{x}_i) > c$, and 0 otherwise. Doing so can better attune each resulting LGD-model to the underlying empirical LGD-distribution, as will be shown later. Nevertheless, we shall call such a dichotomised model $\mathbb{I}\left(w(t,\boldsymbol{x}_i) > c\right)$ a Type B model.

% Sensitivity & specificity, misclassification costs.
In finding this cut-off value $c$, consider first the sensitivity and specificity from ROC-analysis, as examined by \citet{fawcett2006introduction}. Sensitivity is the probability of a true positive (or a write-off event correctly predicted as such), which is defined as $q(c)=\mathbb{P}\left( w(t,\boldsymbol{x}_i) > c \mid \mathcal{C}_1\right)$, where $\mathcal{C}_1$ is the positive class, i.e., an observed write-off event. Similarly, specificity is the probability of a true negative (or a non-event correctly predicted as such), which in turn is defined as $p(c)=\mathbb{P}\left( w(t,\boldsymbol{x}_i) \leq c \mid \mathcal{C}_0 \right)$, where $\mathcal{C}_0$ is the negative class. These two components do not necessarily carry the same weight in classification problems, especially so under IFRS 9 where a bank would rather be over-provided than under-provided in its loss provisions. Put differently, the misclassification cost differs between false positives and negatives, and the latter should be penalised to a greater extent in the interest of risk prudence.

% Generalised Youden Index
Accordingly, let $a>0$ denote a cost multiple (or ratio) of committing a false negative relative to a false positive, and let $\phi$ denote the estimated prevalence of the $\mathcal{C}_1$-event, i.e., $\mathbb{P}(\mathcal{C}_1)$. Both of these quantities are used in the \textit{Generalised Youden Index} (GYI) function, which quantifies the prediction power of a model for a given $c$-value amidst differing misclassification costs.
As introduced and discussed by \citet{geisser1998comparing}, \citet{kaivanto2008maximization}, and \citet{schisterman2008youden}, we define the GYI for a given $c$ as
\begin{equation} \label{eq:gen_Youden_Index}
    J_a(c)= q(c) + \frac{1-\phi}{a\phi} \cdot p(c) - 1 \, ,
\end{equation} whereupon the optimal cut-off $c^*$ is given by \begin{equation} \label{eq:optimalCutoff}
    c^* = \arg \max_c{J_a(c)} \, \nonumber.
\end{equation}
By setting $a=(1-\phi)/\phi$ in \autoref{eq:gen_Youden_Index}, the respective weights of $q(c)$ and $p(c)$ become equal. Though as $a$ increases, the contribution (or weight) of the specificity term $p(c)$ decreases. Since $p(c)$ reflects the avoidance of false positives, a reduced weighting on $p(c)$ implies that the penalty on false positives decreases.
A larger $a$-value therefore produces a lower $c^*$-value, which translates to a more "liberal" model that is more likely to render $\mathcal{C}_1$-predictions than $\mathcal{C}_0$-predictions.
% Implementation
Lastly, the GYI is implemented within the bespoke R-function \texttt{GenYoudenIndex()}, which itself uses the \texttt{JDEoptim()}-function from the \texttt{DEoptimR}-package; see the R-codebase maintained by \citet{gabru2026WriteOffSurvSourcecode}.
We also devise in the appendix (\autoref{app:CostMultipleOptim}) a short procedure for optimising $a$ within the GYI. This procedure is based on minimising the MAE between the empirical and expected term-structures, the latter of which emanates from a dichotomised model given a corresponding $a$-value.

\subsection{Formulating single-stage vs two-stage LGD-models: our model universe}
\label{sec:models_LGD}

While the focus of this study is on write-off risk, we think it prudent to compare the resulting LGD-estimates from the various models in the interest of completeness. But deriving these LGD-estimates requires a loss severity component, in following the two-stage approach to LGD-modelling. For simplicity, we shall reuse the GLM-framework in regressing empirical loss rates (given write-off) $l_i, i=1,\dots,n$ on to a set of input variables. Future work can most certainly refine the estimation of these loss rates from data using more sophisticated models. Following some experimentation with response distributions and appropriate link functions, we select the \textit{compound Poisson} (CP) distribution for our GLM, and we select the (canonical) log link function. In particular, the response variable $Y_i$ is assumed to follow a Tweedie distribution, i.e., $Y_i \sim \text{Tweedie}(\mu_i, \phi,p)$, where $\mu_i$ is the mean, $\phi$ is the dispersion parameter, and $1<p<2$ is the power parameter that determines the CP-structure. This choice is motivated by the fact that the resulting GLM can handle zero-inflated data (such as zero-valued loss rates) rather flexibly, as discussed by \citet{zhang2013likelihood}. 
Its use is quite popular in actuarial science wherein the number of insurance claims can be zero, as well as in estimating the (possibly zero-valued) amount of precipitation during a given period when modelling rainfall.
In fitting a Tweedie CP-GLM, we use the \texttt{cpglm()}-function from the \texttt{cplm}-package in R.

\begin{figure}[ht!]
    \centering
    \includegraphics[width=0.8\linewidth, height=0.3\textheight]{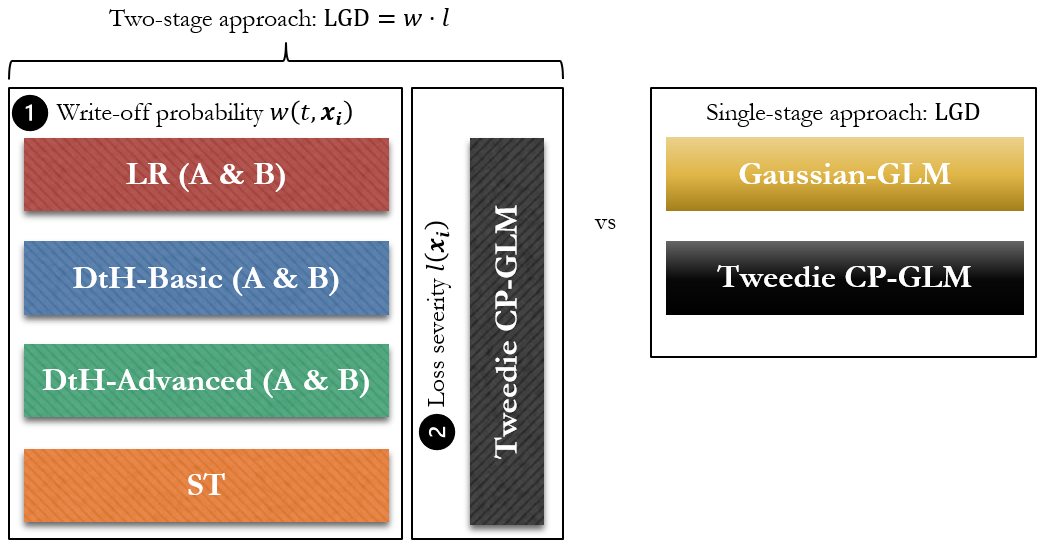}
    \caption{Differentiating the model universe between two-stage and single-stage LGD-modelling approaches.}
    \label{fig:ModelUniv}
\end{figure}

% Advanced & basic
Previous work by \citet{botha2025discTimeSurvTutorial} has shown the merit of developing a baseline model that has an input space that is deliberately stripped bare, which can aid comparisons. We adopt the same design and develop a basic DtH model (labelled as DtH-Basic) that contains but a time factor, in addition to a few rudimentary variables, as selected using expert judgment and statistical significance; see \autoref{app:InputSpace}. This DtH-Basic model is improved upon by its more advanced sibling (labelled as DtH-Advanced) whose variable selection is more rigorous, as will be explained later. The expectation is that the DtH-Advanced should outperform its more simplistic sibling in predicting write-off risk, purely by virtue of having a more enriched input space.
Both of these DtH-models are also dichotomised, thereby availing Types A and B of each model.
In following the two-stage LGD-modelling approach, our model universe therefore includes eight model components: 2x LR-models (A \& B), 2x DtH-Basic models (A \& B), 2x Dth-Advanced models (A \& B), 1x ST-model, and the loss severity, a Tweedie CP-GLM. We summarise these models in \autoref{fig:ModelUniv} for easy reference.

% Single stage LGD-modelling setup
The two-stage LGD-modelling approach can be easily compared to a single-stage approach, wherein the zero-loss cures are included in the modelling dataset. The overall LGD-distribution is now modelled directly as a function of a few input variables. We believe it prudent to do so in positioning our work as a reference point amongst other works that have pursued similar comparisons of LGD-models, such as that of \citet{loterman2012benchmarking}.
Accordingly, and to keep things simple, we reuse the GLM-framework and build two competing single-stage LGD-models: 1) one with a Gaussian distribution and an identity link function; and 2) one with the same Tweedie CP-distribution with a log link function, as in the aforementioned two-stage approach. Both of these single-stage LGD-models form part of our model universe, as summarised in \autoref{fig:ModelUniv}.
The justification for choosing the latter Tweedie CP-GLM is that it should theoretically be able to contend with the probability mass at zero, i.e., the zero-loss cures. The overall expectation is that we can replicate the results from previous studies that showed the two-stage approach to outperform the single-stage approach in general, without using more sophisticated machine learning techniques within a single-stage approach.

\section{An empirical comparison of competing models for write-off risk }
\label{sec:results}

% Variable selection
In building our models, variable selection is performed using an interactive thematic selection process, which is explained as follows. Within a theme, e.g., "macroeconomic variables: inflation", specific questions are posed to be investigated, e.g., \textit{"which lag order of a macroeconomic variable is the `best'?"}. Each question is answered using a variety of model diagnostics, including statistical significance testing, domain knowledge, correlation studies, and goodness-of-fit, as measured using Akaike's Information Criterion (AIC) and McFadden's pseudo $R^2$ from \citet{menard2000coefficients}. For a set of thematically-chosen input variables, we build single-factor models and then evaluate them using the same diagnostics, thereby generating insights on the `best' inputs. This process is complemented by running a stepwise forward selection, as explained by \citet[\S 6]{james2013introduction}, whereafter the final list of input variables is manually curated using expert judgment. This thematic selection process is chiefly followed in estimating our DtH-Advanced model, though the resulting insights are reused in building the other models. See \autoref{app:InputSpace} for a complete list of input variables per model, whilst the details of this selection process are contained within the accompanying R-codebase maintained by \citet{gabru2026WriteOffSurvSourcecode}.
As for evaluating our various models, we shall now discuss and present the results of five main diagnostics.

% Discriminatory power: tROC
Firstly, we analyse the discriminatory power of our LR-models using the \textit{receiver operating characteristic} (ROC), as discussed by \citet{fawcett2006introduction}. Each ROC-analysis is then summarised into the \textit{area under the curve} (AUC) statistic, where greater values indicate greater discriminatory power. For the survival models, a \textit{time-dependent} ROC-analysis (tROC) can be conducted given hazard predictions at a particular spell time point $t=t_{(0)},\dots,t_{(m)}$. This type of ROC-analysis generalises the classical one to a context wherein right-censored cases are adequately treated, as discussed by \citet{heagerty2000} and \citet{bansal2018tutorial}. Since the degree of right-censoring varies across $t$, the discriminatory power of a survival model will itself vary over $t$. As such, the two elements of a tROC-graph can be expressed as functions of both $t$ and the cut-off $p_c$: the true positive rate $T^*(t, p_c)$ and the false positive rate $F^*(t,p_c)$. Under the "cumulative case/dynamic control" framework of \citet{bansal2018tutorial}, both of these quantities are formulated using the conditional survival probability at $t$, itself estimated using the nearest neighbour estimator from \citet{akritas1994}.
For a gentler introduction to tROC-analysis, see \citet{botha2025recurrentEvents} and \citet{botha2025discTimeSurvTutorial}, who have reformulated tROC-analysis within the PD-modelling context using performing spells. This particular spell formulation is similar to default spells within our LGD-modelling context, and so we shall use the exact same setup in conducting tROC-analyses.

\begin{figure}[ht!]
\centering
\begin{subfigure}[b]{0.98\textwidth}
    \caption{DtH-Basic A}
    \centering\includegraphics[width=1\linewidth,height=0.3\textheight]{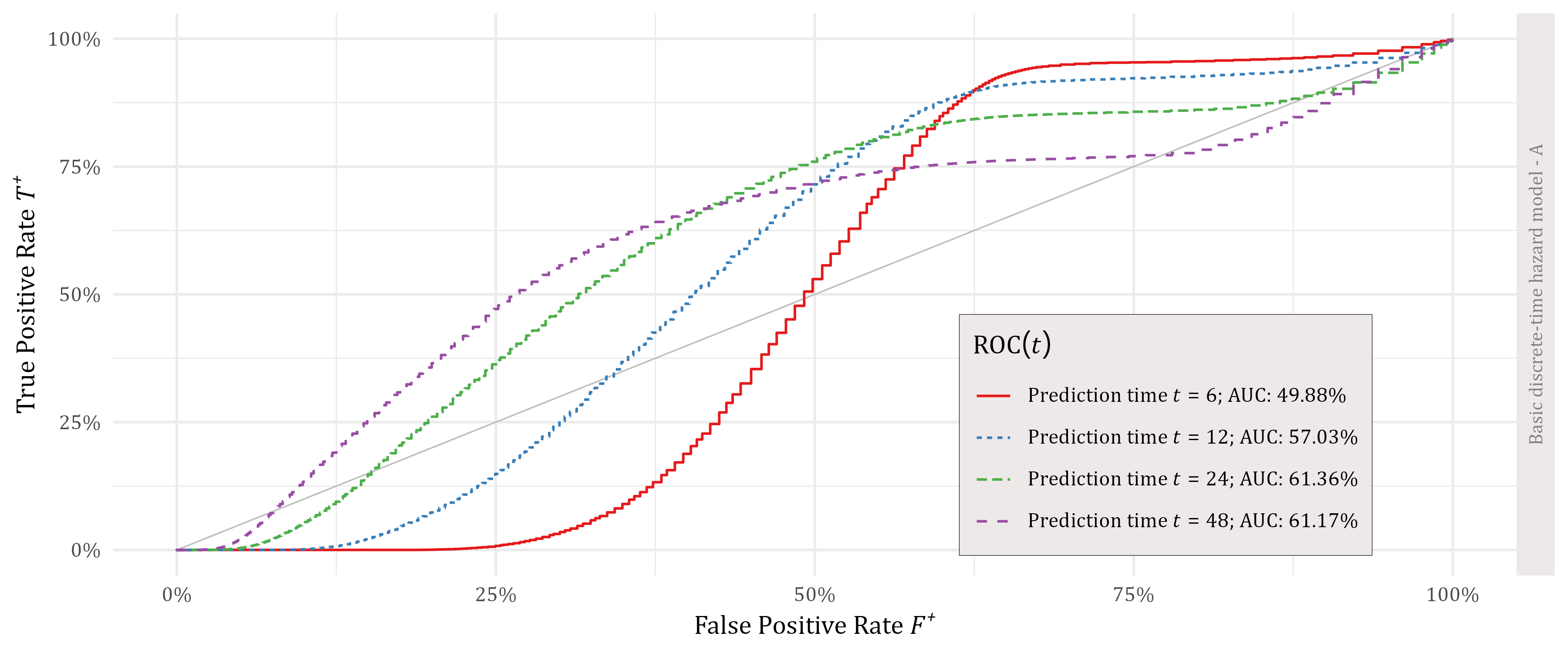}\label{fig:tROC_basic}
\end{subfigure} \\
\begin{subfigure}[b]{0.49\textwidth}
    \caption{DtH-Advanced A}
    \centering\includegraphics[width=1\linewidth,height=0.3\textheight]{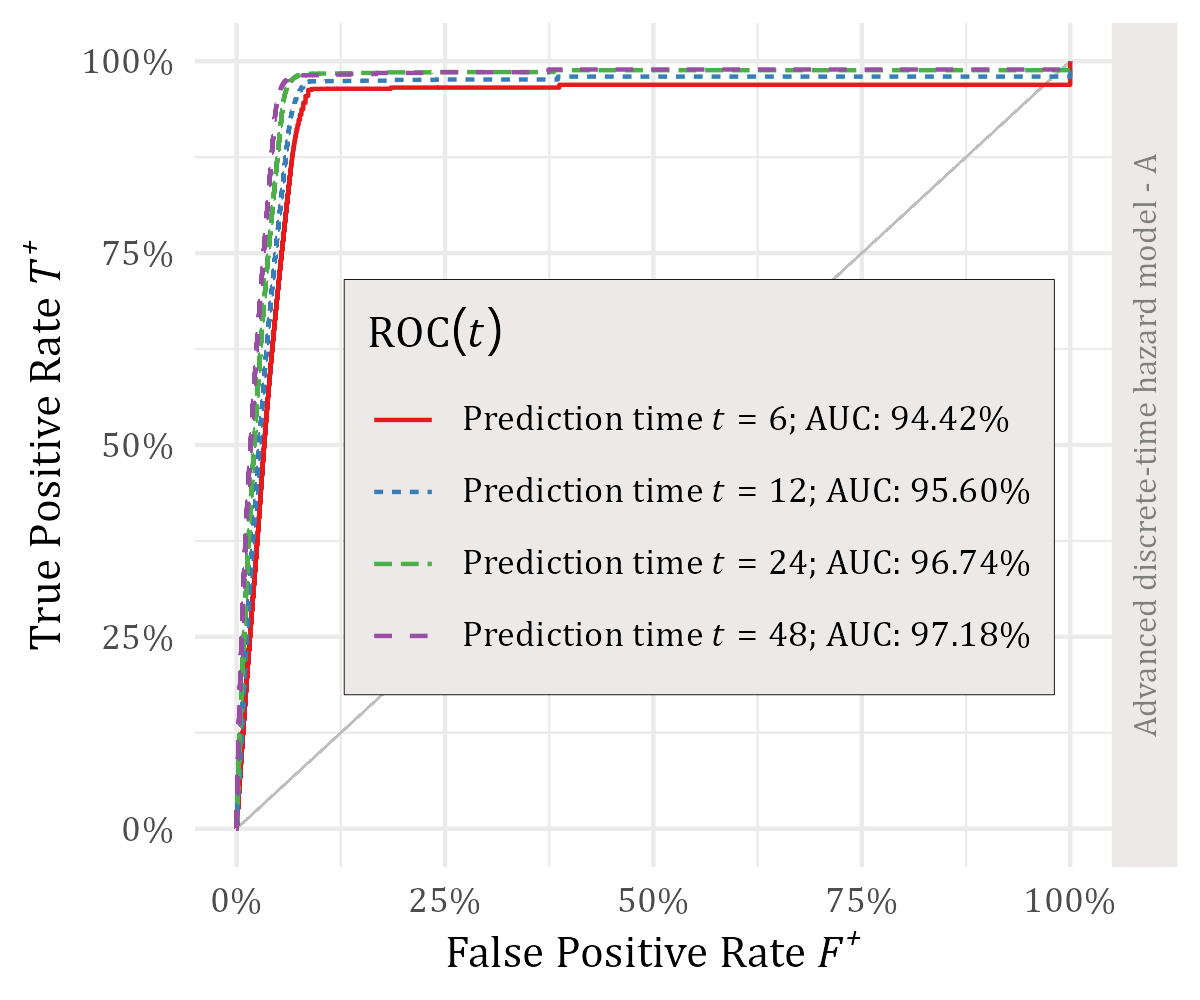}\label{fig:tROC_advanced}
\end{subfigure}
\begin{subfigure}[b]{0.49\textwidth}
    \caption{Survival tree}
    \centering\includegraphics[width=1\linewidth,height=0.3\textheight]{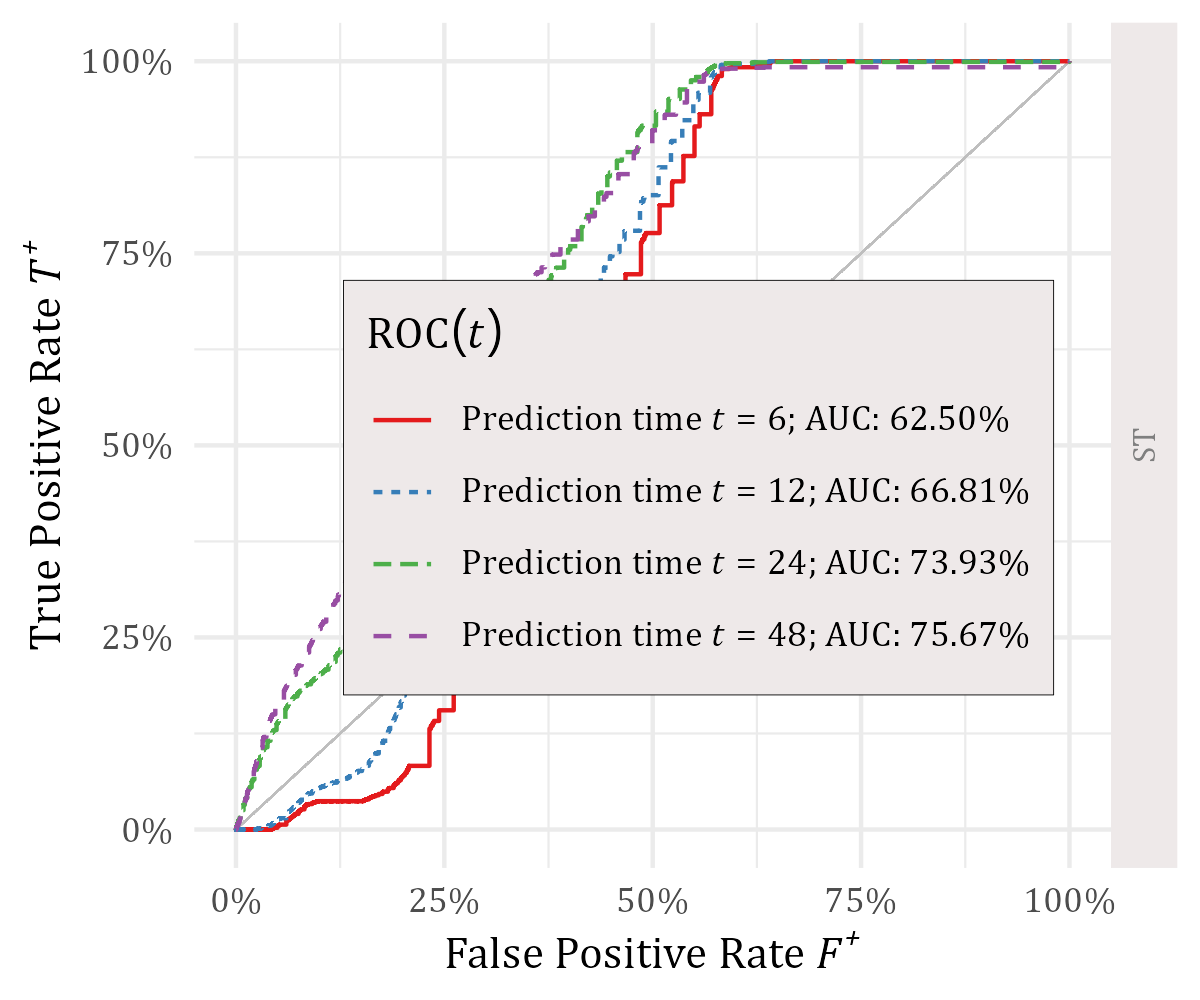}\label{fig:tROC_survTree}
\end{subfigure}%
\hfill%
\caption{Evaluating the discriminatory power of three competing survival models by using the clustered tROC-extension to ROC-analysis at specific time points $t\in \{3,6,24,36\}$. The results per model (Type A-series) are shown respectively in panels \textbf{(a)}-\textbf{(c)}.}\label{fig:tROC}
\end{figure}

% tROC analyses
Regarding the implementation of such a tROC-analysis, we use the bespoke \texttt{tROC.multi()} function that is defined in the accompanying R-codebase. The chosen time points for this tROC-analysis are selected using expert judgment and include $t\in\{6,12,24,48 \}$ months in default. 
Similar to an ROC-analysis, a tROC-analysis can be summarised into a time-dependent AUC-value (tAUC); one for each spell time point $t$. As shown in \autoref{fig:tROC}, the DtH-Advanced model clearly has greater tAUC-values than those of the DtH-Basic model across all time frames. 
Although not shown, this trend holds true for the Type B model variants as well, albeit at lower tAUC-levels across all $t$.
We believe this outperformance of the DtH-Advanced model over the DtH-Basic model attests of the former model having a more comprehensive set of input variables, which underscores the importance of feature engineering.
Furthermore, the ST-model achieved a second-in-class performance, and though it outperformed the DtH-Basic model, it did not best the DtH-Advanced model. This finding is interesting since one would have expected better results from a more sophisticated technique such as survival trees, especially when it shares the same input space of the DtH-Advanced model.

\begin{figure}[!ht]
    \centering
    \includegraphics[width=0.7\linewidth,height=0.4\textheight]{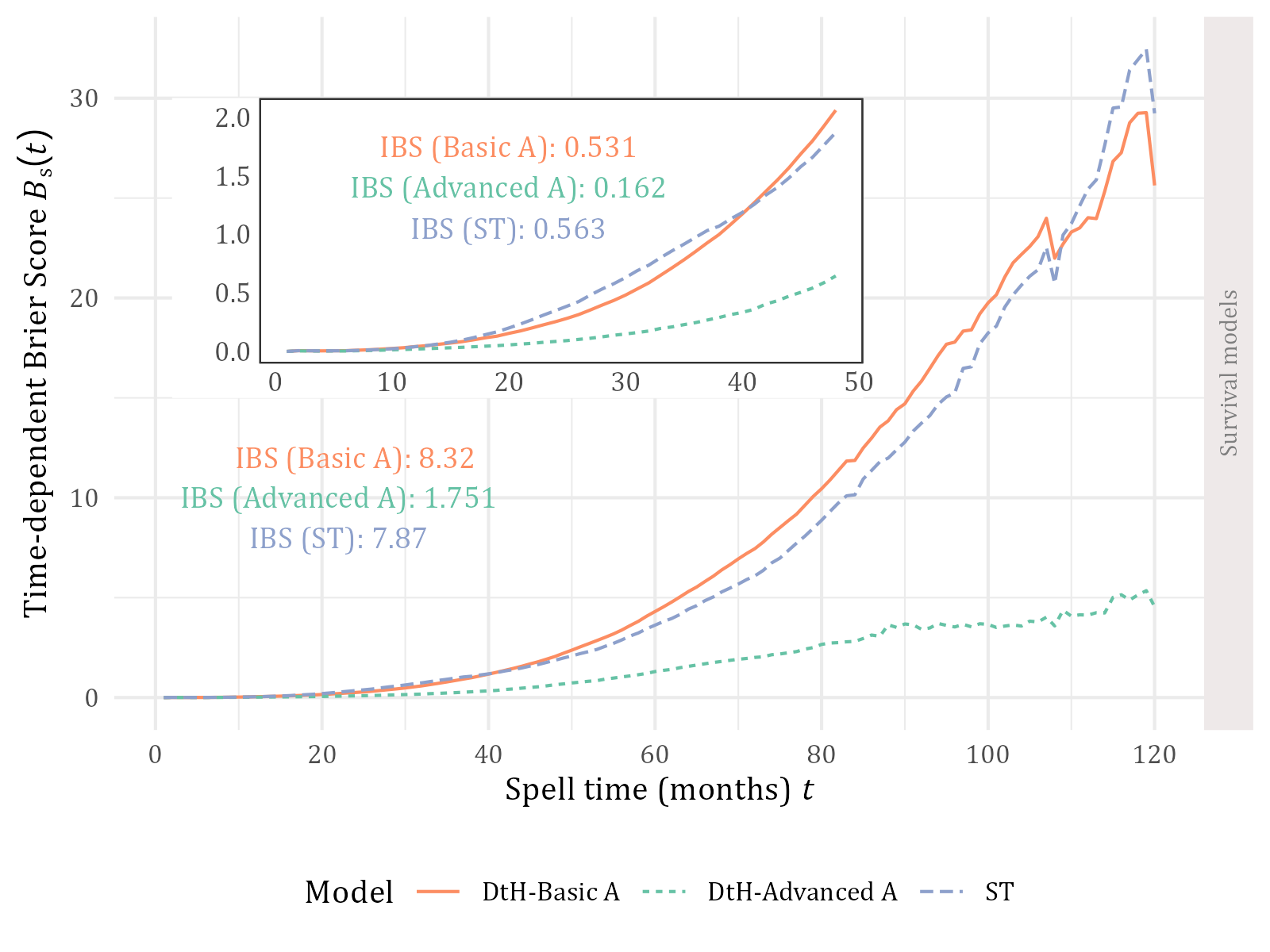}
    \caption{The time-dependent Brier Score (tBS) over spell time $t$ per survival model (Type A-series). The integrated Brier score (IBS) is annotated per survival model in summarising the tBS-values over time.}
    \label{fig:timeBrierScore}
\end{figure}

% Prediction accuracy
Secondly, and in evaluating the prediction accuracy of a survival model over time, one can consider the \textit{time-dependent Brier score} (tBS), as discussed by \citet{Graf1999} and \citet{suresh2022survival}, and implemented by \citet{botha2025discTimeSurvTutorial} within the PD-modelling context. As a measure of prediction error at a particular spell time $t$, this tBS-measure calculates the average squared difference between the predicted survival probabilities and the observed outcomes amidst right-censoring. A lower tBS-value at a particular $t$ indicates lower prediction error and hence greater model performance. A key advantage of the tBS is its model-agnostic nature in that it relies only on the predicted survival probabilities. We use the bespoke \texttt{tBrierScore()} function within the accompanying R-codebase, and graph these tBS-values of each survival model in \autoref{fig:timeBrierScore} over $t$. The results show that the more complex DtH-Advanced model achieves lower tBS-values over all $t$, relative to those of the DtH-Basic model. For both models, the increasing trend in the tBS over $t$ attests of the increasing difficulty of estimating survival probabilities accurately at later spell times, largely due to dwindling sample sizes at those periods. The ST-model achieved tBS-values that are largely similar to those of the DtH-Basic model, which corroborates the surprise finding of the previous tROC-diagnostic.
Moreover, the maximum period $t^*$ over which to calculate the tBS should be meaningfully chosen, particularly given the increasing prevalence of right-censored cases as $t\rightarrow
 t^*$. Our Kaplan-Meier analysis (itself reviewed in \autoref{sec:survFundamentals}) suggested that the vast majority of the dataset is exhausted as $t\rightarrow120$ months in default, though even this period is considered to be extremely long in practice. Having used expert judgment, our inset graph in \autoref{fig:timeBrierScore} shows the tBS over a more realistic workout period of 48 months in default. However, the divergence in the prediction errors seems to worsen exponentially over $t$ between the DtH-Advanced and Dth-Basic/ST models. This result corroborates our previous tROC-results regarding discriminatory power. That is, the DtH-Basic and ST models seem to produce predictions that are less accurate than those of the DtH-Advanced model.

 % IBS
The tBS-values can be aggregated over $t$  into a singular value called the \textit{integrated Brier score} (IBS). Lower IBS-values are deemed as superior in that the predictions of the corresponding survival model agree with reality to a greater extent. Calculating such an IBS-value requires choosing a time-dependent weighting function by which the tBS-values are blended and summed together. We select a uniform weight over spell time in the interest of simplicity, as implemented in the erstwhile \texttt{tBrierScore()} function. For interpreting an IBS-value, \citet{Graf1999} noted a simple rule of thumb of IBS $<0.25$, having made a plausible assumption about the survival probability. In particular, the authors reasoned that, in the absence of any information, one may assign an average survival probability of $\hat{S}(t)=0.5$ to all subject-spells at a given $t$. The corresponding tBS-value of this `model' would then be 0.25, which can be used as an upper limit of sorts when interpreting any other IBS.
Accordingly, and when examining the inset graph in \autoref{fig:timeBrierScore}, it is clear that the IBS-values of both the DtH-Basic model (0.531) and the ST-model (0.563) breach this rule of thumb. In contrast, the IBS of the DtH-Advanced model (0.162) is smaller by far, which attests of the model's superior accuracy when producing predictions.

\begin{figure}[!ht]
    \centering
    \begin{subfigure}[b]{0.75\textwidth}
        \caption{Best fitting models}
        \centering\includegraphics[width=1\linewidth,height=0.4\textheight]{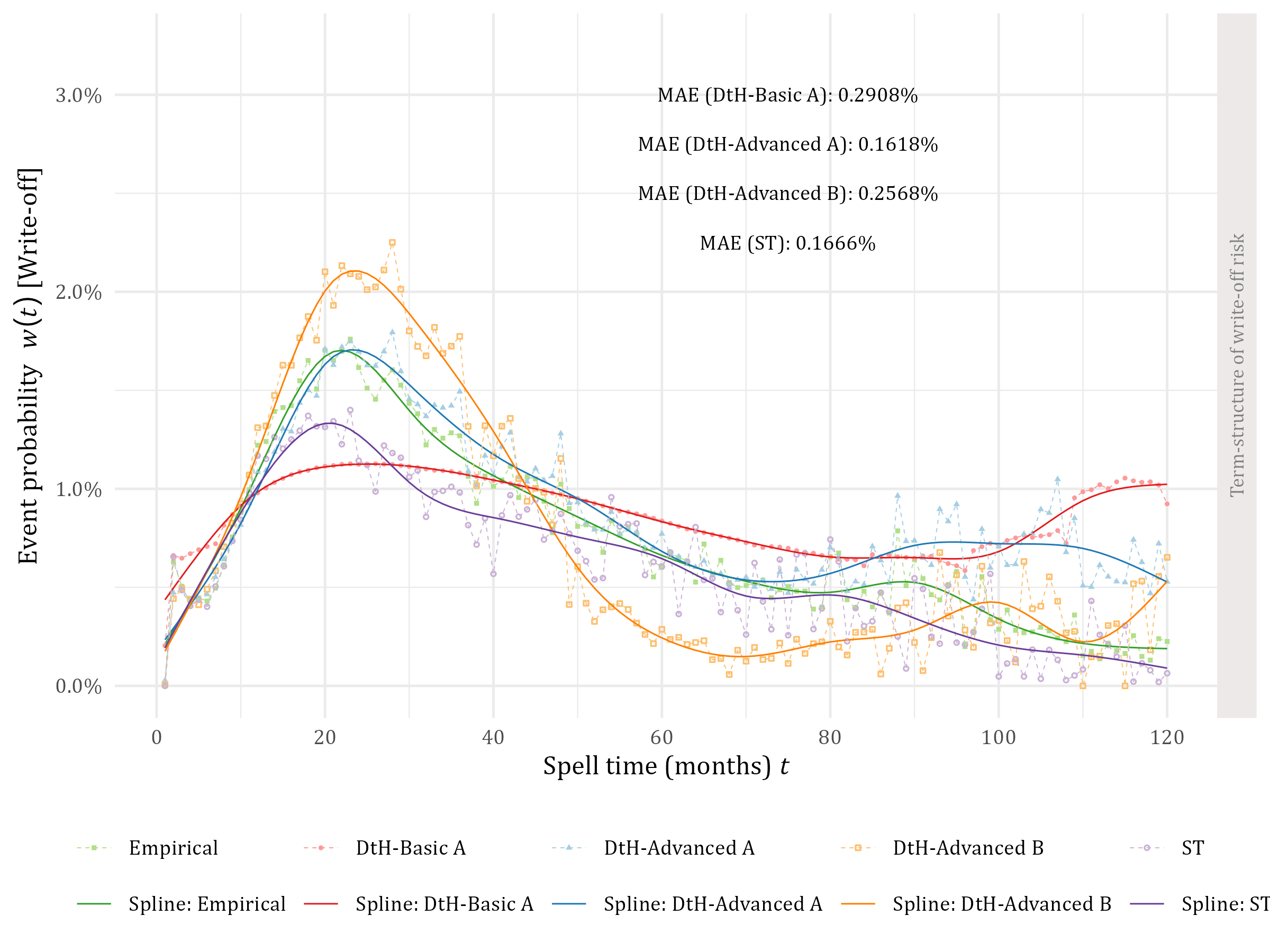}
        \label{fig:TermStructures_a}
    \end{subfigure} \\
    \begin{subfigure}[b]{0.75\textwidth}
        \caption{Worst fitting models}
        \centering\includegraphics[width=1\linewidth,height=0.4\textheight]{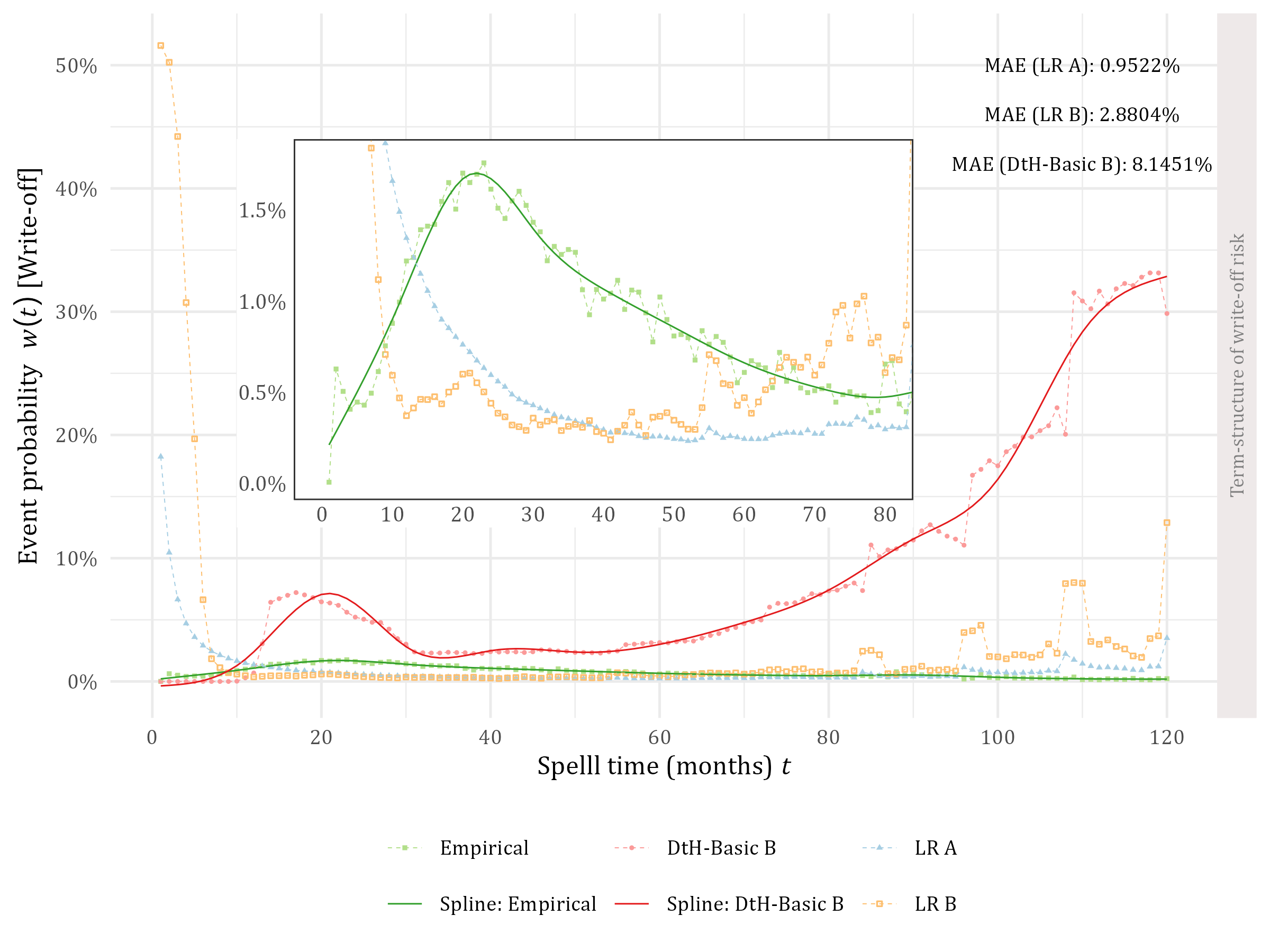}
        \label{fig:TermStructures_b}
    \end{subfigure}    
    \caption{The empirical vs expected term-structures of write-off risk over spell time $t$ per model type. Natural splines are fit in summarising the overall trend. The MAE between the empirical and each expected values are annotated. 
    The best fitting models are shown in panel \textbf{(a)}, whereas panel \textbf{(b)} contains the worst fitting models.
    }
    \label{fig:TermStructures}
\end{figure}

% Term-structures
Our third diagnostic is a comparison of term-structures, which is described as follows.
All of our write-off risk models contain at least one time factor of sorts: time in default spell for the survival models, and default spell age for the LR-model. This model design allows for the derivation of the aggregate write-off probabilities over spell time $t=t_{(1)},\dots, t_{(m)}$, i.e., the write-off term-structure. The \textit{empirical} term-structure is derived using the event probabilities emanating from a Kaplan-Meier analysis, as discussed and illustrated previously in \autoref{sec:survFundamentals}. Against this backdrop, we compare in \autoref{fig:TermStructures} the \textit{expected} term-structures as produced by our models, where the average write-off probability across all applicable spells is obtained for each $t$. Evidently, the term-structure of the LR-model in \autoref{fig:TermStructures_b} diverges dramatically from the empirical term-structure (green) and even differs in shape; a fact that is true for both model Types A and B. This divergence can be quantified by calculating the MAE between any pair of line graphs, with the MAE being 0.952\% (LR Type A) and 2.880\% (LR Type B) respectively. On the other end of the spectrum, both of the DtH-Advanced models in \autoref{fig:TermStructures_a} approximate the empirical term-structure the closest, with the MAE being 0.162\% (Type A) and 0.257\% (Type B) respectively. The survival tree scores a close second place, with an MAE of 0.167\%. It also appears to under-predict the empirical term-structure over most $t$, which is certainly not risk-prudent.
Nonetheless, these results underscore the ability of the survival models to produce predictions that can evolve over time as a default spell unfolds, and which agree more closely with reality. In contrast, and although simplistic, a basic cross-sectional LR-model cannot so easily contend with the element of time in rendering accurate predictions in aggregate.

% Wider comparison: survival models vs LR
As for our fourth diagnostic, we compare the cross-sectional LR-models to the dynamic survival models using the classical AUC-statistic and Brier scores. 
Given the time-dependent nature of survival models, one would need to select a particular time point when comparing them to the LR-model. We choose the point $t=44$ at which the median survival probability (50\%) occurs, based on a Kaplan-Meier analysis of the survival probability.
The following results are obtained, as summarised in \autoref{tab:Diagnostics} across the aforementioned two diagnostics. We note the following points for the Type A models. Firstly, it is by now unsurprising (though still important to note) that the DtH-Advanced model outperforms its basic counterpart yet again. 
Secondly, the LR-model performs better than the DtH-Basic model, but still underperforms against the DtH-Advanced model, despite the LR-model having a very similar set of inputs. 
In comparing Type A models versus their Type B counterparts, the results suggest that dichotomisation can erode model performance. All Type B models achieved similar Brier-scores (6.4--6.62), which are orders of magnitude worse than those of the Type A models. However, we note that dichotomisation inherently leads to information loss, and so the degradation in diagnostics is not entirely surprising. 
Dichotomisation is itself a relatively standard practice for cross-sectional probabilistic classifier models, including our LR-model. 
Yet dichotomising the predictions from a hazard model is largely unstudied amidst censored observations, at least as far as we know. This fact may very well explain the relatively poor performance of the Type B DtH-models. Future research can certainly focus on this aspect (censored observations) when dichotomising the output of hazard models using some threshold.
Lastly, the ST-model performed admirably across both diagnostics, though did not outperform the DtH-Advanced model, which corroborates the model ranking in our previous diagnostics.

\begin{table}[!ht]
\centering
\caption{A summary of various diagnostics by modelling technique, having assessed the survival models at $t=44$ months in default. Underlined figures indicate the best-in-class model.}
\label{tab:Diagnostics}
\begin{tabular}{@{}lll@{}}
\toprule
\textbf{Model}         & \textbf{AUC}     & \textbf{Brier score} \\ \midrule
LR: A            & 86.29\% & 5.48      \\
LR: B            & 74.83\% & 6.55     \\ 
DtH-Basic: A     & 61.18\% & 1.57      \\
DtH-Basic: B     & 49.55\% & 6.4      \\
DtH-Advanced: A  & \underline{97.15\%} & \underline{0.47}      \\
DtH-Advanced: B     & 51.30\% & 6.62      \\
Survival tree & 75.75\% &   1.48          \\ \bottomrule
\end{tabular}
\end{table}

\begin{figure}[!ht]
    \centering
    \includegraphics[width=0.79\linewidth,height=0.5\textheight]{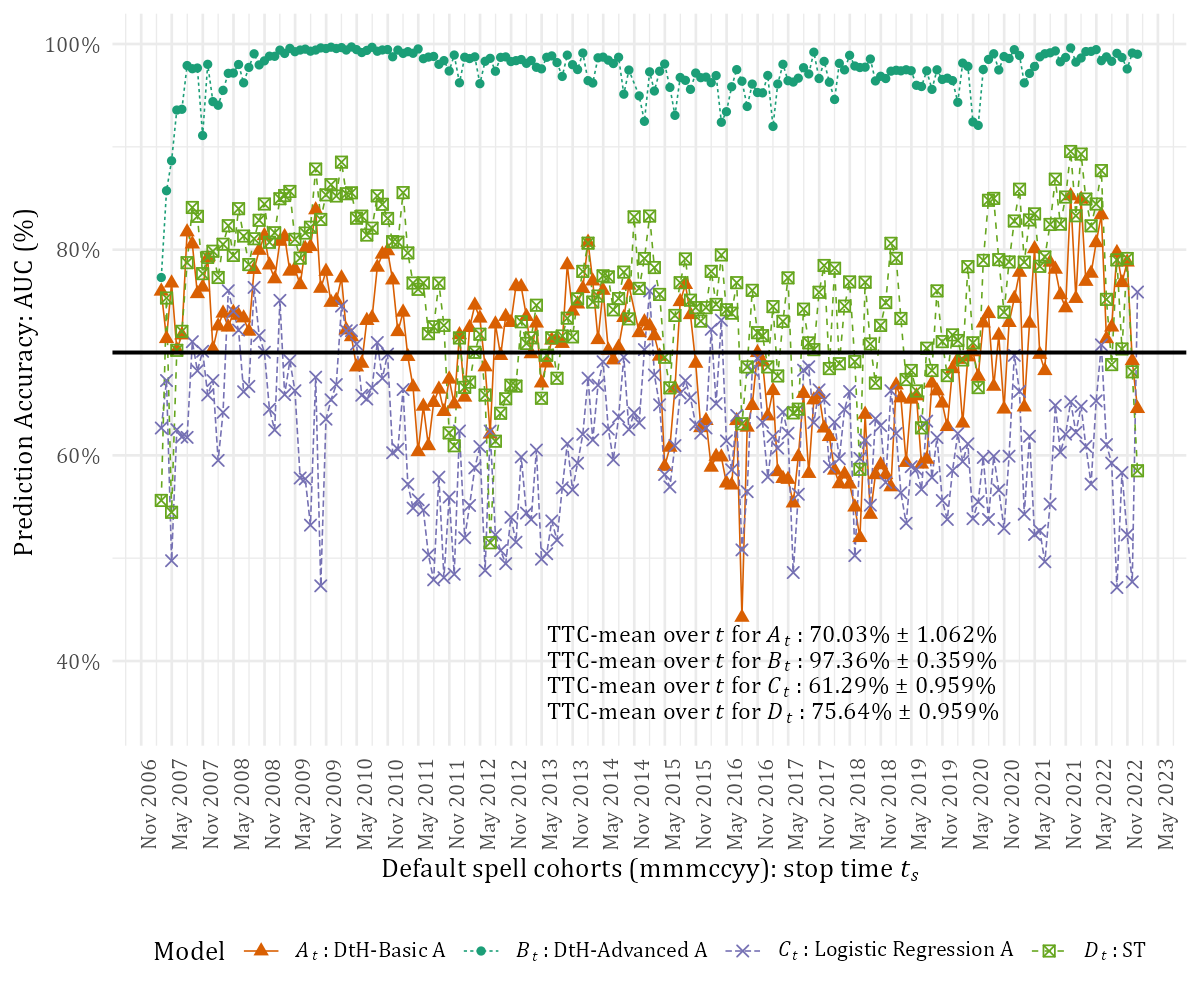}
    \caption{AUC-values over time (resolution date) by model type for the Type A-series. The through-the-cycle (TTC) means of the AUC-values are annotated, with 95\% confidence intervals. The black line (70\%) represents an acceptable level of the AUC over time, based on industry convention.}
    \label{fig:AUC_time}
\end{figure}

% AUC over time
Our fifth and final diagnostic is based on assessing the discriminatory power over time using classical ROC-analysis.
While the AUC-statistic in \autoref{tab:Diagnostics} is calculated across the entire sample, one may also partition the sample by resolution date, and then reperform ROC-analysis within each date-based partition. The AUC is then calculated at each time point, which can be used in assessing the discriminatory power of each model over time; see \autoref{fig:AUC_time} for the Type A-series of models. The results show that the LR-model's discriminatory power is not only the worst over time (with a \textit{through-the-cycle} [TTC] mean AUC-value of 61.29\%), but also varies the most. In contrast, the DtH-Advanced model performs the best (with a TTC-mean AUC-value of 97.36\%), and achieves a remarkably stable set of AUC-values over time. The DtH-Basic and ST models produced similar AUC-values over time, with TTC-means of 70.03\% and 75.64\% respectively. That said, there is a wide disparity in the results of both models, when compared to that of the DtH-Advanced model.
% Type B
Although not shown, we reperform the same analysis for the Type B-series of models, purely in the interest of completeness. The trends reverse in that the LR-model now achieves the greatest AUC over time with a TTC-mean of 67.61\%, followed by the DtH-Basic (51.07\%) and the DtH-Advanced (46.62\%) models. However, we believe it best not to appropriate too much meaning to these Type B results, given the problematic dichotomisation of survival models.

\section{Comparing single-stage and two-stage LGD-models}
\label{sec:results_LGD}

\begin{figure}[ht!]
\centering
\begin{subfigure}[b]{0.49\textwidth}
    \caption{Single-stage: Gaussian GLM}
    \centering\includegraphics[width=1\linewidth,height=0.27\textheight]{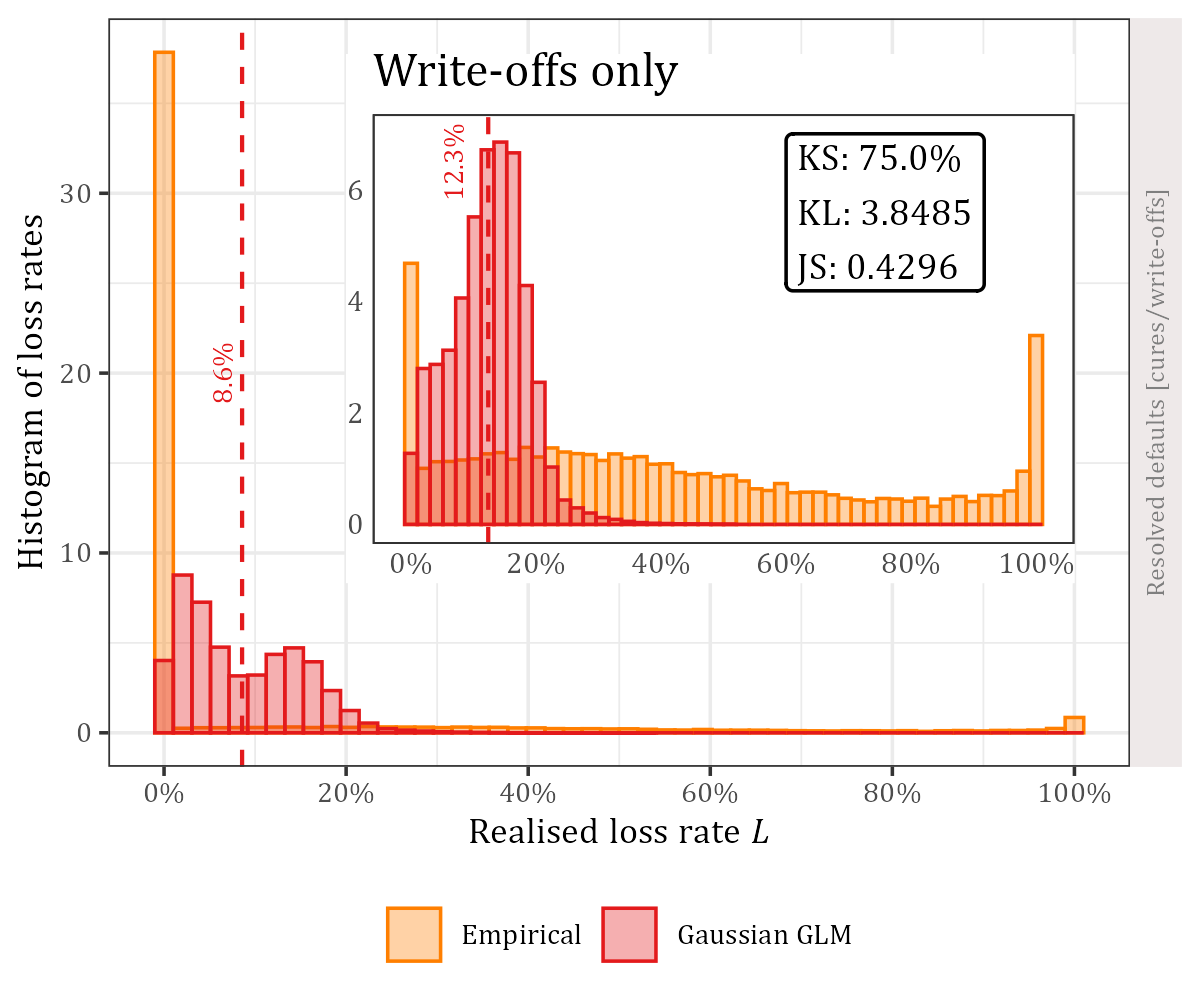}\label{fig:DistrActExp_a}
\end{subfigure}
\begin{subfigure}[b]{0.49\textwidth}
    \caption{Single-stage: Tweedie CP-GLM}
    \centering\includegraphics[width=1\linewidth,height=0.27\textheight]{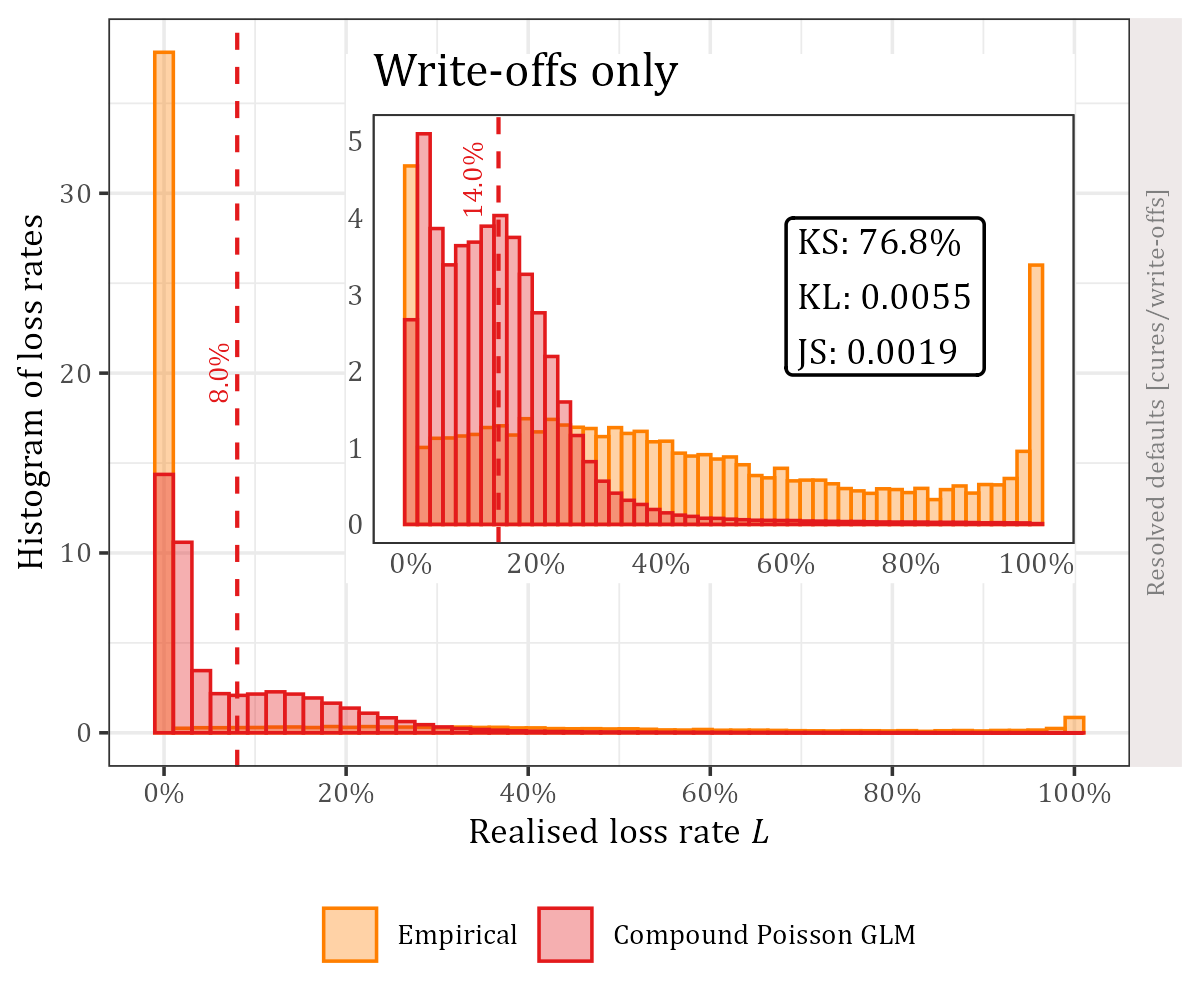}\label{fig:DistrActExp_b}
\end{subfigure} \\
\begin{subfigure}[b]{0.49\textwidth}
    \caption{Two-stage: LR A}
    \centering\includegraphics[width=1\linewidth,height=0.27\textheight]{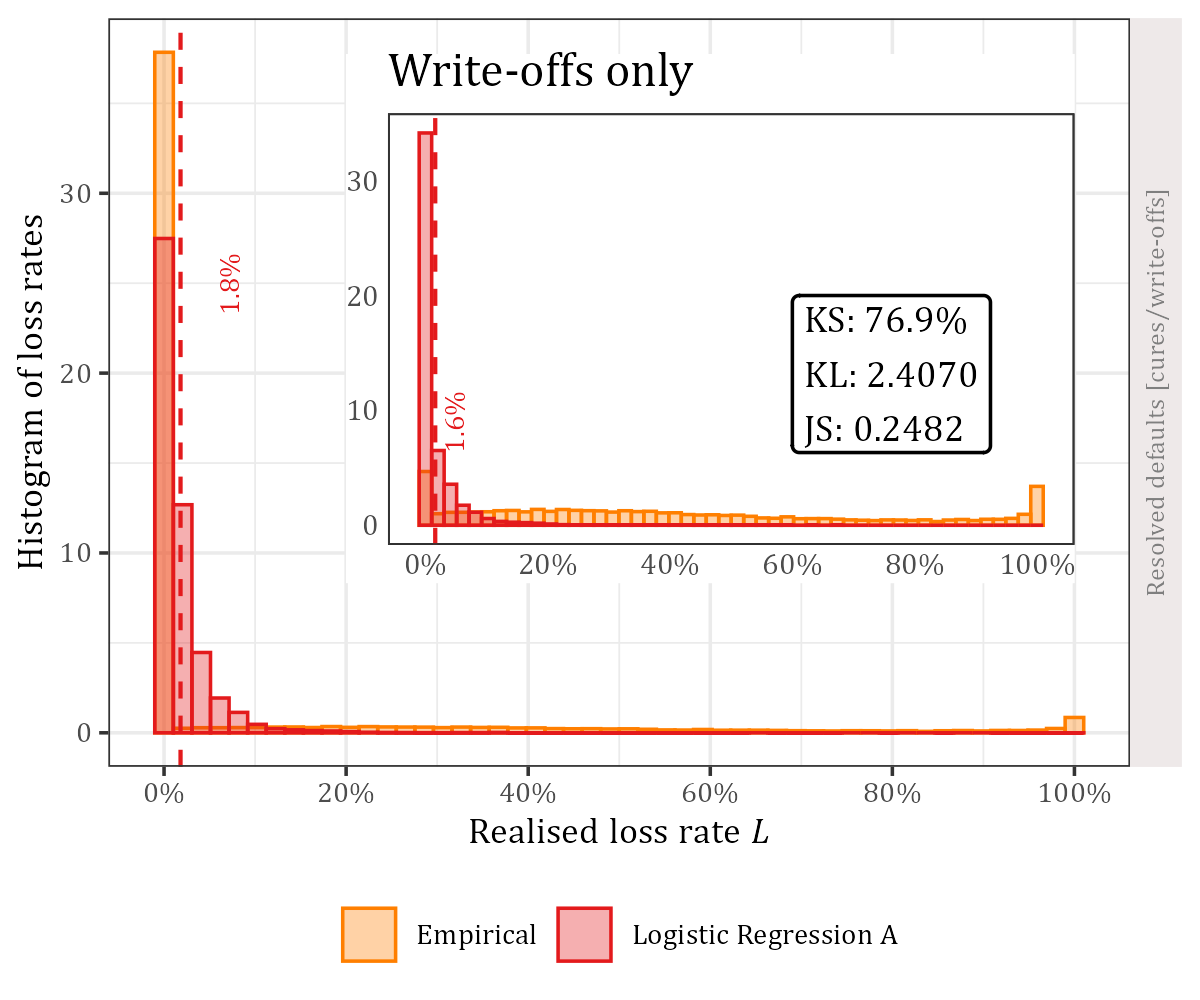}\label{fig:DistrActExp_c}
\end{subfigure}
\begin{subfigure}[b]{0.49\textwidth}
    \caption{Two-stage: LR B}
    \centering\includegraphics[width=1\linewidth,height=0.27\textheight]{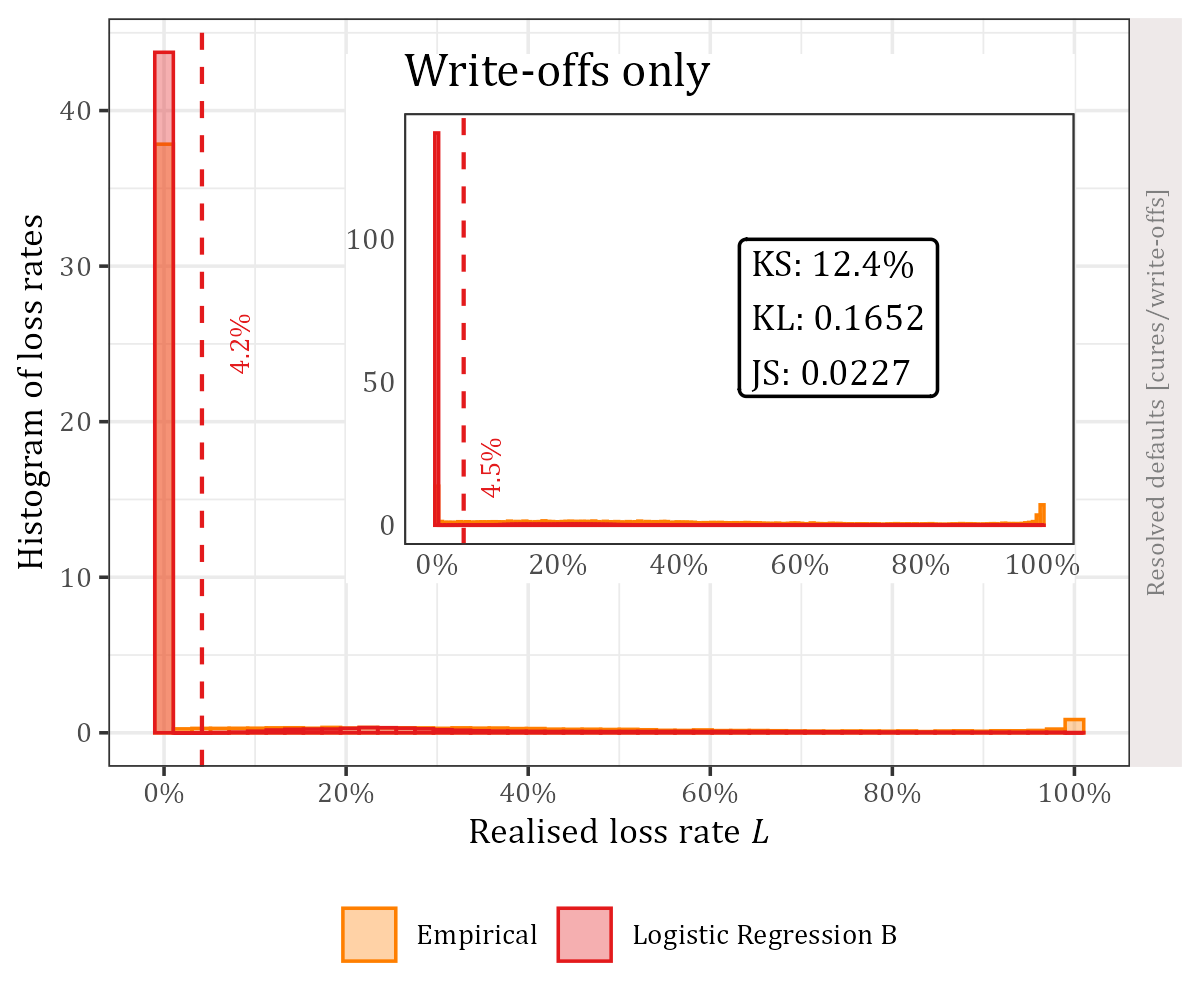}\label{fig:DistrActExp_d}
\end{subfigure}\\
\begin{subfigure}[b]{0.49\textwidth}
    \caption{Two-stage: DtH-Basic A}
    \centering\includegraphics[width=1\linewidth,height=0.27\textheight]{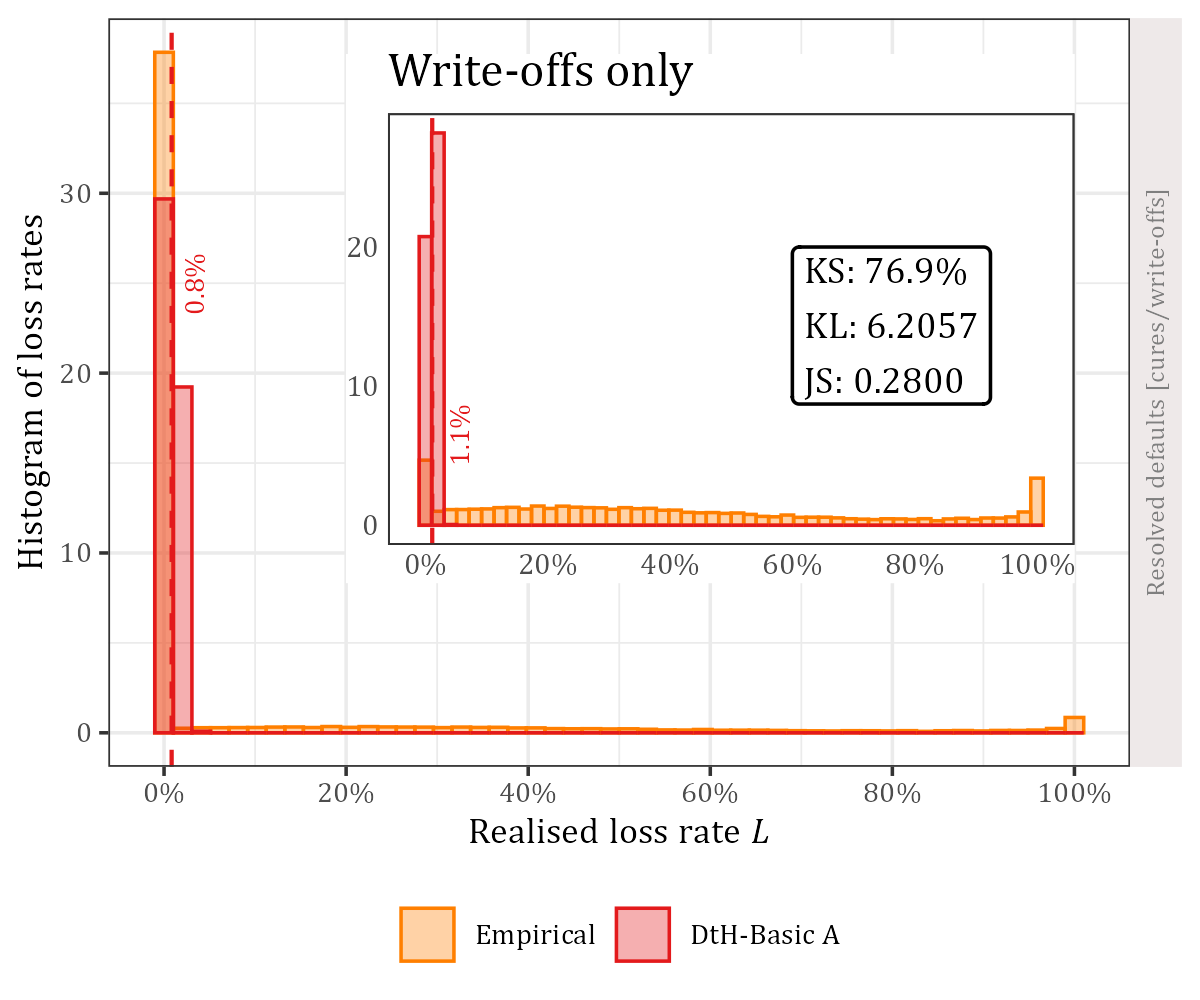}\label{fig:DistrActExp_e}
\end{subfigure}
\begin{subfigure}[b]{0.49\textwidth}
    \caption{Two-stage: DtH-Basic B}
    \centering\includegraphics[width=1\linewidth,height=0.27\textheight]{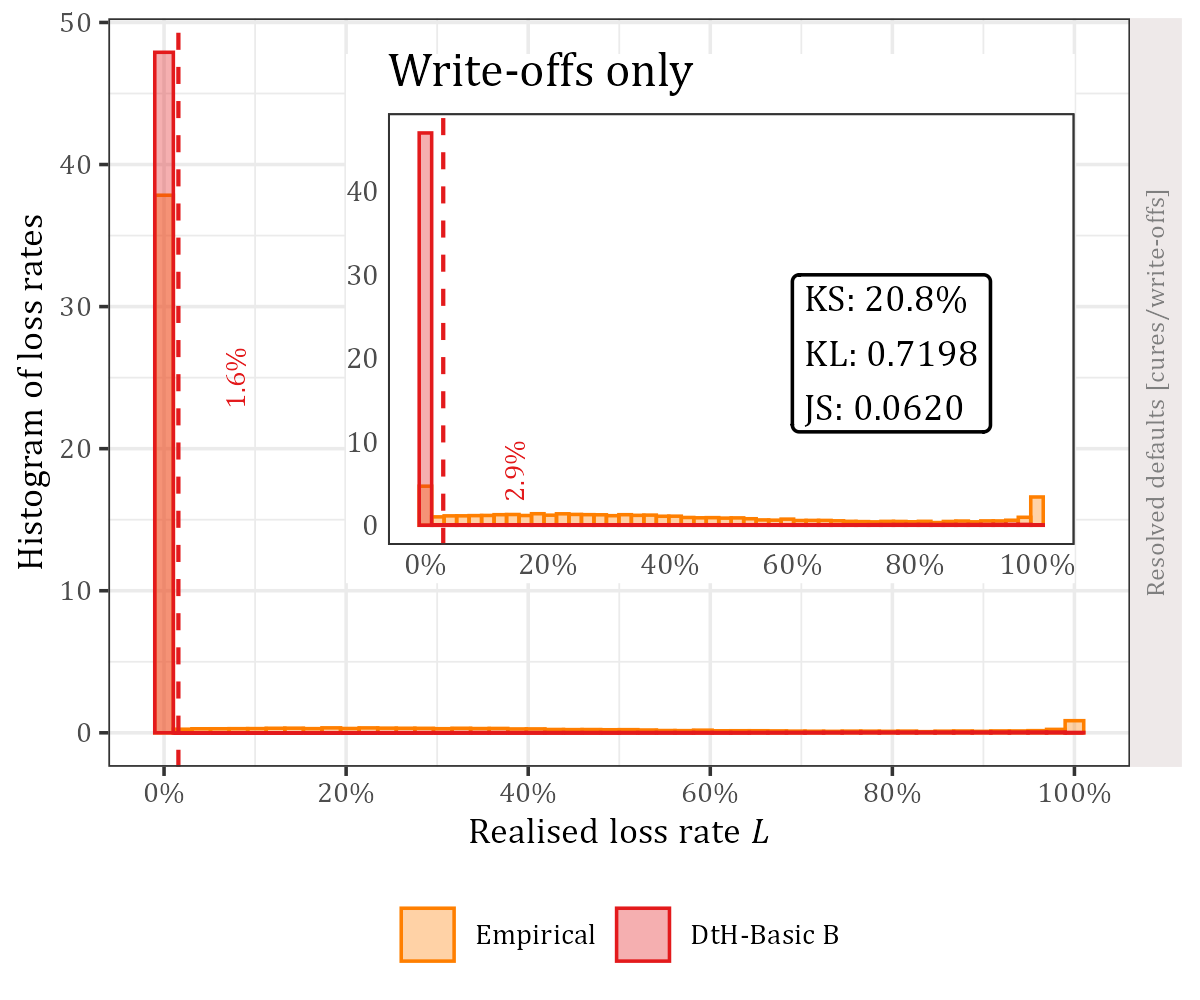}\label{fig:DistrActExp_f}
\end{subfigure}
\hfill%
\caption{LGD-distributions of empirical vs expected loss rates per model. Distributional diagnostics include the Kolmorogov-Smirnov (KS) test statistic, Kullback-Leibler (KL), and the Jensen-Shannon (JS) divergence. The mean expected loss rate is overlaid (in red) per model and in each inset graph.}\label{fig:DistrActExp}
\end{figure}

\begin{figure}[ht!]
\centering
\begin{subfigure}[b]{0.49\textwidth}
    \caption{Two-stage (write-off risk): DtH-Advanced A}
    \centering\includegraphics[width=1\linewidth,height=0.27\textheight]{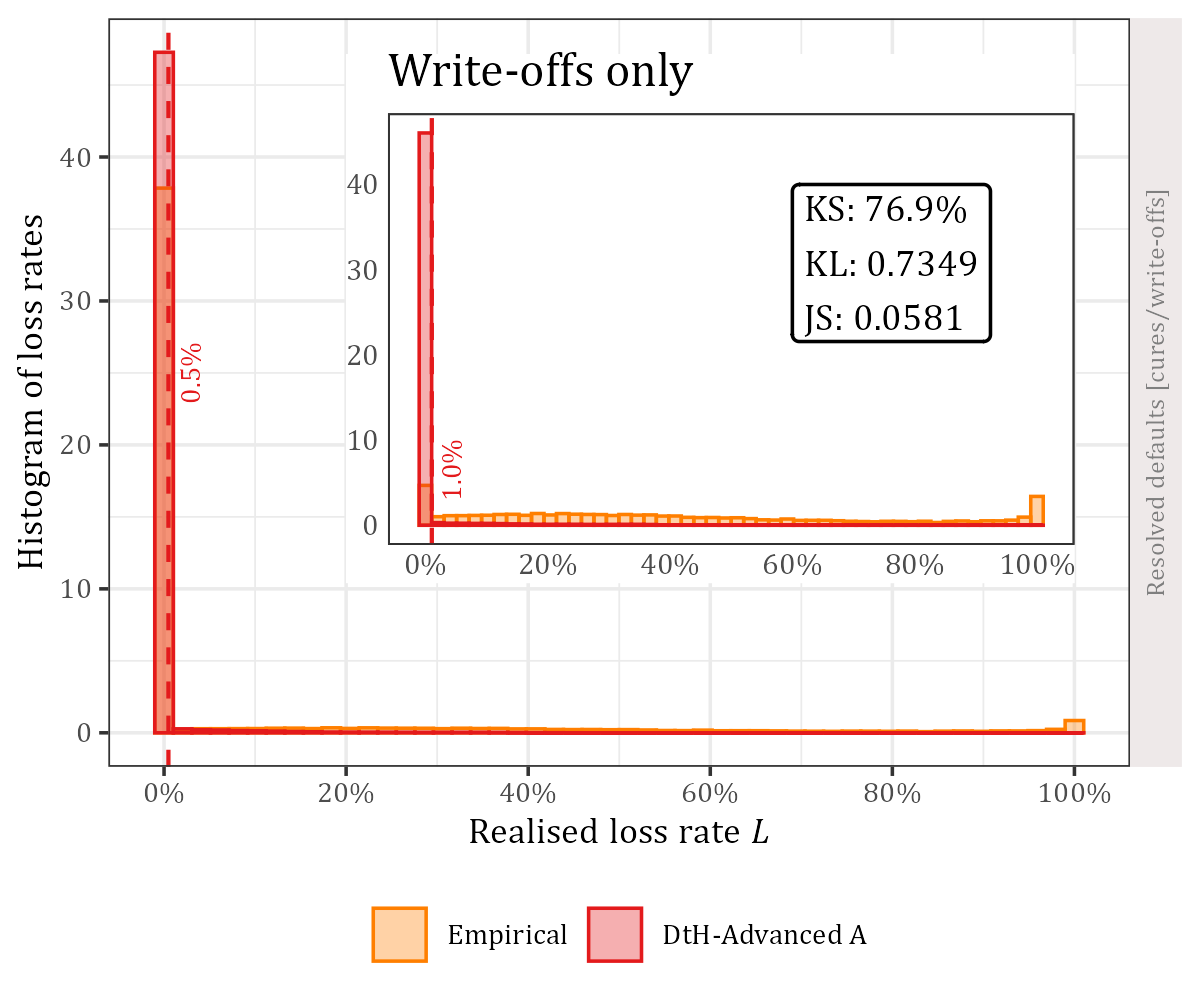}\label{fig:DistrActExp_g}
\end{subfigure}
\begin{subfigure}[b]{0.49\textwidth}
    \caption{Two-stage (write-off risk): DtH-Advanced B}
    \centering\includegraphics[width=1\linewidth,height=0.27\textheight]{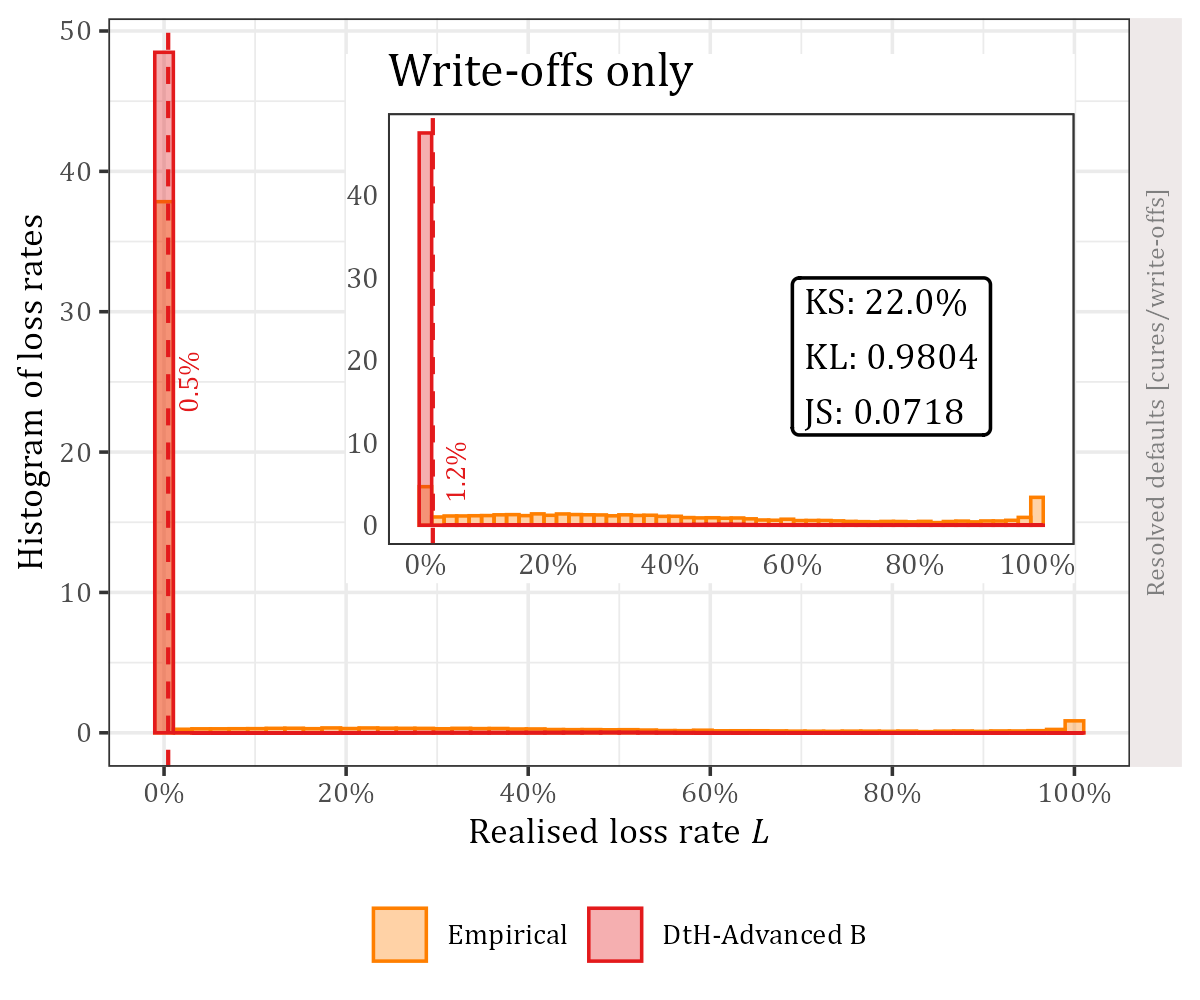}\label{fig:DistrActExp_h}
\end{subfigure}
\begin{subfigure}[b]{0.49\textwidth}
    \caption{Two-stage (write-off risk): Survival tree}
    \centering\includegraphics[width=1\linewidth,height=0.27\textheight]{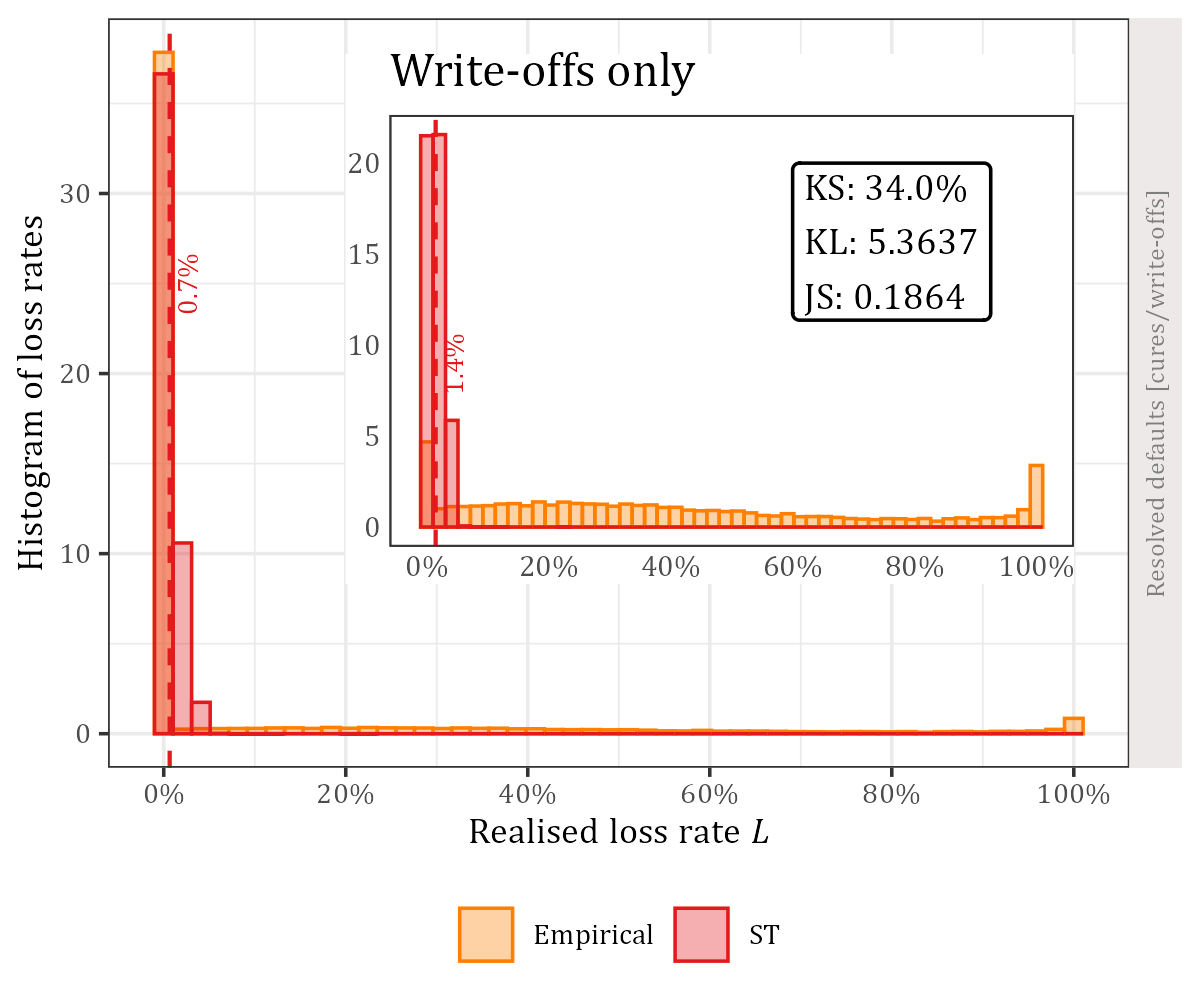}\label{fig:DistrActExp_i}
\end{subfigure}
\begin{subfigure}[b]{0.49\textwidth}
    \caption{Two-stage (loss severity): Tweedie CP-GLM}
    \centering\includegraphics[width=1\linewidth,height=0.27\textheight]{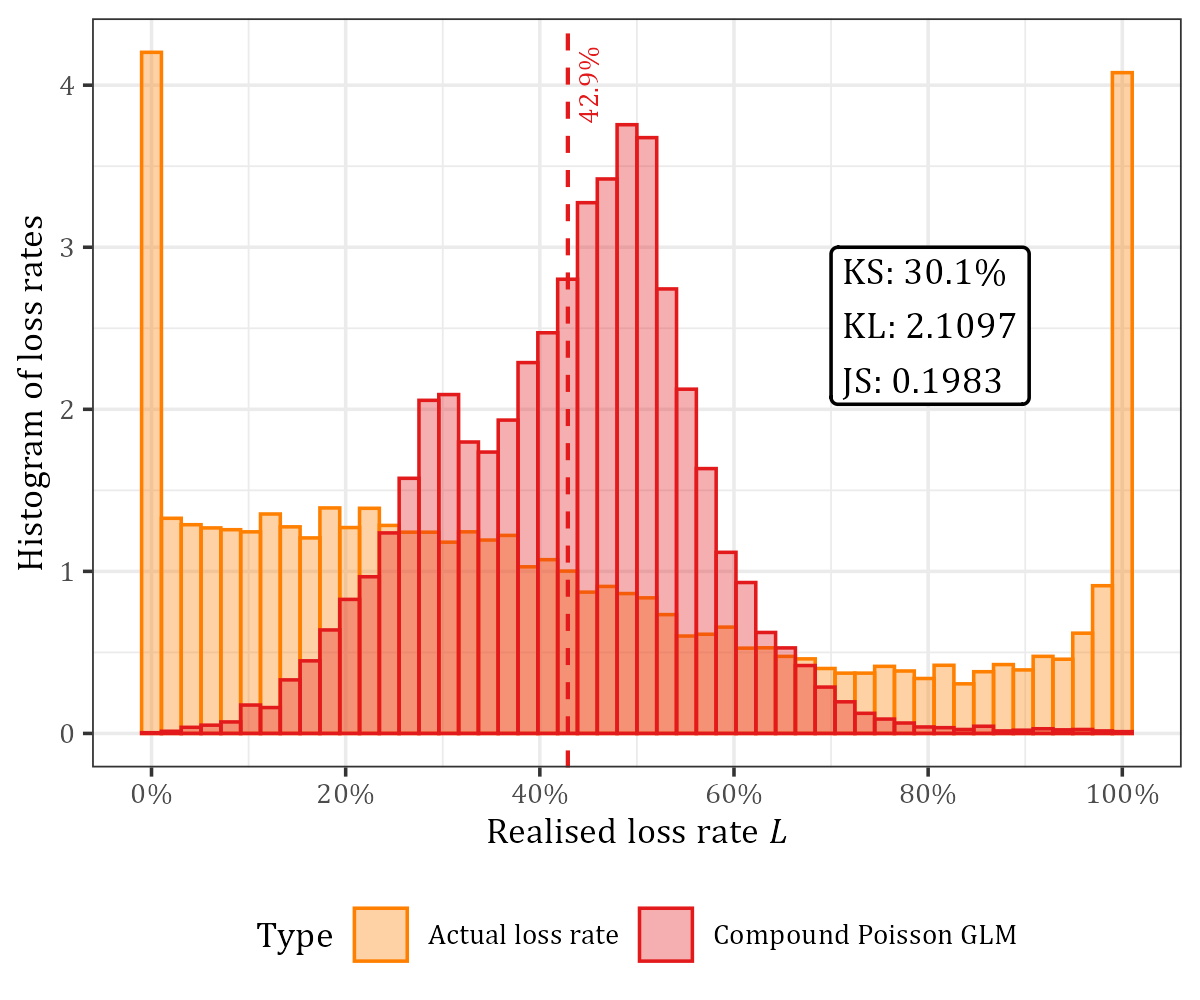}\label{fig:DistrActExp_j}
\end{subfigure}
\hfill%
\caption{LGD-distributions of empirical vs expected loss rates per model (continued).}\label{fig:DistrActExp2}
\end{figure}

In evaluating the various LGD-models, we opt for a simple distributional analysis wherein the empirical LGD-distribution of realised LGD-values is compared with its expected counterpart, as produced by each LGD-model. The premise hereof is that the distribution of the predicted LGD-values ought to resemble that of the realised LGD-values as closely as possible. In analysing the similarity between either distribution, we employ the well-known \textit{Kolmogorov-Smirnov} (KS) test statistic, as well as two divergence measures from \citet{zeng2013metric}: \textit{Kullback-Leibler} (KL) and \textit{Jensen-Shannon} (JS). Smaller values in any of these measures indicate greater similarity between two distributions (empirical vs expected), and hence a more accurate model. We show the distributional analyses in \crefrange{fig:DistrActExp}{fig:DistrActExp2}, and note the following three main points.

Firstly, and as shown in \autoref{fig:DistrActExp_j}, the underlying loss severity model within the two-stage approach struggles to capture the salient aspects of the empirical distribution of realised loss rates. These aspects include the two modes near zero and one, as well as the slight right-skew. The model itself achieved a coefficient of determination of only $R^2=22.45\%$, which is relatively weak in explaining the variance. While the focus of our work remains on write-off risk models, the influence of the loss severity component is unmistakeable within the composite LGD-model. Accordingly, the underperformance of the two-stage LGD-models is noticeable across most similarity metrics, at least when compared to their single-stage counterparts. For example, the Tweedie CP-GLM model achieved a KL of 0.0055 and a JS of 0.0019 (see \autoref{fig:DistrActExp_b}), which are much smaller/better than those metrics of the best-in-class two-stage Type A LGD-model (see \autoref{fig:DistrActExp_g}). I.e., the DtH-Advanced Type A model has a KL of 0.7349 and a JS of 0.0581. This result is corroborated by the fact that the mean loss rates of the single-stage LGD-models (8.6\% and 8.0\% respectively) are much closer to the empirical mean of about 10\%, relative to those means of the two-stage LGD-models.
However, we posit that this result is highly dependent on the shape of the underlying empirical LGD-distribution, which resembles an `L'-shape in our case rather than the typical `U'-shape of most LGD-distributions. Modelling an `L'-shaped LGD-distribution proves to be demonstrably difficult, largely as a result of an abundance of zero-valued cures. For the two-stage LGD-modelling approach to triumph over its single-stage sibling, one would likely have to produce a fairly decent loss severity model as well (at least better than at present), which should match the quality of the write-off risk models.

% Type A survival models
Secondly, it would appear from \crefrange{fig:DistrActExp}{fig:DistrActExp2} that the DtH-Advanced Type A model outperforms the other two-stage models (Type A). In particular, the former model scores a KL-statistic that improves more than three-fold upon that of the LR Type A model (see \autoref{fig:DistrActExp_c}), and more than four-fold in the JS-statistic. We believe that this result underscores the dynamicity of survival models over their cross-sectional counterparts, which ultimately renders the LGD-predictions as more accurate. Interestingly, the more exotic ST-model (see \autoref{fig:DistrActExp_i}) did not outperform either of the two Dth-Advanced models across the KL and JS statistics.
% Dichotomisation
Thirdly, and as troublesome as it is, dichotomisation did improve the similarity metrics for both the LR and DtH-Basic models. As an example, the KL of the LR-model strengthened almost 15 times from 2.47070 to 0.1652 (see \autoref{fig:DistrActExp_d}) following dichotomisation, whereas the DtH-Basic model's KL improved almost nine times from 6.2057 to 0.7198 (see \autoref{fig:DistrActExp_f}). However, we achieve the opposite result for the DtH-Advanced models in that dichotomisation yielded greater dissimilarity between the distributions. It is possible that this result attests yet again of the issue of dichotomising the output of survival models, as previously discussed in \autoref{sec:results}. Nonetheless, it seems that greater accuracy in LGD-predictions can be attained via dichotomisation, especially so for the cross-sectional LR-model.

\section{Conclusion}
\label{sec:conclusion}

% Generic problem at hand
It is notoriously difficult to model the LGD risk parameter in credit risk modelling. This is particularly true when considering the two aspects that characterise most LGD-distributions: bimodality and skewed tails. These two aspects can complicate the direct modelling of the realised LGD-values as a function of a set of input variables (or predictors).
As a result, these aspects have inspired a \textit{two-stage} LGD-modelling approach in literature, whereby the LGD is commonly decomposed into a write-off risk component and a loss severity (given write-off) component. Each component can then be separately modelled, whereafter the product of the two model outputs are multiplied together in forming an LGD-estimate. Various studies have highlighted the success of the two-stage LGD-modelling approach. However, we note that the majority of these studies have contended with a so-called `U-shaped' LGD-distribution, where the two modes are each respectively located at the tail-ends of the distribution.
In contrast, our dataset exhibits more of an `L-shaped' LGD-distribution, with one major mode located at zero (representing the cures), and an extremely minor mode at one. We believe that this shape may be more characteristic of a secured portfolio (e.g., mortgages) in general, whereas a U-shape better reflects an unsecured portfolio (e.g., credit cards). Regardless, a peculiar distributional shape will of course have consequences for the success of any particular modelling strategy. In turn, an inappropriate modelling strategy may introduce unnecessary bias into the ECL-estimates under IFRS 9, which can compromise a bank's loss provisions.

% Contributions: setup + term-structure
We have explored the use of survival analysis in modelling the write-off risk component, having used South African mortgage data. One of the most important input variables within these survival models is that of the time spent in a default spell. Its inclusion allows us to approximate the empirical term-structure of write-off risk, i.e., the collection of write-off probabilities over default spell time $t$. This empirical term-structure is itself generated using the Kaplan-Meier (KM) estimator from survival analysis, thereby leveraging right-censored observations in constructing an `actual/empirical' curve that represents reality. By itself, this KM-based method already represents a more efficient use of data.
The output of our various models (including survival models) were then duly aggregated and compared to this empirical term-structure; itself forming a reusable and simple diagnostic for assessing dynamic LGD-models. Our results showed that a particular type of survival model -- a \textit{discrete-time hazard} (Dth) model -- outperformed other two-stage contenders across most metrics, particularly when comparing the expected vs the empirical term-structures. Other metrics included time-dependent varieties of both discriminatory power (tROC-analysis) and the Brier score (tBS).

% Contributions: 
As inspired by previous studies, we ensconced our write-off risk models within a two-stage LGD-modelling approach, such that we can conduct a benchmark study amongst different modelling techniques. Doing so required building a bespoke loss severity model, which was itself estimated using a Tweedie compound Poisson (CP) GLM. In fact, a Tweedie CP-GLM also drove one of our two candidate single-stage LGD-models, with the other one being a Gaussian GLM.
However, in modelling the loss severity within the two-stage approach, we only achieved a mediocre fit to the data. The downstream results showed that a single-stage LGD-model (CP-GLM) outperformed all of the two-stage LGD-models, including the more exotic conditional inference survival tree (ST) model. Though rather unexpected, we ascribe this result to the peculiar shape of the LGD-distribution. The implication is that a two-stage LGD-modelling approach can only triumph over its single-stage counterpart when \textit{both} of its components perform adequately, which is not the case at present.
Another result is that the dichotomisation of the write-off probability into 0/1-values improved the accuracy of most two-stage LGD-models. In so doing, zero-valued cures can be predicted more accurately, and the resulting mode at zero within the LGD-distribution is better captured.
Overall, our work serves both as a benchmark study and a tutorial to modelling the write-off risk component in LGD-modelling, especially when considering the ancillary material in the appendix.

% Future work
Future research can focus on an alternative way of handling competing risks within our survival models (i.e., the cured outcomes). Doing so can theoretically reduce the latent bias introduced by such risks, which are currently treated as right-censored observations. E.g., each cause (or outcome, such as write-off and cure) actually has a \textit{cumulative incidence function} (CIF) in the competing risks literature, where such a CIF is denoted as $F_j(t)$, and signifies the probability of experiencing failure from a specific cause $j$ by time $t$. One may estimate each cause-specific CIF by using the Aalen-Johansen estimator, which accounts for the possibility of other causes occurring first.
Another research avenue is experimenting with machine learning forms of survival analysis, e.g., random survival forests. Doing so may deliver greater performance than what we have obtained with the slightly less exotic ST-model.
Future effort may also be dedicated to building models that more accurately predict the loss severity component. This route may very well redeem the two-stage LGD-modelling strategy, especially when coupled with decent write-off risk models, such as those that we have developed in this study.
One can also explore the appropriate dichotomisation of the output from survival models within the present context, which we believe is currently understudied.

\appendix
\section{Appendix}
\label{app:appendix}

In \autoref{app:SurvivalTrees}, we review the fundamentals of a particular type of survival trees, which includes its in-depth formulation and application in the R-programming language.
An illustration is given in \autoref{app:dataStructure} of how one should structure the underlying credit data towards conducting survival analysis in estimating the write-off term-structure.
In affirming the representativeness of resampled datasets, the \textit{resolution rate} is defined and illustrated in \autoref{app:resolutionRate}.
We formulate a brief optimisation procedure in \autoref{app:CostMultipleOptim} for finding the cost multiple $a$ within the Generalised Youden Index (GYI), which is used in dichotomising our model outputs.
Lastly, \autoref{app:InputSpace} contains a description of the input variables within our models.

\subsection{An overview of conditional inference survival trees}
\label{app:SurvivalTrees}

Survival trees extend \textit{recursive binary partitioning} (RBP) methods to right-censored time-to-event data. In particular, the covariate (or input) space is recursively partitioned into increasingly homogeneous subgroups, such that observations within a terminal node exhibit similar behaviour in the subsequent survival distributions. In credit risk applications, survival trees can serve as a flexible tool for uncovering heterogeneity in write-off timing (the response variable) when given various input variables. These trees can help with discovering complex interactions and non-linear effects, without imposing restrictive (semi-)parametric assumptions such as proportional hazards. For an overview of survival trees in the credit risk context, see \citet{frydman2022random}.

In this study, we focus on survival trees constructed within the \textit{conditional inference} framework of \citet{hothorn2006unbiased}, abbreviated as \textit{HHZ}-trees and implemented in the \texttt{partykit} R-package from \citet{hothorn2015partykit}. The construction of such HHZ-trees deliberately separates the variable selection step from the subsequent splitting step, wherein the space of the selected variable is split into two regions. Separating these two steps avoids the selection bias that is typically associated with classical RBP-methods. To this point, a typical RBP-method jointly optimises a splitting criterion over all inputs variables and across all admissible cut points. This implies the disproportionate selection of those inputs with many potential split points (due to different measurement scales), or of those categorical inputs with many categories.
This selection bias exists even when the association between such inputs and the outcome is weak. While post-estimation tree pruning is commonly employed to mitigate overfitting, it does not address this inherent selection bias and instead introduces additional tuning parameters, as discussed by \citet{hothorn2006unbiased}. In contrast, HHZ-trees assess the association between each input and the outcome using the conditional distribution of a linear test statistic, as derived using a permutation-based test procedure. The optimal binary split is only determined \textit{after} identifying the variable that exhibits the strongest association with the outcome according to this statistical criterion, thereby eliminating the aforementioned selection bias. We shall now discuss various aspects of HHZ-trees over the next few subsections.

\subsubsection{Broad steps for implementing recursive binary partitioning (RBP) using censored data}

% Generic mathematical setup
Let $T_{ij}$ denote the spell duration for default spell $j\in\{1,\dots,n_i\}$ of loan $i\in\{1,\dots,N_p\}$, where $n_i$ denotes the maximum number of spells endured by $i$, and $N_p$ is the total number of loans. Let $\delta_{ij}\in\{0,1\}$ be the event indicator such that $T_{ij}$ represents the write-off time if $\delta_{ij}=1$. Otherwise if $\delta_{ij}=0$, then $T_{ij}$ signifies the right-censoring time. Together, $T_{ij}$ and $\delta_{ij}$ form the bivariate survival response $Y_{ij}=(T_{ij},\delta_{ij})$ with a sample space of $\mathcal{Y}$.
For notational convenience, let us re-index the $n$ observed spells generically by $k=1,\dots,n$, which are treated as independent observations for tree construction, yielding $Y_k=(T_k,\delta_k)$.
We shall investigate the conditional distribution of this response $Y=(Y_1,\dots,Y_k)$ given the observations of $m$ input variables $\boldsymbol{X}=(X_1,\dots,X_m)$; themselves measured at arbitrary scales and taken from the input (or covariate) sample space $\boldsymbol{\mathcal{X}}=\mathcal{X}_1\times\cdots\times\mathcal{X}_m$. In a tree-based framework, the conditional distribution $D(Y\mid \boldsymbol{X})$ of a generic response $Y_k$ is assumed to depend on a function $f$ of the inputs, i.e., 
\begin{equation}
    D(Y_k\mid \boldsymbol{X}_k)=D(Y_k\mid f(X_{1k},\dots,X_{mk})) \, . \nonumber     
\end{equation}
This function $f$ maps each input vector to one of $r$ regions (or terminal nodes), thereby partitioning the input space into the disjoint cells $B_1,\dots,B_r$ such that $\boldsymbol{\mathcal{X}}=\bigcup_{l=1}^r B_l$. Within each terminal node $B_l$, the conditional distribution $D(Y \mid \boldsymbol{X} \in B_l)$ is assumed to be homogeneous, as discussed by \citet{hothorn2006unbiased}. Ultimately, our dataset $\mathcal{D}_n$ that we will use in tree-construction is defined as 
\begin{equation}
    \mathcal{D}_n=\left\{(Y_k,X_{1k},\dots,X_{mk});k=1,\dots,n \right\} \, . \nonumber
\end{equation}

% Case weights and RBP-steps
Consider the case weights $\boldsymbol{w}=(w_1,\dots,w_n)$ that are associated with the $n$ observations. Observations belonging to a particular terminal node $B_l$ will receive non-negative weights, whereas observations outside of $B_l$ will receive weight zero. The following are then generic steps towards implementing RBP within the HHZ-framework of \citet{hothorn2006unbiased}. Firstly, and given $\boldsymbol{w}$, we test the global hypothesis of independence between any of the $m$ covariates and the response $Y$. If we reject this hypothesis, then we select the $v^*$th input with the strongest association with $Y$, as measured by the corresponding test statistic (itself explained later). Secondly, choose a set $A^*\subset \mathcal{X}_{v^*}$ for splitting $\mathcal{X}_{v^*}$ into two disjoint sets $A^*$ and $\mathcal{X}_{v^*} \setminus A^*$. Membership to the resulting two child nodes is defined via updated case weights respective to each node, i.e.,  
\begin{equation} \label{eq:CaseWeights}
    \boldsymbol{w}_{\mathrm{left},k}= w_k \mathbb{I}\left(X_{v^*k} \in A^* \right)
    \, \quad \text{and} \quad \boldsymbol{w}_{\mathrm{right},k}=w_k \mathbb{I}\left(X_{v^*k} \notin A^* \right) \quad \text{for all} \ k=1,\dots,n \, ,
\end{equation}
where $\mathbb{I}(\cdot)$ is the indicator function. Steps 1--2 are then recursively repeated with modified case weight vectors $\boldsymbol{w}_\mathrm{left}$ and $\boldsymbol{w}_\mathrm{right}$ until the global null hypothesis of independence can no longer be rejected at a pre-specified nominal level $\alpha$.

% Best variable identification
\subsubsection{Identifying the `best' input variable using a log-rank type test statistic within the HHZ-framework}

% Variable selection primaries
In Step 1 of constructing an HHZ-tree, \citet{hothorn2006unbiased} advised that one would generally need to assess whether any information about the response $Y$ is contained in the $m$ input variables, i.e., performing variable selection.
Consider the following $m$ partial null hypotheses 
\begin{equation}
    H^v_0: D(Y\mid X_v)=D(Y) \quad v=1,\dots,m \ , \nonumber
\end{equation}
each of which asserts distributional independence between $Y$ and the input variable $X_v$ within the current parent node, as identified and represented by the case weights $\boldsymbol{w}$. From these partial hypotheses, one can form the global null hypothesis $H_0=\bigcap_{v=1}^mH_0^v$, which states that $Y$ is jointly independent of all inputs. However, we shall restrict our attention to the partial hypotheses since they form the basis of variable selection within an HHZ-tree. 

% Testing 
In testing one of these partial hypotheses $H^v_0$, \citet{hothorn2006unbiased} and \citet{fu2017survival} proposed that the association between $Y$ and the input variable $X_v$ is tested using linear test statistics $\boldsymbol{T}_v\left(\mathcal{D}_n, \boldsymbol{w}\right)$ of the form
\begin{equation} \label{eq:linearStatistics}
    \boldsymbol{T}_v\left(\mathcal{D}_n, \boldsymbol{w}\right) = \mathrm{vec}\left( \sum_{k=1}^n w_kg_v(X_{vk})h(Y_k)^\mathrm{T} \right) \in \mathbb{R}^{p_vq} \, .
\end{equation}
In \autoref{eq:linearStatistics}, the "vec" operator stacks the columns of the resulting $p_v\times q$ matrix into a $p_vq$-dimensional column vector. The function $g_v:\mathcal{X}_v\rightarrow\mathbb{R}^{p_v}$ is a non-random transformation of $X_v$ reflecting its measurement scale, where $p_v$ denotes the dimension of the transformed input, i.e., the length of the vector produced by $g_v(\cdot)$. This function is commonly chosen to be the identity $g_v(X_v)=X_v$ for continuous inputs, though other choices certainly exist such as binning schemes or basis expansions (e.g., spline bases).
Furthermore, $h:\mathcal{Y}\times \mathcal{Y}^n\rightarrow\mathbb{R}^q$ is an influence function of the responses $Y_1,\dots,Y_n$ that depends on the observed sample (i.e., the $\mathcal{Y}^n$ argument) and is treated as fixed when assessing the distribution of the test statistic under the null hypothesis $H_0^v$.

% Influence function: log-rank score
As for choosing the influence function $h$ in \autoref{eq:linearStatistics}, it is common to use the \textit{log-rank score} statistic within a survival analysis setting, as described by \citet{fu2017survival}, given the appealing non-parametric nature of this statistic. Having ordered the distinct event times as $t_{(1)} <\cdots t_{(s)} <\cdots < t_{(S)}$, we define $d_s$ as the number of observed write-off events at time $t_{(s)}$ and let $n_s$ denote the size of the risk set immediately prior to time $t_{(s)}$. The log-rank score associated with each bivariate survival observation $Y_k=(T_k,\delta_k)$ is then given by
\begin{equation} \label{eq:logrankScore}
    h(Y_k) =\delta_k -\sum_{s\, : \, t_{(s)}\leq T_k} {  \frac{d_s}{n_s} } \, . 
\end{equation}
Each score $h(Y_k)$ represents the difference between the observed event indicator $\delta_k$ and its expected value under the null hypothesis that all observations share a common survival distribution; i.e., the survival response is independent of any input variable before splitting. In particular, the hazard contributions $d_s/n_s$ are calculated using the global (pooled) risk sets $\left\{ n_s\right\}_{s=1}^S$ formed prior to any splitting. These contributions are then summed across all event times $t_{(s)}\leq T_k$, thereby yielding the expected number of write-off events for the $k^\mathrm{th}$ observation under the aforementioned null model.
Each $h(Y_k)$ is essentially an "observed-minus-expected" contribution, which underpins the classical log-rank score paradigm.

% Log-rank test statistic
We then compute a weighted sum of these log-rank scores from \autoref{eq:logrankScore} for each candidate input variable $X_v$ within the current parent node, thereby calculating the linear statistic
\begin{equation} \label{eq:linearStatistic_logrank}
    T_v = \sum_{k=1}^n {w_k g_v(X_{vk})h(Y_k) } \, .
\end{equation}
This statistic $T_v$ measures the association (or degree of covariation) between the survival response $Y$ and each transformed $g_v(X_v)$ within the current node. But \citet{hothorn2006unbiased} explained that the joint distribution between $Y$ and $X_v$ is generally unspecified under the null hypothesis $H_0^v$ of independence. As such, the sampling distribution of $T_v$ cannot be derived analytically under $H^v_0$ since no parametric model is assumed. This prohibits directly assessing whether the observed value of $T_v$ is unusually large.
As a remedy, one can use a \textit{permutation test} procedure to approximate this sampling distribution. Under $H^v_0$, assume that the survival outcomes $Y_k,k=1,\dots,n$, and hence the log-rank scores $h(Y_k)$, are \textit{exchangeable} with respect to $X_v$ amongst its observations with positive case weights within the current node. That is, any re-ordering of these scores amongst the fixed observations is equally likely to occur under $H^0_v$.
Consequently, one can obtain the conditional distribution of $T_v$ under $H^v_0$ by repeatedly permuting the response values.  
In this context, "permuting the response" means that the log-rank scores $h(Y_k)$ are randomly re-allocated amongst the observations of $X_v$ within the current parent node, whilst keeping fixed the input values $X_{vk}$ and case weights $w_k$. Doing so destroys the association between $Y$ and $X_v$, though it still preserves the marginal distributions of both. By repeatedly recalculating \autoref{eq:linearStatistic_logrank} after each such a permutation, one obtains the permutation distribution $\mathcal{L}(T_v | X_v, \boldsymbol{w})$ of $T_v$ within a parent node, without requiring parametric assumptions.

% Standardised log-rank test statistic
However, it is not necessary to enumerate all permutations of the response relative to the input variable $X_v$, which can quickly become computationally infeasible. In particular, \citet{hothorn2006unbiased} provided expressions of the conditional mean and variance of the linear statistic $T_v$, denoted respectively by $\mu_v$ and $\sigma^2_v$ for the univariate case where $T_v$ is a scalar. Both $\mu_v$ and $\sigma^2_v$ are evaluated under the null hypothesis $H^v_0$ of independence between the survival response $Y$ and the transformed input $g_v(X_v)$. The statistic $T_v$ can therefore be standardised as
\begin{equation} \label{eq:linearStat_Standardised}
    Z_v = \frac{T_v - \mu_v}{ \sigma_v} \, ,
\end{equation}
which is asymptotically normal under $H^v_0$. Such standardisation allows for calculating a corresponding $p$-value, thereby measuring all inputs on the same scale and allowing for unbiased variable selection. Specifically, we calculate the $p$-value of $Z_v$ from \autoref{eq:linearStat_Standardised} under $H^v_0$ as
\begin{equation}
    p_v = 2\Phi (-|Z_v|) \, , \nonumber
\end{equation}
where $\Phi(\cdot)$ is the standard normal cumulative distribution function. Ultimately, larger absolute values of $Z_v$, and hence smaller values of $p_v$, serve as greater evidence against $H^v_0$. The variable $X_{v^*}$ with the smallest $p_{v^*}$ is then selected for splitting in Step 2 of the HHZ-framework. More formally, and given a pre-specified significance level $\alpha$, we have $v^*=\arg\min_{v=1,\dots,m}p_v$ provided that $p_{v^*} \leq \alpha$; otherwise, the algorithm stops and the current node becomes terminal.

\subsubsection{Split point selection within a chosen input variable}

Step 2 in the HHZ-framework from \citet{hothorn2006unbiased} concerns split point estimation for a chosen input variable $X_{v^*}$, whereby its domain $\mathcal{X}_{v^*}$ is partitioned into two disjoint regions. Consider now all admissible subsets $A\subset \mathcal{X}_{v^*}$, where admissibility depends on the variable type. For example, if $X_{v^*}$ is an ordered (numeric) variable, then admissible splits are restricted to threshold-type partitions of the form 
\begin{equation} 
    A=\left\{x\in X_{v^*} : x\leq c \right\} \nonumber
\end{equation}
for some cut-point $c$. 
If $X_{v^*}$ is a categorial variable with $K$ levels, then admissible splits correspond to all non-empty, proper subsets of the category set, i.e., $A\subset \{1,2, \dots, K \}$ with $A\ne \emptyset$ and $A\ne \mathcal{X}_{v^*}$.
Admissibility may also impose further constraints, such as the minimum number of observations within a node. Each candidate subset $A$ induces two non-empty child nodes, corresponding to $A$ and its complement $A^c = \mathcal{X}_{v^*} \setminus A$, which can again be represented via updated case weights as in \autoref{eq:CaseWeights}.

% linear statistic
For each admissible subset $A$, a special case of the linear statistic from \autoref{eq:linearStatistics} is then computed as the weighted sum of the log-rank scores within one of the two induced subgroups, i.e., 
\begin{equation} \label{eq:logRankScores_sum}
    T_{v^*}(A) = \sum_{k=1}^n{ w_k \mathbb{I}\left( X_{v^*k} \in A \right) h(Y_k)} \, ,
\end{equation}
This statistic corresponds to the classical two-sample log-rank statistic that compares two survival distributions; one computed from the observations with $X_{v^*k} \in A$, and one calculated from those in the complement set $A^c$. 
It suffices to calculate the statistic for one subgroup only (instead of calculating it for both groups) since the total weighted sum of log-rank scores sum to zero within the current parent node, i.e.,
\begin{equation}
    T_{v^*}(A)+T_{v^*}(A^c)=0 \, .\nonumber
\end{equation}

% Variable selection
Following the calculation of $T_{v^*}(A)$ from \autoref{eq:logRankScores_sum} for a given split $A$, we proceed again to a permutation test procedural setup, as in Step 1. The aim is to approximate the conditional permutation distribution $\mathcal{L}\left(T_{v^*}(A) | X_{v^*}, \boldsymbol{w} \right)$ of the linear statistic under the null hypothesis $H^{v^*}_0$ of independence between $X_{v^*}$ and the survival response $Y$.
As before, this permutation distribution is again obtained under $H^{v*}_0$ by randomly permuting $Y$ (or equivalently, the log-rank scores $h(Y_k)$) relative to the fixed input values $X_{v^*k}$ and fixed case weights $w_k$ within the current parent node.
\citet{hothorn2006unbiased} provided analytic expressions for estimating the conditional mean and variance of $T_{v^*}(A)$ under $H^{v^*}_0$ for a given split $A$, denoted respectively as the mean $\mu_{v^*}(A)$ and variance $\sigma^2_{v^*}(A)$. The test statistic of each candidate split $A$ is then standardised as
\begin{equation}
    Z_{v^*}(A) = \frac{T_{v^*}(A)-\mu_{v^*}(A)}{\sigma_{v^*}(A)} \, .
\end{equation}
Finally, the optimal subset $A^*$ is chosen as
\begin{equation} \label{eq:bestSubset}
    A^*=\arg\max_A{\left| Z_{v^*}(A) \right|} \, ,
\end{equation}
which is the split that yields the most extreme two-sample log-rank statistic. Put differently, and amongst all admissible binary partitions of $\mathcal{X}_{v^*}$, the algorithm selects the subset $A^*$ that exhibits the largest standardised deviation away from the centre of its conditional permutation distribution.

%Conditional on selecting a covariate $X_k$, the split point is chosen by maximising a two-sample log-rank statistic \citep{segal1988regression} over all admissible binary partitions induced by $X_k$. Consider a candidate split that divides the observations in $\mathcal{N}$ into left and right child nodes. Let $d_{Lt}$ and $d_{Rt}$ denote the number of write-offs at time $t$ in the left and right nodes, and let $Y_{Lt}$ and $Y_{Rt}$ be the corresponding numbers at risk. Defining $d_t=d_{Lt}+d_{Rt}$ and $Y_t=Y_{Lt}+Y_{Rt}$, the log-rank statistic is
%\[
%U = \sum_t \Big(d_{Lt}-Y_{Lt}\frac{d_t}{Y_t}\Big), \qquad
%V = \sum_t \frac{Y_{Lt}Y_{Rt}d_t(Y_t-d_t)}{Y_t^2(Y_t-1)}, \qquad
%Z = \frac{U}{\sqrt{V}},
%\]
%with $Z$ approximately standard normal under the null hypothesis of equal survival functions. The split that maximises $|Z|$ is selected.

\subsubsection{Estimating the survivor function within each terminal node}

Having induced an HHZ-tree as described by \citet{hothorn2006unbiased}, each terminal node $B_l$ defines a region of the input space within which the survival distribution is assumed to be homogeneous. As such, and using the observations within a $B_l$, the conditional survivor function $S_l(t)$ is estimated nonparametrically using
the Kaplan--Meier (KM) estimator. Let 
\begin{equation}
    \mathcal{I}_l = \{ k \in \{1,\dots,n\} : \boldsymbol{X}_k \in B_l \} \nonumber
\end{equation}
denote the index set of observations resorting into node $B_l$. Within each $B_l$, we define $d_{ls}$ as the number of write-off events at ordered failure time $t_{(s)}$ amongst the observations in $B_l$. Similarly, let $n_{ls}$ signify the size of the local risk set just prior to $t_{(s)}$ within $B_l$. The quantities $d_{ls}$ and $n_{ls}$ are then respectively defined as
\begin{align}
    d_{ls} &= \sum_{k\, \in \,\mathcal{I}_l}{ \mathbb{I}\left(T_k=t_{(s)}, \delta_k=1\right) } \, ,
    \nonumber \\
    n_{ls} &= \sum_{k\, \in \,\mathcal{I}_l}{ \mathbb{I}\left(T_k \geq t_{(s)} \right)} \, .\nonumber
\end{align}

% Kaplan-Meier
Now consider estimating the survival probability $\mathbb{P}(T>t\mid \boldsymbol{X} \in B_l)$ within node $B_l$, where $T$ denotes a random variable that represents the non-negative lifetimes of default spells. The node-specific KM-estimator of this survival probability is then specified as
\begin{equation} \label{eq:KaplanMeier_node}
    \hat{S}_l(t)=\prod_{t_{(s)} \, \leq \, t}{\left( 1-\frac{d_{ls}}{n_{ls}} \right)} \, .
\end{equation}
Using \autoref{eq:KaplanMeier_node}, one may then derive the hazard rates $h_l(t)$ and the associated write-off event probabilities $f_l(t)$ within node $B_l$, which are respectively estimated as
\begin{align}
    h_l(t) = 1 - \frac{\hat{S}_l(t)}{\hat{S}_l(t-1)}
    \, , \nonumber \\
    \hat{f}_l(t) = \hat{S}_l(t-1)\hat{h}_l(t)\nonumber \, .
\end{align}

%\subsubsection{Random survival forests.}
%Random survival forests aggregate predictions from many survival trees grown on bootstrap samples and random subsets of covariates to improve predictive performance \citep{ishwaran2008random}. While such ensembles can reduce variance relative to a single tree, they typically do so at the cost of reduced interpretability and increased computational complexity. Since conditional inference survival trees already provide unbiased variable selection and controlled tree complexity, the marginal benefits of further ensembling are likely to be limited in settings where interpretability and structural insight are primary objectives. Ensemble methods may nevertheless be valuable in applications where predictive accuracy is the dominant concern.

\subsubsection{Practical implementation of an HHZ-tree in the \texttt{R}-programming language}

For completeness and reproducibility, we provide an R-based implementation of an HHZ-tree using the \texttt{partykit} R-package from \citet{hothorn2015partykit}. This implementation estimates the HHZ-based survival tree that is described in \autoref{sec:survModels_ST}, using a few dummy input variables. Note however that an HHZ-tree simultaneously evaluates the association between the response and multiple candidate inputs within each node, which implies a multiple hypothesis testing problem. Within the HHZ-framework, this multiplicity is addressed by testing the global null hypothesis of independence between the response and all inputs, though using multiplicity-adjusted $p$-values such as a Bonferroni correction. % or resampling-based min-$p$ procedures.
In the interest of simplicity, our implementation also uses a Bonferroni correction when evaluating the permutation-based $p$-values across inputs. Variable selection proceeds by choosing the input with the smallest adjusted $p$-value, provided that it satisfies the significance threshold implied by \texttt{mincriterion = 0.99}.

\newpage

\begin{lstlisting}
    modSurvTree <- ctree(Surv(DefSpell_Age, DefSpell_Event) ~
                     Balance + pmnt_method + InterestRate_Nom + M_DTI_Growth_6,
                     data=datTrain,
                     control=ctree_control(mincriterion=0.99, minsplit=1000,
                                           minbucket=50, testtype="Bonferroni",
                                           maxdepth=4))
\end{lstlisting}

\subsection{Illustrating the necessary data structure for LGD survival modelling}
\label{app:dataStructure}

Consider the data structure in \autoref{tab:dataStructure_defSpells} of the longitudinal credit dataset $\mathcal{D} = \left\{i, t_i, j, t_{ij}, \tau_d, \tau_s, \mathcal{R}_{ij}^\mathrm{D}, T_{ij}, e_{ijt} \right\}$, as defined in \autoref{sec:survival_concepts}. We identify each row using $(i,j, t_{ij})$ as the composite key across the month-end observations $t_{ij}$ within a particular default spell $(i,j)$.
Loan 1 ($i=1$) had a single default spell that ended in write-off at time $t_i=6$, whereas Loan 2 ($i=2$) became right-censored at $t_i=14$, which presumably coincides with the study-end.
Loan 3 ($i=3$) had two default spells; the first spell cured while the second spell ended in write-off, both having spent two and three months in default respectively. Loan 4 ($i=4$) had a delayed entry (i.e., it was left-truncated) such that observation only started at month $t_i=13$, which is why its default entry-time is duly adjusted -- assuming that it was in still default prior. It cured $T_{ij}=3$ months later at $t_i=15$, followed by two successive default spells; the last of which became right-censored at time $t_i=41$. Note that these examples ignore the imposition of any probation period within the default definition, simply in the interest of brevity.

\begin{table}[ht!]
\centering
\caption{Illustrating the structure of the raw panel dataset $\mathcal{D}$, filtered for default spells. The alternating grey-shaded rows indicate loan-level history, while the alternating colour-shaded cells signify different default spell-level histories of each loan; the remaining unshaded cells denote period-level information. Inspired by \citet{botha2025recurrentEvents}.}
\begin{tabular}{p{1cm} p{1.1cm} p{1.6cm} p{1.3cm} p{1.2cm} p{1.2cm} p{1.9cm} p{1.3cm} p{1.3cm}}
\toprule
\textbf{Loan} $i$ & \textbf{Period} $t_i$ & \textbf{Spell number} $j$ & \textbf{Spell period} $t_{ij}$ & \textbf{Default time} $\tau_d$ & \textbf{Stop time} $\tau_s$ & \textbf{Resolution type} $\mathcal{R}_{ij}^\mathrm{D}$ & \textbf{Spell age} $T_{ij}$ & \textbf{Event} $e_{ij}$ \\ \midrule
\cellcolor[HTML]{EFEFEF}1 & 5 & \cellcolor[HTML]{ECF4FF}1 & 1 & \cellcolor[HTML]{ECF4FF}0 & \cellcolor[HTML]{ECF4FF}2 & \cellcolor[HTML]{ECF4FF}1: Write-off & \cellcolor[HTML]{ECF4FF}2 & \cellcolor[HTML]{ECF4FF} 0 \\
\cellcolor[HTML]{EFEFEF}1 & 6 & \cellcolor[HTML]{ECF4FF}1 & 2 & \cellcolor[HTML]{ECF4FF}0 & \cellcolor[HTML]{ECF4FF}2 & \cellcolor[HTML]{ECF4FF}1: Write-off & \cellcolor[HTML]{ECF4FF}2 & \cellcolor[HTML]{ECF4FF} 1 \\
\cellcolor[HTML]{C0C0C0}2 & 12 & \cellcolor[HTML]{C0DAFE}1 & 1 & \cellcolor[HTML]{C0DAFE}0 & \cellcolor[HTML]{C0DAFE}3 & \cellcolor[HTML]{C0DAFE}3: Censored & \cellcolor[HTML]{C0DAFE}3 & \cellcolor[HTML]{C0DAFE} 0 \\
\cellcolor[HTML]{C0C0C0}2 & 13 & \cellcolor[HTML]{C0DAFE}1 & 2 & \cellcolor[HTML]{C0DAFE}0 & \cellcolor[HTML]{C0DAFE}3 & \cellcolor[HTML]{C0DAFE}3: Censored & \cellcolor[HTML]{C0DAFE}3 & \cellcolor[HTML]{C0DAFE} 0 \\
\cellcolor[HTML]{C0C0C0}2 & 14 & \cellcolor[HTML]{C0DAFE}1 & 3 & \cellcolor[HTML]{C0DAFE}0 & \cellcolor[HTML]{C0DAFE}3 & \cellcolor[HTML]{C0DAFE}3: Censored & \cellcolor[HTML]{C0DAFE}3 & \cellcolor[HTML]{C0DAFE} 0 \\
\cellcolor[HTML]{EFEFEF}3 & 6 & \cellcolor[HTML]{ECF4FF}1 & 1 & \cellcolor[HTML]{ECF4FF}0 & \cellcolor[HTML]{ECF4FF}2 & \cellcolor[HTML]{ECF4FF}2: Cured & \cellcolor[HTML]{ECF4FF}2 & \cellcolor[HTML]{ECF4FF} 0 \\
\cellcolor[HTML]{EFEFEF}3 & 7 & \cellcolor[HTML]{ECF4FF}1 & 2 & \cellcolor[HTML]{ECF4FF}0 & \cellcolor[HTML]{ECF4FF}2 & \cellcolor[HTML]{ECF4FF}2: Cured & \cellcolor[HTML]{ECF4FF}2 & \cellcolor[HTML]{ECF4FF} 0 \\
\cellcolor[HTML]{EFEFEF}3 & 24 & \cellcolor[HTML]{E6FFE6}2 & 1 & \cellcolor[HTML]{E6FFE6}0 & \cellcolor[HTML]{E6FFE6}3 & \cellcolor[HTML]{E6FFE6}1: Write-off & \cellcolor[HTML]{E6FFE6}3 & \cellcolor[HTML]{E6FFE6} 0 \\
\cellcolor[HTML]{EFEFEF}3 & 25 & \cellcolor[HTML]{E6FFE6}2 & 2 & \cellcolor[HTML]{E6FFE6}0 & \cellcolor[HTML]{E6FFE6}3 & \cellcolor[HTML]{E6FFE6}1: Write-off & \cellcolor[HTML]{E6FFE6}3 & \cellcolor[HTML]{E6FFE6} 0 \\
\cellcolor[HTML]{EFEFEF}3 & 26 & \cellcolor[HTML]{E6FFE6}2 & 3 & \cellcolor[HTML]{E6FFE6}0 & \cellcolor[HTML]{E6FFE6}3 & \cellcolor[HTML]{E6FFE6}1: Write-off & \cellcolor[HTML]{E6FFE6}3 & \cellcolor[HTML]{E6FFE6} 1 \\
\cellcolor[HTML]{C0C0C0}4 & 13 & \cellcolor[HTML]{C0DAFE}1 & 13 & \cellcolor[HTML]{C0DAFE}12 & \cellcolor[HTML]{C0DAFE}15 & \cellcolor[HTML]{C0DAFE}2: Cured & \cellcolor[HTML]{C0DAFE}3 & \cellcolor[HTML]{C0DAFE} 0 \\
\cellcolor[HTML]{C0C0C0}4 & 14 & \cellcolor[HTML]{C0DAFE}1 & 14 & \cellcolor[HTML]{C0DAFE}12 & \cellcolor[HTML]{C0DAFE}15 & \cellcolor[HTML]{C0DAFE}2: Cured & \cellcolor[HTML]{C0DAFE}3 & \cellcolor[HTML]{C0DAFE} 0 \\
\cellcolor[HTML]{C0C0C0}4 & 15 & \cellcolor[HTML]{C0DAFE}1 & 15 & \cellcolor[HTML]{C0DAFE}12 & \cellcolor[HTML]{C0DAFE}15 & \cellcolor[HTML]{C0DAFE}2: Cured & \cellcolor[HTML]{C0DAFE}3 & \cellcolor[HTML]{C0DAFE} 0 \\
\cellcolor[HTML]{C0C0C0}4 & 24 & \cellcolor[HTML]{B5FFB5}2 & 1 & \cellcolor[HTML]{B5FFB5}0& \cellcolor[HTML]{B5FFB5}4 & \cellcolor[HTML]{B5FFB5}2: Cured & \cellcolor[HTML]{B5FFB5}4 & \cellcolor[HTML]{B5FFB5} 0 \\
\cellcolor[HTML]{C0C0C0}4 & 25 & \cellcolor[HTML]{B5FFB5}2 & 2 & \cellcolor[HTML]{B5FFB5}0 & \cellcolor[HTML]{B5FFB5}4 & \cellcolor[HTML]{B5FFB5}2: Cured & \cellcolor[HTML]{B5FFB5}4 & \cellcolor[HTML]{B5FFB5} 0 \\
\cellcolor[HTML]{C0C0C0}4 & 26 & \cellcolor[HTML]{B5FFB5}2 & 3 & \cellcolor[HTML]{B5FFB5}0 & \cellcolor[HTML]{B5FFB5}4 & \cellcolor[HTML]{B5FFB5}2: Cured & \cellcolor[HTML]{B5FFB5}4 & \cellcolor[HTML]{B5FFB5} 0 \\
\cellcolor[HTML]{C0C0C0}4 & 27 & \cellcolor[HTML]{B5FFB5}2 & 4 & \cellcolor[HTML]{B5FFB5}0 & \cellcolor[HTML]{B5FFB5}4 & \cellcolor[HTML]{B5FFB5}2: Cured & \cellcolor[HTML]{B5FFB5}4 & \cellcolor[HTML]{B5FFB5} 0 \\
\cellcolor[HTML]{C0C0C0}4 & 40 & \cellcolor[HTML]{FFE1BD}3 & 1 & \cellcolor[HTML]{FFE1BD}0 & \cellcolor[HTML]{FFE1BD}2 & \cellcolor[HTML]{FFE1BD}3: Censored & \cellcolor[HTML]{FFE1BD}2 & \cellcolor[HTML]{FFE1BD} 0 \\
\cellcolor[HTML]{C0C0C0}4 & 41 & \cellcolor[HTML]{FFE1BD}3 & 2 & \cellcolor[HTML]{FFE1BD}0 & \cellcolor[HTML]{FFE1BD}2 & \cellcolor[HTML]{FFE1BD}3: Censored & \cellcolor[HTML]{FFE1BD}2 & \cellcolor[HTML]{FFE1BD} 0 \\ \bottomrule
\end{tabular}
\label{tab:dataStructure_defSpells}
\end{table}
\subsection{A diagnostic measure for testing the representativeness of resampled LGD datasets}
\label{app:resolutionRate}

% Resolution rate
As discussed by \citet{botha2025discTimeSurvTutorial}, the training and validation sets $\left\{\mathcal{D}_T, \mathcal{D}_V \right\}$ should not exhibit undue sampling bias over time. We can measure this bias using the \textit{resolution rate}, which can be calculated for either set and compared to that of the panel dataset $\mathcal{D}$.
Consider now that $\mathcal{D}$ can be partitioned into a series of non-overlapping monthly spell cohorts $\mathcal{D}(t')$ over reporting time $t'\in\{t_1',\dots,t_l',\dots,t_n'\}$, e.g., Jan-2008 to Dec-2022. Let $r_\psi\left(t'_l,\mathcal{D}\right)$ denote the resolution rate at which spells resolve at any time $t'_l$ into a specified type $\psi$ within a given dataset $\mathcal{D}$. The type $\psi\in\{1,2,3\}$ refers respectively to write-off (1), cured (2), or censored (3) outcomes. Suppose that $n_{t'}$ denotes the size of $\mathcal{D}(t')$ at $t'$. Now suppose that $\mathcal{D}(t')$ contains all spells that commonly stop  at $t'$, i.e., the so-called `cohort-end'-definition from \citet{botha2025recurrentEvents}, which resolves into the resolution date. We then define the resolution rate of type $\psi$ at each $t'$ as 
\begin{equation} \label{eq:ResolRate}
    r_\psi^\mathrm{D}(t',\mathcal{D}) = \frac{1}{n_{t'}} \sum_{(i,j) \, \in \, \mathcal{D}(t')} \mathbb{I}(\mathcal{R}_{ij}^\mathrm{D} = \psi) \quad \forall \  \mathcal{D}(t') \subset \mathcal{D} \ \text{and for} \ \psi \in \mathcal{R}^\mathrm{D} \, ,
\end{equation}
where $\mathbb{I}(\cdot)$ is an indicator function, and $\mathcal{R}^\mathcal{D}$ is a nominal-valued random variable with realisations $\mathcal{R}_{ij}^\mathrm{D},i=1,\dots,N_d,j=1,\dots,n_i$. This resolution rate is indeed similar to the one introduced by \citet{botha2025discTimeSurvTutorial} for performing spells.

\begin{figure}[ht!]
    \centering
    \includegraphics[width=0.8\linewidth, height=0.47\textheight]{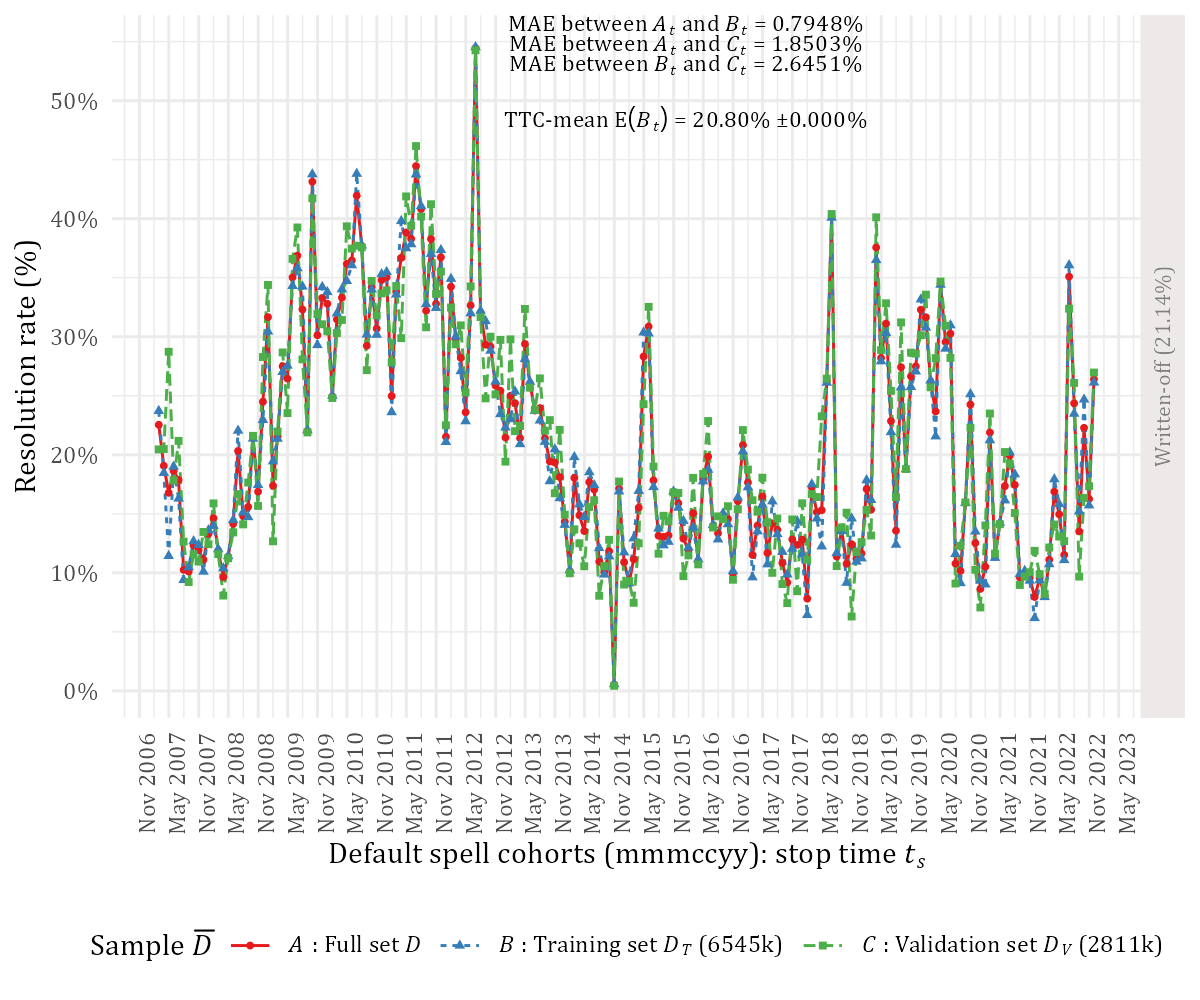}
    \caption{Comparing the resolution rates of type $\psi=1$ (Write-off) over time across the various datasets, having used residential mortgage data. The MAE-based AD-measure from \autoref{eq:ResolRate_MAE} summarises the discrepancies over time for each dataset-pair. Inspired by \citet{botha2025discTimeSurvTutorial}.}
    \label{fig:ResolRate_WOff}
\end{figure}

% Resolutation rate: illustration
We can now check the sets $\left\{\mathcal{D}_T, \mathcal{D}_V \right\}$ for time-dependent sampling bias using \autoref{eq:ResolRate}. In particular, the resolution rates $r_\psi^\mathrm{D}(t',\mathcal{D})$, $r_\psi^\mathrm{D}(t',\mathcal{D}_T)$, and $r_\psi^\mathrm{D}(t',\mathcal{D}_V)$ are duly calculated and compared over $t'$ towards screening for large discrepancies. This exercise is formalised by using the MAE-based \textit{average discrepancy} (AD) measure as a diagnostic from \citet{botha2025discTimeSurvTutorial}, which is expressed between any two non-overlapping sets $\mathcal{D}_1$ and $\mathcal{D}_2$ as
\begin{equation} \label{eq:ResolRate_MAE}
     \text{AD: } \quad \bar{r}_\psi^\mathrm{D} \left(\mathcal{D}_1, \mathcal{D}_2 \right) = \frac{1}{n} \sum_{t'}{\big\vert r_\psi^\mathrm{D}(t',\mathcal{D}_1) - r_\psi^\mathrm{D}(t',\mathcal{D}_2) \big\vert} \quad \forall \ t' \ \text{and for} \ \psi \in \mathcal{R}^\mathrm{D}\, .
\end{equation}
Smaller AD-values signify greater representativeness between two subsampled sets. One can calculate the AD-measure for various combinations of our datasets, which would include $\bar{r}_\psi^\mathrm{D}(\mathcal{D},\mathcal{D}_T)$, $\bar{r}_\psi^\mathrm{D}(\mathcal{D},\mathcal{D}_V)$, and $\bar{r}_\psi^\mathrm{D}(\mathcal{D}_T,\mathcal{D}_V)$.
% Illustration
We demonstrate and compare the write-off resolution rate ($\psi=1$) in \autoref{fig:ResolRate_WOff} across different samples of data. These rates clearly track major macroeconomic phenomena such as the 2008 financial crisis and the Covid-19 pandemic, during which times the resolution rates spiked dramatically. For the 2008 crisis, there is clearly a delayed effect in the resolution rate, which is simply a function of the lengthy workout process of defaults. The last spike in the resolution rates during the 2022-2023 period is ascribed to the increases in the policy rate, as set by the central bank in response to rising inflation rates at the time. This effect caused instalments to become unaffordable for many borrowers, thereby inducing default and straining collection efforts.
% Representativeness
Nevertheless, one can observe that all rates are reasonably close to another, with an AD-value of about 0.8\% between $\mathcal{D}$ and $\mathcal{D}_T$. Based on these results, we consider the resampled sets as fairly representative of $\mathcal{D}$, which bodes well for the generalisation ability of the eventual models, beyond the training data.

\subsection{A short procedure for optimising the cost multiple \texorpdfstring{$a$}{L} within the Generalised Youden Index using the MAE}
\label{app:CostMultipleOptim}

In dichotomising a given probability score $w(t,\boldsymbol{x}_i)$ into a 0/1-decision, consider the indicator function $\mathbb{I}\left( w(t,\boldsymbol{x}_i) > c^* \right)$ that outputs 1 if $w(t,\boldsymbol{x}_i) > c^*$, and 0 otherwise. The optimal cut-off $c^*$ is obtained by finding the maximum value of the Generalised Youden index $J_a(c)$ from \autoref{eq:gen_Youden_Index} across a range of possible cut-off values $c\in[0,1]$, given the entire set of probability scores across $i=1,\dots,N$. Having dichotomised the underlying probabilistic model using $c^*$, we calculate the empirical and expected dichotomised term-structures. The empirical term-structure is estimated as the collection of the event rates $f\left(t \right)$ across failure times $t=t_{(1)},\dots, t_{(m)}$, where $f(t)$ is itself estimated using Kaplan-Meier analysis; see \autoref{sec:survFundamentals}. The expected term-structure is found by first obtaining  the hazard-predictions $\hat{h}(t, \boldsymbol{x}_i) $ from a fitted model given a set of input variables $\boldsymbol{x}_i=\left\{ \boldsymbol{E}_{ij}, \boldsymbol{x}_{ij}, \boldsymbol{x}_{ij}(t), \boldsymbol{x}(t) \right\}$, as explained in \autoref{sec:survModels}. These hazard-predictions are then related to the corresponding event rates $f(t,\boldsymbol{x}_i)=\hat{S}(t-1,\boldsymbol{x}_i) \cdot \hat{h}(t,\boldsymbol{x}_i)$, where $\hat{S}(t,\boldsymbol{x}_i)$ is the predicted account-level survival probability from \autoref{eq:KaplanMeier}.

% Expected term-structure
The expected term-structure is then the aggregated form of these event rates $f(t,\boldsymbol{x}_i)$, denoted by $f_\mathrm{E}(t)$ and expressed using the mean as 
\begin{equation} \label{eq:term_structure_exp}
    f_\mathrm{E}(t) = \frac{1}{\left|\mathcal{D}_t\right|} \sum_{i \, \in \, \mathcal{D}_t} { f(t,\boldsymbol{x}_i) } \quad \text{for} \ t=t_{(1)},\dots,t_{(m)} \, ,
\end{equation}
where each $\mathcal{D}_t$ is a set that contains those subject-spells that have survived up to $t$. Similarly, the dichotomised version of the expected term-structure given $c^*$ is
\begin{equation} \label{eq:term_structure_exp_dich}
    f_\mathrm{E}^*(t,c^*) = \frac{1}{\left|\mathcal{D}_t\right|} \sum_{i \, \in \, \mathcal{D}_t} { \mathbb{I}\left( f(t,\boldsymbol{x}_i) > c^* \right)} \quad \text{for} \ t=t_{(1)},\dots,t_{(m)} \,.
\end{equation}
Between each collection $\left\{ f\left(t_{(k)} \right) \right\}_{k=1}^{m}$ and $\left\{ f_\mathrm{E}^*\left(t_{(k)}, c^* \right) \right\}_{k=1}^{m}$, we calculate the MAE given a particular cost multiple $a$ and resulting $c^*$ as
\begin{equation} \label{eq:MAE_termStructures}
    \mathrm{MAE}(a) = \frac{1}{m} \sum_{k=1}^m{ \left|  f\left(t_{(k)} \right) - f_\mathrm{E}^*\left(t_{(k)}, c^* \right) \right|} \,.
\end{equation}
where $c^*=\arg \min_c{J_a(c)}$. \autoref{eq:MAE_termStructures} is then repeatedly calculated using a range of $a$-values, thereby resulting in a collection of corresponding MAE-values. The minimum hereof should indicate the chosen $a$-value, thereby concluding the optimisation procedure.

\begin{figure}[!ht]
    \centering
    \includegraphics[width=0.7\linewidth,height=0.4\textheight]{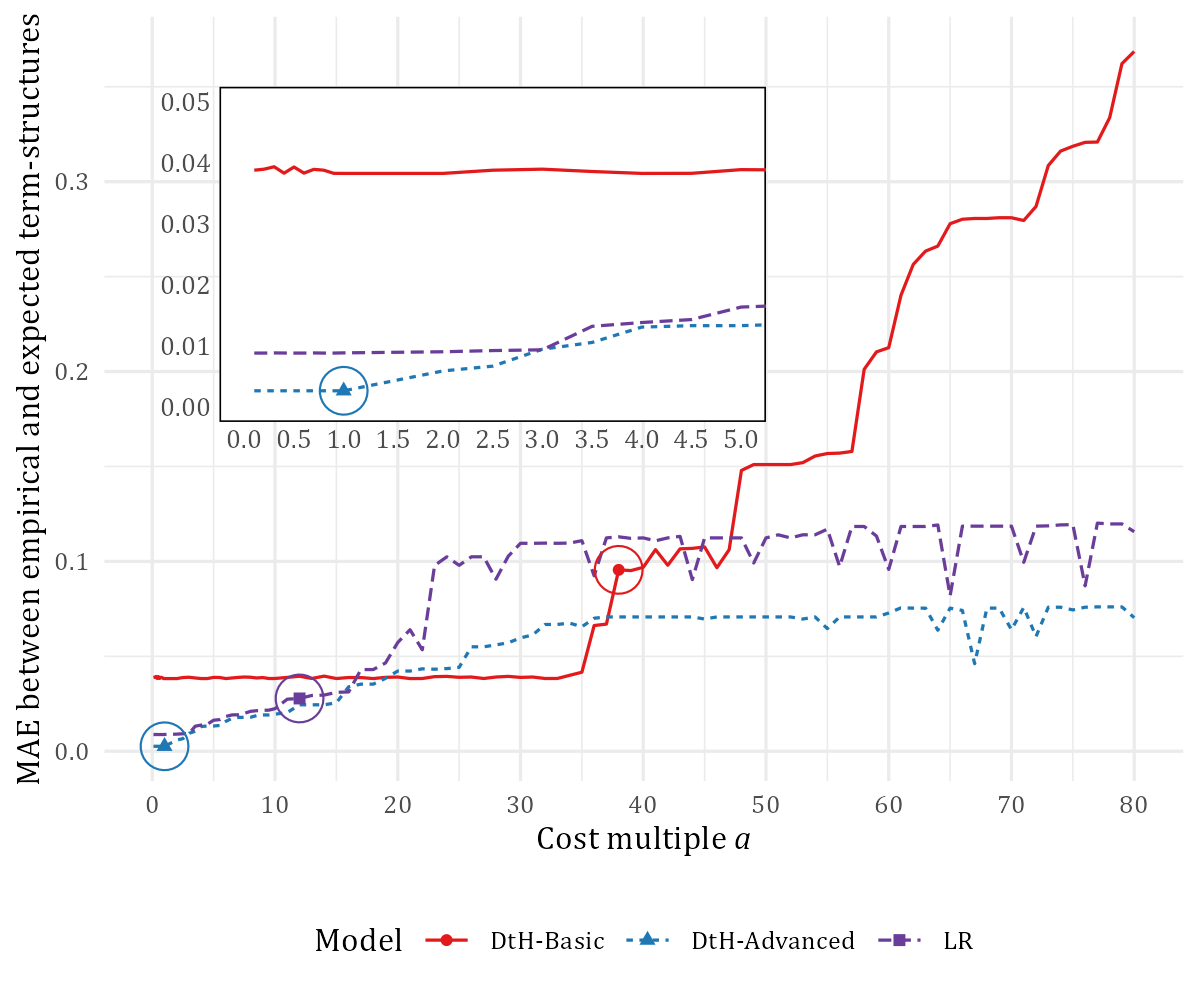}
    \caption{Optimisation results of the cost multiple $a$ in minimising the MAE between the empirical and expected term-structure. The latter emanates from a dichotomised model, having imposed a particular cut-off $c^*$; itself obtained from calculating the Generalised Youden index $J_a$ given $a$. The encircled points indicate chosen $a$-values.}
    \label{fig:CostMultiple_Optima}
\end{figure}

% Optimisation rsults
We show the results of the aforementioned optimisation procedure in \autoref{fig:CostMultiple_Optima}. For computational expediency, limits are imposed on the search space of $c$ for each model during the optimisation. The upper bound hereof is found following a distributional analysis on the underlying event rates $f(t,\boldsymbol{x}_i), i=1,\dots,N$ per model, whereafter the upper bound is chosen such that it equals the 99\% quantile of event rates. Accordingly, these limits are $c\in[0,0.025]$ for the DtH-Basic model, $c\in[0,0.3]$ for the DtH-Advanced model, and $c\in[0,0.4]$ for the LR-model. We shall now discuss the optimisation results for each model, and present the corresponding $c^*$-threshold.

% General
The MAE generally increases as $a$ increases in \autoref{fig:CostMultiple_Optima}, whilst the corresponding $c^*$-value decreases as $a$ increases.
% DtH-Basic model + LR models: problem
Instead of being a single point, the minimum MAE-values present across a range of $a$-values for both the DtH-Basic and LR models. However, the resulting term-structures are unrealistic when imposing the corresponding $c^*$-values at any of these $a$-values. In particular, these term-structures appear as flat zero-valued lines over most of $t$. This result implies that all of the model predictions are zero-valued, and that the corresponding $c^*$-threshold is too high; despite yielding a minimum MAE-value in the downstream term-structure diagnostic. Moreover, such severe under-prediction of the event probability would not be in the interest of risk prudence. 
% DtH-Basic model + LR models: solution
As a remedy, we chose the lowest $a$-value (and corresponding $c^*$-threshold) that started to yield a more credible term-structure. In this rather manual process, we used expert judgment in adjusting the optimisation results, such that the resulting term-structure will at least over-predict the event probability across most of $t$. Accordingly, this process yielded $a=38$ for the DtH-Basic model with a corresponding threshold of $c^*=0.0152$; and $a=12$ for the LR-model with an associated threshold of $c^*=0.0651$. Credible term-structures, as shown in \autoref{fig:TermStructures_b}, are obtained using both of these $c^*$-choices.
% DtH-Advanced model: results
For the DtH-Advanced model, the minimum MAE is achieved at $a=1$ and the corresponding threshold is $c^*=0.295$. This $c^*$-choice also yielded a realistic term-structure, as shown in \autoref{fig:TermStructures_a}, and no special adjustment is required.
\subsection{A description of selected input variables within each LGD-model}
\label{app:InputSpace}

In \autoref{tab:featuresDescription}, the selected input variables of the finalised LGD-models are described. This description includes a mapping between variables and the specific LGD-model, whilst relegating the fitting procedure and its diagnostics to the codebase maintained by \citet{gabru2026WriteOffSurvSourcecode}, purely in the interest of brevity.

\begin{longtable}{p{3.7cm} p{9.3cm} p{2cm}}
\caption{The selected input variables mapped across the various LGD-models. Write-off risk models include the single-stage Gaussian-GLM (1s-Gaussian), single-stage Compound-Poisson-GLM (1s-CP), two-stage logistic regression (2s-LR), two-stage DtH-basic (2s-DtH-Bas), two-stage DtH-advanced (2s-DtH-Adv), and the two-stage survival tree (2s-ST). The loss severity is modelled with a Compound-Poisson-GLM (LS-CP).
Subscripts $[\mathrm{a}]$ denote loan account-level variables, $[\mathrm{p}]$ are portfolio-level inputs, and $[\mathrm{m}]$ represent macroeconomic covariates.}
\label{tab:featuresDescription} \\
\toprule
\textbf{Variable} & \textbf{Description} & \textbf{Models} \\ 
\midrule
\endfirsthead
\caption[]{(continued)} \\
\toprule
\textbf{Variable} & \textbf{Description} & \textbf{Models} \\ 
\midrule
\endhead
\midrule \multicolumn{3}{r}{\textit{Continued on next page}} \\
\endfoot
\bottomrule
\endlastfoot
\footnotesize{\texttt{AgeToTerm\_Avg}$_{[\mathrm{p}]}$}  & \footnotesize{Mean value across the portfolio of the ratio between a loan's age and its term.} & \footnotesize{1s-Gaussian; LS-CP} \\
\footnotesize{\texttt{AgeToTerm}$_{[\mathrm{a}]}$}  & \footnotesize{Ratio between a loan's age and its term.} & \footnotesize{1s-CP} \\
\footnotesize{\texttt{ArrearsDir\_3}$_{[\mathrm{a}]}$}  & \footnotesize{The trending direction of the arrears direction, qualitatively obtained by comparing the current arrears-level to that of 3 months ago, binned as: 1) increasing; 2) milling; 3) decreasing (reference); and 4) missing.} & \footnotesize{LS-CP} \\
\footnotesize{\texttt{Arrears\_Med}$_{[\mathrm{p}]}$} & \footnotesize{Median-imputed amount in arrears.} & \footnotesize{1s-Gaussian; 1s-CP} \\
\footnotesize{\texttt{ArrearsToBal\_Avg\_1}$_{[\mathrm{p}]}$} & \footnotesize{Mean value across the portfolio of the ratio between the arrears amount and the outstanding balance, lagged by 1 month.} & \footnotesize{1s-CP; 2s-LR} \\
\footnotesize{\texttt{Balance\_Real\_1}$_{[\mathrm{a}]}$} & \footnotesize{Inflation-adjusted outstanding balance of the loan, lagged by 1 month} &  \footnotesize{1s-Gaussian; 1s-CP; 2s-LR; 2s-DtH-Adv; 2s-ST; LS-CP}  \\
\footnotesize{\texttt{DefSpell\_Age}$_{[\mathrm{a}]}$} & \footnotesize{Default spell age (months).} &  \footnotesize{1s-Gaussian; 1s-CP; 2s-LR} \\
\footnotesize{\texttt{DefSpell\_Age\_Mean}$_{[\mathrm{a}]}$} & \footnotesize{Mean default spell age (months) across the portfolio at the time.} &  \footnotesize{1s-Gaussian;} \\
\footnotesize{\texttt{DefaultStatus\_Avg}$_{[\mathrm{p}]}$} & \footnotesize{Fraction of the portfolio in default.} &  \footnotesize{1s-Gaussian} \\
\footnotesize{\texttt{DefaultStatus\_Avg\_12}$_{[\mathrm{p}]}$} & \footnotesize{12-month lagged version of \texttt{DefaultStatus\_Avg}.} &  \footnotesize{1s-CP; 2s-LR; 2s-DtH-Adv; 2s-ST} \\
\footnotesize{\texttt{g0\_Delinq\_1}$_{[\mathrm{a}]}$}  & \footnotesize{Delinquency-level (1-month lag), or the number of payments in arrears as measured by the $g_0$-measure; see the $g_0$-measure from \citet{botha2021paper1}.} & \footnotesize{2s-DtH-Bas; 2s-DtH-Adv} \\
\footnotesize{\texttt{g0\_Delinq\_Avg}$_{[\mathrm{p}]}$}  & \footnotesize{Non-defaulted average delinquency $g_0$ across the portfolio at the time.} & \footnotesize{1s-CP; 2s-LR; 2s-DtH-Adv; 2s-ST}\\
\footnotesize{\texttt{g0\_Delinq\_Any\_Avg\_1}$_{[\mathrm{p}]}$}  & \footnotesize{Non-defaulted fraction of the portfolio with any degree of delinquency beyond $g_0=0$, lagged by 1 month.} & \footnotesize{2s-DtH-Bas } \\
\footnotesize{\texttt{g0\_Delinq\_Any\_Avg\_12}$_{[\mathrm{p}]}$}  & \footnotesize{12-month lagged version of \texttt{g0\_Delinq\_Any\_Avg}.} & \footnotesize{1s-Gaussian } \\
\footnotesize{\texttt{g0\_Delinq\_Num}$_{[\mathrm{a}]}$} & \footnotesize{Number of times that the $g_0$-measure changed in value over loan life so far.} &  \footnotesize{1s-Gaussian; 1s-CP; 2s-LR} \\
\footnotesize{\texttt{g0\_Delinq\_SD\_6}$_{[\mathrm{a}]}$} & \footnotesize{The sample standard deviation of the $g_0$-measure [\texttt{g0\_Delinq}] over a rolling 6-month window.} &  \footnotesize{1s-Gaussian; LS-CP } \\
\footnotesize{\texttt{Instalment\_Real}$_{[\mathrm{a}]}$} & \footnotesize{Inflation-adjusted expected instalment of the loan.} &  \footnotesize{LS-CP}  \\
\footnotesize{\texttt{InterestRate\_Nominal}$_{[\mathrm{a}]}$} & \footnotesize{Nominal interest rate per annum of a loan.} & \footnotesize{1s-Gaussian; 1s-CP; 2s-LR; 2s-DtH-Adv} \\
\footnotesize{\texttt{InterestRate\_Mar\_Med\_2}$_{[\mathrm{p}]}$} & \footnotesize{Median value across the portfolio of the nominal interest rates of loans at the time, lagged by 2 months.} & \footnotesize{2s-LR; 2s-DtH-Adv; 2s-ST} \\
\footnotesize{\texttt{M\_DebtToIncome\_3}$_{[\mathrm{m}]}$}  &\footnotesize{Debt-to-Income: Average household debt expressed as a percentage of household income per quarter, interpolated monthly, lagged by 3 months.} & \footnotesize{LS-CP} \\
\footnotesize{\texttt{M\_DebtToIncome\_6}$_{[\mathrm{m}]}$} &\footnotesize{6-month lagged version of DTI growth rate, \texttt{M\_DebtToIncome}.} & \footnotesize{2s-LR; 2s-ST} \\
\footnotesize{\texttt{M\_DebtToIncome\_12}$_{[\mathrm{m}]}$} &\footnotesize{12-month lagged version of DTI growth rate, \texttt{M\_DebtToIncome}.} & \footnotesize{2s-DtH-Adv} \\
\footnotesize{\texttt{M\_Inflation\_Growth\_3}$_{[\mathrm{m}]}$} & \footnotesize{Year-on-year growth rate in inflation index (\textit{Consumer Price Index} [CPI]) per month, lagged by 3-months.} & \footnotesize{2s-LR} \\
\footnotesize{\texttt{M\_Inflation\_Growth\_9}$_{[\mathrm{m}]}$} & \footnotesize{9-month lagged version of the CPI growth rate, \texttt{M\_Inflation\_Growth}.} & \footnotesize{2s-DtH-Bas} \\
\footnotesize{\texttt{M\_Inflation\_Growth\_12}$_{[\mathrm{m}]}$} & \footnotesize{12-month lagged version of the CPI growth rate, \texttt{M\_Inflation\_Growth}.} & \footnotesize{1s-Gaussian; 2s-DtH-Adv} \\
\footnotesize{\texttt{M\_RealGDP\_Growth}$_{[\mathrm{m}]}$}  & \footnotesize{Year-on-year growth rate in the 4-quarter moving average of real GDP per quarter, interpolated monthly, lagged by 12 months.} & \footnotesize{1s-CP; 2s-LR} \\
\footnotesize{\texttt{M\_RealIncome\_Growth\_9}$_{[\mathrm{m}]}$}  & \footnotesize{Year-on-year growth rate in the 4-quarter moving average of real income per quarter, interpolated monthly, lagged by 9 months.} & \footnotesize{1s-CP; 2s-DtH-Adv} \\
\footnotesize{\texttt{M\_RealIncome\_Growth\_12}$_{[\mathrm{m}]}$}  & \footnotesize{12-month lagged version of real income growth, \texttt{M\_RealIncome\_Growth}.} & \footnotesize{LS-CP} \\
\footnotesize{\texttt{M\_Repo\_Rate\_2}$_{[\mathrm{m}]}$} & \footnotesize{Prevailing repurchase (or policy) rate set by the South African Reserve Bank (SARB), lagged by 2 months.} & \footnotesize{2s-LR} \\
\footnotesize{\texttt{M\_Repo\_Rate\_6}$_{[\mathrm{m}]}$} & \footnotesize{6-month lagged version of \texttt{M\_Repo\_Rate}} & \footnotesize{1s-Gaussian} \\
\footnotesize{\texttt{M\_Repo\_Rate\_9}$_{[\mathrm{m}]}$} & \footnotesize{9-month lagged version of \texttt{M\_Repo\_Rate}.} & \footnotesize{1s-CP} \\
\footnotesize{\texttt{M\_Repo\_Rate\_12}$_{[\mathrm{m}]}$} & \footnotesize{12-month lagged version of \texttt{M\_Repo\_Rate}.} & \footnotesize{2s-DtH-Adv} \\
\footnotesize{\texttt{NewLoans\_Pc}$_{[\mathrm{a}]}$} & \footnotesize{Fraction of the portfolio that constitutes new loans.} & \footnotesize{1s-CP; 2s-DtH-Adv; LS-CP} \\
\footnotesize{\texttt{PayMethod}$_{[\mathrm{a}]}$} & \footnotesize{A categorical variable designating different payment methods: 1) debit order (reference); 2) salary; 3) payroll or cash; and 4) missing.} & \footnotesize{2s-LR; 2s-DtH-Adv; 2s-ST; LS-CP} \\
\footnotesize{\texttt{Principal\_Real}$_{[\mathrm{a}]}$} & \footnotesize{Inflation-adjusted principal loan amount.} &  \footnotesize{2s-LR; 2s-Dth-Adv; LS-CP} \\
\footnotesize{\texttt{PrevDefaults}$_{[\mathrm{a}]}$} & \footnotesize{Indicating whether the loan experienced previous default spells.} & \footnotesize{1s-Gaussian; 1s-CP; 2s-LR; 2s-ST; LS-CP} \\
\footnotesize{\texttt{SpellNum\_Bn}$_{[\mathrm{a}]}$} & \footnotesize{The current default spell number, or total number of visits to the default state over loan life, binned as ("1", "2", "3", "4+") spells.} & \footnotesize{1s-CP; 2s-LR; 2s-DtH-Bas; 2s-DtH-Adv; 2s-ST} \\ 
\footnotesize{\texttt{TimeSpell}$_{[\mathrm{a}]}$} & \footnotesize{Logarithm of the time spent in a default spell towards embedding the baseline hazard.} & \footnotesize{2s-DtH-Bas} \\
\footnotesize{\texttt{TimeSpell\_Bn}$_{[\mathrm{a}]}$} & \footnotesize{Binned version of the time spent in a default spell.} & \footnotesize{2s-DtH-Adv} \\ 
\footnotesize{\texttt{TimeSpell*SpellNum\_Bn}$_{[\mathrm{a}]}$} & \footnotesize{An interaction effect between the logarithm of the time spent in a default spell, and \texttt{SpellNum\_Bn}.} & \footnotesize{2s-DtH-Adv} \\
\end{longtable}

%--------------------------------------------------------%
%	REFERENCE LIST
%--------------------------------------------------------%

%TC:ignore

\singlespacing
\printbibliography % using biblatex
%\section*{References}
%\bibliographystyle{newapa}
% see http://texdoc.net/texmf-dist/doc/latex/natbib/natbib.pdf for more styles
%\bibliography{bibliography} 
\onehalfspacing

%TC:endignore

%--------------------------------------------------------%
%	END OF DOCUMENT
%--------------------------------------------------------%

\end{document}